\documentclass[universe,article,accept,pdftex,oneauthor]{Definitions/mdpi} 
\usepackage{graphicx,pstricks,epsfig,rotating,amsmath,amssymb}
\usepackage{soulutf8}
\firstpage{1} 
\pubvolume{1}
\issuenum{1}
\articlenumber{0}
\pubyear{2025}
\copyrightyear{2025}
\externaleditor{Firstname Lastname}
\datereceived{31 March 2025} 
\daterevised{26 June 2025 } 
\dateaccepted{ }                                
\datepublished{ } 
\hreflink{https://doi.org/} 

\Title{Properties of the Object HESS J1731-347 as a Twin Compact Star}

\TitleCitation{Properties of the Object HESS J1731-347 as a Twin Compact Star}

\Author{David E. Alvarez-Castillo $^{1,2,3}$\orcidA{}}

\AuthorNames{David E. Alvarez-Castillo}

\AuthorCitation{Alvarez-Castillo, D.E.}

\address{%
$^{1}$ \quad Institute of Nuclear Physics, Polish Academy of Sciences, Radzikowskiego 152, 31-342 Cracow, Poland; dalvarez@ifj.edu.pl\\
$^{2}$ \quad Incubator of Scientific Excellence---Centre for Simulations of Superdense Fluids, plac Maksa Borna 9, 50-204 Wroclaw, Poland\\
$^{3}$ \quad  Facultad de Ciencias Físico Matemáticas, Universidad Autónoma de Nuevo León, 
 Av. Universidad S/N, C.U.,  San Nicolás de los Garza 66455, NL, Mexico 
}

\abstract{By consideration of the compact object HESS J1731-347 as a hybrid twin compact star, i.e., a more compact star than its hadronic twin of the same mass, its stellar properties are derived. In addition to showing that the properties of compact stars in this work are in good agreement with state-of-the-art constraints both from measurements carried out in laboratory experiments as well as by multi-messenger astronomy observations, the realization of an early strong hadron--quark first-order phase transition as implied by the twins is discussed.}


\newcommand{\beq}{\begin{equation}}
\newcommand{\eeq}{\end{equation}}
\newcommand{\ba}{\begin{array}}
\newcommand{\ea}{\end{array}}
\newcommand{\bea}{\begin{eqnarray}}
\newcommand{\eea}{\end{eqnarray}}
\newcommand{\bi}{\begin{itemize}}  
\newcommand{\ei}{\end{itemize}}
\newcommand{\ben}{\begin{enumerate}} 
\newcommand{\een}{\end{enumerate}}
\newcommand{\bc}{\begin{center}}
\newcommand{\ec}{\end{center}}

\newcommand{\De}{\Delta}
\newcommand{\ep}{\varepsilon}

\newcommand{\ptrans}{p_{\rm trans}}

\newcommand{\etrans}{\varepsilon_{\rm trans}}

\newcommand{\destab}{\De\ep_{\rm crit}}

\newcommand{\dd}{\textrm{d}}

\def\openone{\leavevmode\hbox{\small1\kern-3.8pt\normalsize1}}%

\def\mf{{\mbox{\tiny\em MFA}}}

\def\bea{\begin{eqnarray}}
\def\eea{\end{eqnarray}}
\def\beq{\begin{equation}}
\def\eeq{\end{equation}}

\begin{document}

\section{Introduction}

Neutron stars are massive and small stellar objects, therefore, matter in their interiors can be as dense as several times nuclear saturation density $n_0\approx 0.16 \, \text{fm}^{-3}$, the typical density value in atomic nuclei. There are several possibilities for the state of matter  at extreme densities, like hyperon-rich matter, meson condensates, or deconfined quark matter (QM). The latter considers deconfined quarks interacting with gluons in a plasma-like state as opposed to hadronic matter where quarks are confined inside nucleons or mesons. This fact has motivated various studies that regard the equation of state (EoS) of compact stars from theoretical, experimental and astrophysical grounds. On the one hand, laboratory experiments can probe nuclear properties of isospin symmetric matter most precisely at nuclear saturation density, whereas observations of neutron stars provide information about integral quantities like stellar mass, radius, moment of inertia, etc.

The nuclear symmetry energy $E_{s}$ function, which is related to the deviation in the symmetry between the number of neutrons and protons is an important quantity in the description of neutron star matter; therefore, it has been object of studies in nuclear experiments as well as an input for EoS models. {The value of the symmetry energy  saturation density $S=E_{s}(n_0)$ has been estimated to be around $30$ MeV by an analysis that takes into consideration diverse measurements like neutron skin thicknesses and dipole polarizabilities, the symmetry energy at saturation falling within the range $30$ MeV$<S<32$ MeV~\cite{Lattimer:2023rpe}. Its slope also at saturation is characterized by the parameter $L$, which is quite uncertain due to discrepancies between different measurements and analyses, roughly falling within a range of $40$ MeV$<L<80$ MeV when taking into account different analyses; for example, it has been particularly reported by the aforementioned analysis~\cite{Lattimer:2023rpe} that $L=51\pm31$ MeV. In fact, the uncertainty in the estimation of symmetry energy parameters stems from the different empirical approaches. In \cite{Carlson:2022nfb}, the authors implement relativistic mean field and nonrelativistic Skyrme-type interactions to reproduce the ground state binding energies, the charge radii, and the giant monopole resonances of a set of spherical nuclei in order to assess how well they can reproduce the observed properties of compact stars. They obtain the best values of $S=31.8 \pm 0.7$ MeV and $L=58.1 \pm 9.0$ MeV. It is important to note that in their analysis they do not consider phase transitions in dense matter, which we do specially consider in this work. Another estimate comes from the second Lead Radius EXperiment (PREX-II), the focus of which is to measure the neutron skin thickness of the $^{208}$Pb nucleus by means of parity-violating electron scattering, in order to determine how much the neutrons in the nucleus extend beyond the protons. Using PREX-II data, ref.
~\cite{Reed:2021nqk} reports
$S =38.1\pm4.7$ MeV and $L =106\pm37$ MeV, which are remarkably high.}

Matter 
 inside compact stars is indeed very different from that produced in relativistic heavy ion collisions (RHICs). Matter in the core of a compact star is homogeneous, contains leptons, is electrically charge neutral, and most importantly is in full thermodynamic equilibrium---conditions that are very different from those encountered in colliding nuclei. Matter in RHICs is extremely hot,  temperatures reach $T \sim 10^{12}$ K $(\sim 100{-}500$ MeV), the baryon chemical potential $\mu_B$ is small and this regime corresponds to the quark--gluon plasma (QGP) phase, where quarks and gluons are deconfined.  Neutron star matter is extremely dense, central densities can be several times nuclear saturation density ($\sim 5-10 \, n_0$), the temperature is relatively low compared to QCD scales (T $\sim 10^8{-}10^{10}$ K), and it is essentially treated as zero-temperature matter in most models. In addition, this matter is composed of degenerate, confined hadronic or possibly deconfined quark matter in a strongly interacting, high-$\mu_B$, low-T regime. When it comes to equilibrium and timescales, in RHICs, the system is far from equilibrium initially and rapidly evolves over $\sim10^{-23}$ seconds, with only a small, transient droplet of hot QCD matter created. Therefore, it is difficult for the system to achieve full thermodynamic equilibrium. The theoretical descriptions often rely on hydrodynamics, kinetic theory, or statistical models. Conversely, neutron star matter is in long-term equilibrium, both thermal and chemical (especially in the core). The system is static or quasi-static on astrophysical timescales (thousands to millions of years). In RHICs, matter is dominated by deconfined quarks and gluons, and usually the system exhibits chiral symmetry restoration and deconfinement. In compact stars, matter is typically dominated by confined hadrons (mostly neutrons, some protons, electrons, and muons), though inner cores may contain hyperons (strange baryons), deconfined quark matter (e.g., 2SC, CFL phases), pion/kaon condensates, or other exotic phases.

The most recent multi-messenger astronomical observations of compact stars include gravitational wave (GW) detections, as has been the case with GW170817~\cite{LIGOScientific:2017ync}, involving two compact stars, or GW190814, possibly containing at least one compact star and a black hole \cite{LIGOScientific:2020zkf}.
Radio detection of pulsars has allowed the most accurate measurements of compact star masses, which include massive objects like J0740+6620~\cite{NANOGrav:2019jur}. In addition, for this particular object, the Neutron Star Composition Explorer (NICER) has successfully provided an additional mass-radius measurement. NICER is focused on 
X-ray pulsating objects, and in this case, a joint radio and X-ray study has narrowed the uncertainties in the compact star mass and radius determination, commonly presented as a region in the so called mass-radius diagram, where sequences of compact stars are displayed. 
~\ref{MvsR}. The most massive star of a sequence is determined by the EoS and is a quantity that is sought to be measured; for instance, a recent combined analysis with multi-messenger data of neutron stars has inferred it to be 2.25$^{+0.08}_{-0.07}$ M$_{\odot}$ at about 3\% precision~\cite{Fan:2023spm}. A useful reference~\cite{Ascenzi:2024wws} is a recent and complete review of multi-messenger observations of compact stars. It is expected that the next generation of interferometers used as gravitational waves detectors will considerably advance our knowledge on the compact star EoS; see also the recent blue paper on the Einstein telescope collaboration for the big picture and concept~\cite{Abac:2025saz}.

The twin compact stars scenario, in which there exists two compact stars of the same mass but of different sizes due to their different internal composition, provides a way to study the properties of dense nuclear matter by means of astrophysical observations. Within this realization, a strong first-order phase transition inside compact stars will give rise to a disconnected branch in their mass-radius diagram. The concept of compact star twins was introduced in seminal works like~\cite{Gerlach:1968zz,Kampfer:1981yr,Glendenning:1998ag,Schertler:1998cs} followed by a period of low activity perhaps due to the lack of observational constraints. The phenomenon started to gain more interest as shown in~\cite{Alvarez-Castillo:2013cxa,Benic:2014jia,Alvarez-Castillo:2017qki,Montana:2018bkb,Blaschke:2019tbh,Zacchi:2016tjw,Espino:2021adh} up to the multi-messenger era. The deconfinement transition could be located in the low-temperature, high-density axis of the isospin asymmetric Quantum Chromodynamics (QCD) phase diagram. Furthermore, the strong first-order phase transition could also be located at finite temperatures, implying that thermal twins might exist~\cite{Carlomagno:2024vvr} and be formed during core-collapse supernovae explosions or in the process of binary star mergers that comprise at least one compact star. Recently, the estimated values of the millisecond pulsars PSR J0740+6620, PSR J0437-4715, J0030+0451, and PSR J1231-1411 have been re-analyzed under geometrical considerations of the topology of the X-ray stellar emitting area~\cite{Vinciguerra:2023qxq,Salmi:2024aum,Salmi:2024bss,Choudhury:2024xbk}. Importantly, the study presented in~\cite{Li:2024sft} has found the emission configurations that best favor compact star twins. 

The object HESS J1731-347 has been reported to be a very light compact star with a mass of 0.77 M$_{\odot}$ and a radius of 10.4 km~\cite{Doroshenko:2022nwp}. Understanding the EoS is, therefore, challenging, with many propositions for explanation available. {These ideas range from models considering pure hadronic compact stars~\cite{Kubis:2023gxa,Li:2023vso,Brodie:2023pjw}, kaon condensation inside the compact stars~\cite{Veselsky:2024eae}, strange quark stars~\cite{Horvath:2023uwl,DiClemente:2022wqp,Ju:2025mig,Rather:2023tly,Oikonomou:2023otn,Yuan:2025mmn}, hybrid compact stars~\cite{Li:2024lmd,Gao:2024chh,Laskos-Patkos:2024yxx,Laskos-Patkos:2023tlr}, slow stable compact stars~\cite{Mariani:2024gqi}, modifications arising from the existence of dark matter~\cite{Sagun:2023rzp,Hong:2024sey}, or alternative theories of gravity like in the work of~\cite{ElHanafy:2024cti}, which considers the quadratic Rastall gravity. Bayesian results comparing some of the above possibilities for this stellar object are presented in~\cite{Tewari:2024qit}.}

The purpose of this study is to highlight the astrophysical properties of HESS J1731-347 as a hybrid star, namely, a twin compact star. This scenario could be indeed corroborated with upcoming multi-messenger observations. Without a loss of generalities, the most characteristic properties of compact star twins are shared with any other observed neutron star that are in order with current the state-of-the-art constraints.

The rest of this manuscript is organized follows. In Section 2, the equations of state under consideration describing the HESS pulsar are introduced. In Section 3, all the astrophysical properties are derived together with the computation methodologies. Finally, the summary, conclusions, and outlook are presented.

\section{Equations of State for Compact Star Twins describing HESS J1731-347}

\subsection{Hadronic EoS}

In this work, two equations of state models are considered, resulting in three realizations of compact stars sequences. The resulting compact stars are described by a hadronic mantle that surrounds a quark matter core. The hadronic EoS is the density dependent relativistic mean field (RMF) model DD2 in its original version as well as the DD2F, whose stiffness is adjusted above saturation density $n_0$ to reproduce the pressure constraint by the matter flow in relativistic heavy ion collisions~\cite{Danielewicz:2002pu}. Moreover, both approached are equipped with the excluded volume correction that takes into account the stiffness of matter above saturation density from the interactions between the quarks inside nucleons~\cite{Typel:2016srf}.  The DD2MEVp80 and DD2FMEVp80 EoSs feature this effect by considering the available volume fraction $\Phi_N$ for the motion of nucleons as density dependent in a Gaussian form\vspace{6pt}

\begin{equation}
\Phi_N = \exp\left[-v^{2}(n-n_0)^2/2\right],~{\rm for}~ n > n_0~,
\end{equation}
and $\Phi_N=1$ if $n\le n_0$.
Here, $v = 16\pi r_N^3/3$ is the van der Waals excluded volume
for a nucleon with a hard-core radius $r_N$, and $n_0=0.15$ fm$^3$ is the saturation density of infinite, symmetric nuclear matter. 
The index ``p80'' with the DD2 and DD2F parametrizations denotes a positive excluded volume parameter of $v=8$ fm$^3$. In order to describe compact stars, the matter in their interiors described by these hadronic excluded volume EoSs should undergo a phase transition at a lower density value than the one breaching causality $c_s<c$, which happens at energy density values of $\varepsilon$ = 277.963 MeV/fm$^3$ for DD2FMEVp80, and  $\varepsilon$ = 272.77 MeV/fm$^3$ for DD2MEVp80. These nuclear EoSs have been extensively used in systematic studies of hybrid stars, see for instance,~\cite{Benic:2014jia,Alvarez-Castillo:2016oln,Kaltenborn:2017hus}.

\subsection{Quark Matter EoS}
\subsubsection{CSS Model}
Furthermore, the cores of compact stars in this work are expected to be comprised of deconfined quark matter; therefore, they are either described by the constant speed of sound (CSS) parameterization~\cite{Zdunik:2005kh,Alford:2013aca,Pal:2025skz} or the non-local NJL model (nlNJL), as first introduced in~\cite{Blaschke:2007ri} or recently extended to include three quark flavors  in~\cite{Ivanytskyi:2024zip}. The CSS parameterization has been found to successfully describe quark matter equations of state~\cite{Zdunik:2012dj,Shahrbaf:2021cjz}, whereas the latter is an effective model that explicitly incorporates all the effects of the quark interactions and is regulated by several parameters. Hadronic EoSs are connected to QM EoSs via a Maxwell construction. The CSS parametrization is given by:
\beq
\varepsilon(p) = \left\{\!
\begin{array}{ll}
\varepsilon_{H}(p) & p<\ptrans \\
\varepsilon_{H}(\ptrans)+\De \varepsilon+c_{s}^{-2} (p-\ptrans) & p>\ptrans
\end{array}
\right.\ .
\label{EoSqm}
\eeq
Here, 
 $\varepsilon_{H}$ corresponds to hadronic matter, $p_{trans}$ is the pressure value at the phase transition, $c_{\rm s}$ is the speed of sound in quark matter, and $\De\ep$ is the energy density jump typical of a Maxwell-constructed first-order phase transition. Table~\ref{table_CSS_parameters} shows the choice of parameters for the description of the HESS J1731-347 star within this approach.
\begin{table}[H]
\small
\caption{\label{table_CSS_parameters}EoS 
 parameters for CSS EoS labeled as DD2MEVp80-CSS($c_s$ = 0.9), characterized by the speed of sound $c_s$.}
\begin{tabularx}{\textwidth}{Cccccc}
\noalign{\hrule height 1pt} 
\textbf{Model} &	\boldmath{$M_{onset}$} &	 	\boldmath{$n_{trans}$} & 		\boldmath{$\varepsilon_{trans}$} &	\boldmath{$p_{trans}$}  &	\boldmath{$\Delta \varepsilon$}\\
&					\textbf{[M$_{\odot}$]} &			\textbf{[1/fm$^{3}$]} &			\textbf{[MeV/fm$^{3}$]}  &		\textbf{[MeV/fm$^{3}$]}  & \textbf{[MeV/fm$^{3}$]}\\
\midrule
DD2MEVp80-CSS ($c_s$ = 0.9) & 0.7 &0.193 &185.220 &10.313 & 268.573 \\
\noalign{\hrule height 1pt}
\end{tabularx}


\end{table}
\subsubsection{Non-Local NJL model}

The non-local NJL model is a more involved approach to two light-flavored quarks matter in which matter is computed under compact star conditions, i.e., electric and color charge neutrality, as well as in beta equilibrium. The starting point is the Euclidean action~\cite{Blaschke:2007ri}
\begin{eqnarray}
S_E &=& \int d^4 x \ \left\{ \bar \psi (x) \left(- i \rlap/\partial + m_c
\right) \psi (x) - \frac{G_S}{2} j^f_S(x) j^f_S(x) 
\right.\nonumber\\
&& \left. 
- \frac{H}{2}
\left[j^a_D(x)\right]^\dagger j^a_D(x) {-}
\frac{G_V}{2} j_V^{\mu}(x)\, j_V^{{\mu}}(x) \right\} \, .
\label{action}
\end{eqnarray}
Here, $m_c$ is the current quark mass, which is considered to be equal for $u$
and $d$ quarks. The non-local currents $j_{S,D,V}(x)$ include
operators introduced on a separable approximation of the effective one gluon
exchange model (OGE) of QCD. These currents are\vspace{6pt}

\begin{eqnarray}
j^f_S (x) &=& \int d^4 z \  g(z) \ \bar \psi(x+\frac{z}{2}) \ \Gamma_f\,
\psi(x-\frac{z}{2})\,,
\\
j^a_D (x) &=&  \int d^4 z \ g(z)\ \bar \psi_C(x+\frac{z}{2}) \ \Gamma_D \ \psi(x-\frac{z}{2}) 
\\
j^\mu_V (x) &=& \int d^4 z \ g(z)\ \bar \psi(x+\frac{z}{2})~ i\gamma^\mu
\ \psi(x-\frac{z}{2}). 
\label{cuOGE}
\end{eqnarray}
with $\psi_C(x) = \gamma_2\gamma_4 \,\bar \psi^T(x)$, $\Gamma_f=(\openone,i\gamma_5\vec\tau)$, and $\Gamma_D=i \gamma_5 \tau_2 \lambda_a$, whereas $\vec \tau$ and $\lambda_a$, with $a=2,5,7$, stand for Pauli and Gell-Mann  matrices acting on flavor and color spaces, respectively.  $g(z)$ in Equation~(\ref{cuOGE}) is a covariant form factor for the nonlocality of the effective quark interactions~\cite{GomezDumm:2005hy}. Moreover, a dimensionless vector coupling strength is defined as $\eta=G_V/G_S$, which is treated as a free parameter responsible for the stiffness of quark matter EoS at nonzero densities. Further steps on the EoS computation include bosonization of the theory under the framework of the mean field approximation (MFA) and consideration of the Euclidean action at zero temperature and finite baryon chemical potential $\mu_B$. Eventually, one arrives to the gap equations of the theory, which involves derivatives of the mean field grand canonical thermodynamic potential per unit volume $\Omega^\mf$ with respect to the mean field values of isospin zero fields and diquark mean fields. After imposing the compact star matter conditions, one arrives to the QM pressure as a function of baryon chemical potential:
\begin{equation}
\label{eq:P-mu}
p(\mu)=p(\mu;\eta(\mu),B(\mu))=-\Omega^\mf(\eta(\mu)) - B(\mu)~,
\end{equation}
where a chemical potential-dependent bag pressure shift stemming, for instance, from a medium dependence of the gluon sector is allowed, with both parameters $\eta$ and $B$ also dependent on the baryon chemical potential. Motivated by the fact that the value of the vector coupling strength parameter $\eta$ may actually vary as a function of the chemical potential~\cite{Blaschke:2012dba}, an interpolating function is introduced within the QM description in order to model the EoS in a certain range of chemical potentials. The interpolation allows (a) the modeling of the unknown density dependence of the quark confinement by interpolating the bag pressure contribution between zero and a finite value $B$ at low densities near the hadron-to-quark matter transition, and (b) modeling the density dependent stiffening of quark matter at high  density by interpolating between QM for two values of the vector coupling strength. This is achieved by utilizing two smooth switch-off functions, one that changes from $1$ to $0$ at a lower chemical potential $\mu_<$ related to a width $\Gamma_<$, 
\begin{equation}
f_{<}(\mu)=\frac{1}{2}\left[1-\tanh\left(\frac{\mu-\mu_<}{\Gamma_<}\right)\right],
\end{equation}
and one that does the same at a higher chemical potential $\mu_\ll$ with a width $\Gamma_\ll$, 
\begin{equation}
f_{\ll}(\mu)=\frac{1}{2}\left[1-\tanh\left(\frac{\mu-\mu_\ll}{\Gamma_\ll}\right)\right].
\end{equation}
These functions are complemented with the corresponding switch-on expressions,
\begin{equation}
f_{>}(\mu) = 1 - f_<(\mu)~,~~f_{\gg}(\mu) = 1 - f_{\ll}(\mu),
\end{equation}
which help to define the doubly interpolated QM pressure:
\begin{eqnarray}
p(\mu) &=& [f_<(\mu) p(\mu;\eta_<,B) + f_>(\mu) p(\mu;\eta_<,0)] f_{\ll}(\mu)
\nonumber\\
&& +f_{\gg}(\mu)p(\mu;\eta_>,0).
\end{eqnarray}
Alternatively, the above expression can be  equivalently written in terms of  a chemical potential-dependent  vector mean-field coupling $\eta(\mu)$ and a chemical potential-dependent bag pressure $B(\mu)$. In the highest density region, the non-local NJL model is substituted by the CSS parametrization in order to avoid an EoS causality violation. The CSS formulation in terms of the chemical baryon potential $\mu$ is as follows:
\begin{eqnarray}
	p(\mu) & = & p_0 + p_1 \left({\mu}/{\mu_x} \right)^\beta , \hspace{2cm} \mathrm{for~} \mu>\mu_x,\\
	\varepsilon(\mu) & = &-p_0 + p_1 (\beta -1) \left({\mu}/{\mu_x} \right)^\beta ,\hspace{6mm}  \mathrm{for~} \mu>\mu_x,\\
	n_B(\mu) & = & p_1 \displaystyle \frac{\beta}{\mu_x} \left({\mu}/{\mu_x} \right)^{\beta-1}, \hspace{2cm}  \mathrm{for~} \mu>\mu_x,
	\label{eq:ccs}
	\end{eqnarray}
	where $\mu_x$ is the matching chemical potential to the CSS QM description. The squared speed of sound is given in terms of the $\beta$ exponent:
	\begin{equation}
	c_s^2=\frac{\partial p/ \partial \mu}{\partial \varepsilon / \partial \mu} = \frac{1}{\beta - 1}~,
	\label{eq:css}
	\end{equation}
	or equivalently,
	\begin{equation}
	\beta = 1 + \displaystyle \frac{1}{c_s^2} ~.
	\label{eq:css_beta}
	\end{equation}
	Thus, the condition that $c_s^2 \le 1$ implies that $\beta \ge 2$. 
	The coefficients $p_0$ and $p_1$ in Equation~(\ref{eq:ccs}) are defined as:
	\begin{eqnarray}
	p_0 & = & \left[(\beta - 1) p_x - \varepsilon_x \right]/\beta\\
	p_1 & = & \left(p_x + \varepsilon_x \right) / \beta~,
	\label{eq:css_pp}
	\end{eqnarray}
	with $p_x=p(\mu_x)$ and $\varepsilon_x=\varepsilon(\mu_x)$. For the CSS EoS segments supplementing the nlNJL QM in this work, $\beta=2.034$. Other parameters of this hybrid star EoS model are listed in Table~\ref{tab-nlNJL-param}. Further details on the full derivation of the nlNJL QM model can be found in~\cite{Alvarez-Castillo:2018pve,Blaschke:2020qqj}. 
\begin{table}[H]
\caption{ Properties of the DD2MEVp80-nlNJL models, set 2 successfully describes the HESS J1731-347 compact object as a twin compact star. The EoS share all the parameter values but $\mu_{<}$ which is equal to 860 MeV for set 1 and 955 MeV for set 2. The resulting critical values for physical quantities at the onset of the first-order phase transition density like the the critical chemical potential $\mu_c$, critical baryon density $n_c$, critical energy density $\varepsilon_c$, and critical pressure $p_c$ and compact star mass $M_{onset}$, are also presented.
		\label{tab-nlNJL-param}
}  
	\begin{tabularx}{\textwidth}{LCC}
		\noalign{\hrule height 1pt}

		\textbf{DD2FMEVp80-nlNJL} & \textbf{Set 1} & \textbf{Set 2} \\
\midrule
		$\mu_{<}$ [MeV] &  860 & 955\\
		$\Gamma_{< }$ [MeV] & 150 & 150\\
		$\mu_{\ll}$ [MeV] & 1700 & 1700\\
		$\Gamma_{\ll}$ [MeV] & 350 & 350\\
		$B$ [MeV/fm$^3$] & 65 & 65\\
		$\eta_<$ & 0.05 & 0.05 \\
		$\eta_>$ & 0.12 & 0.12 \\
		\midrule
		$\mu_c$ [MeV] & 993.030 & 1048.010 \\
		$p_c$[MeV/fm$^{3}$] & 6.151 & 17.019 \\ 
		$\epsilon_c$ [MeV/fm$^{3}$] & {169.274} & 207.269 \\
		$n_c$ [fm$^{-3}]$ & 0.177 & 0.214\\
		\midrule
		$M_{onset}$ [$M_\odot$] &0.45 & 1.00\\
		\noalign{\hrule height 1pt}
	\end{tabularx}

\end{table}
\subsubsection{Maxwell Construction and Seidov Criterion for Stellar Stability}

\begin{figure}[H]
		\centering
		\vspace{-5mm}
		\includegraphics[width=0.48\textwidth]{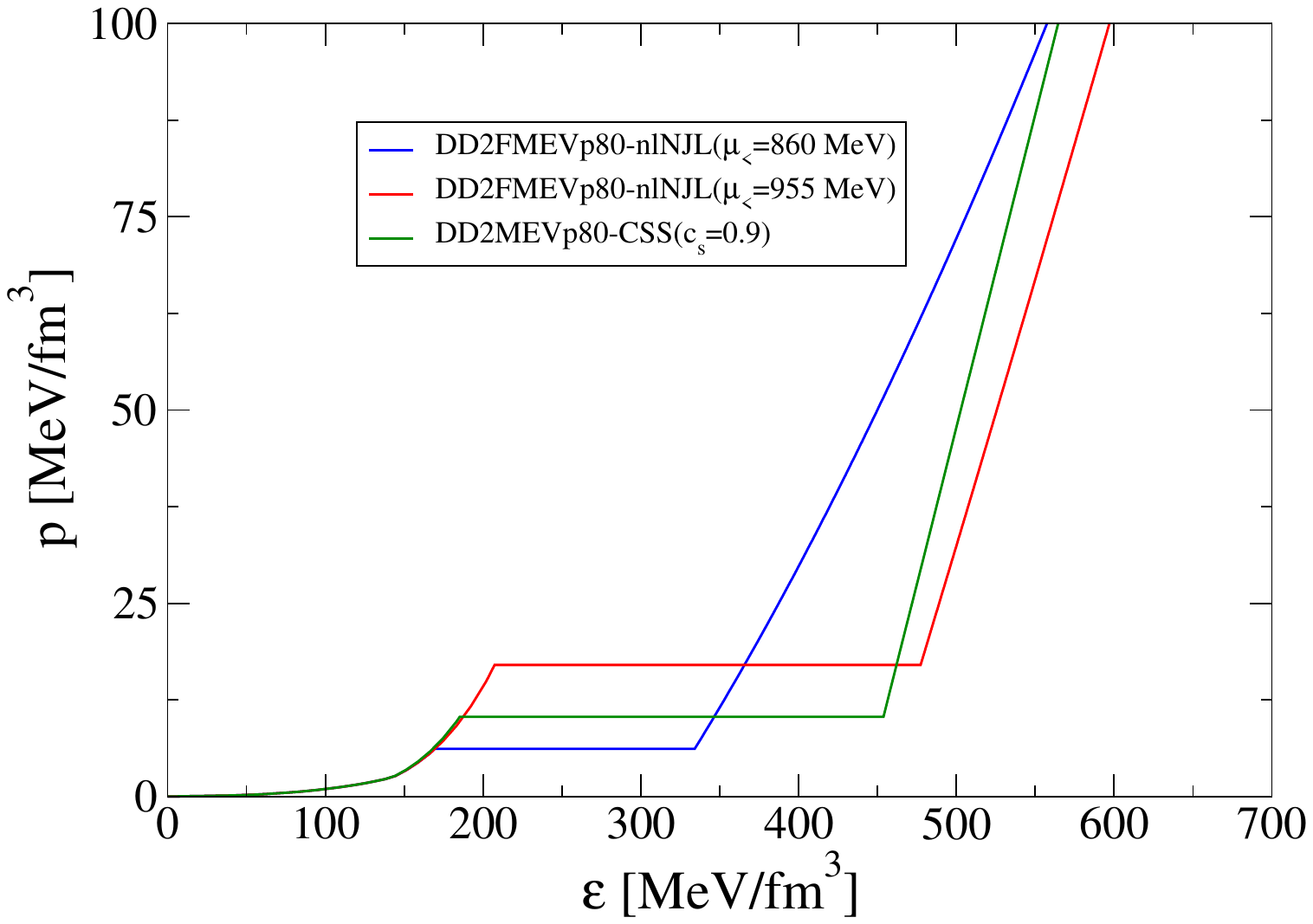} 
		\includegraphics[width=0.48\textwidth]{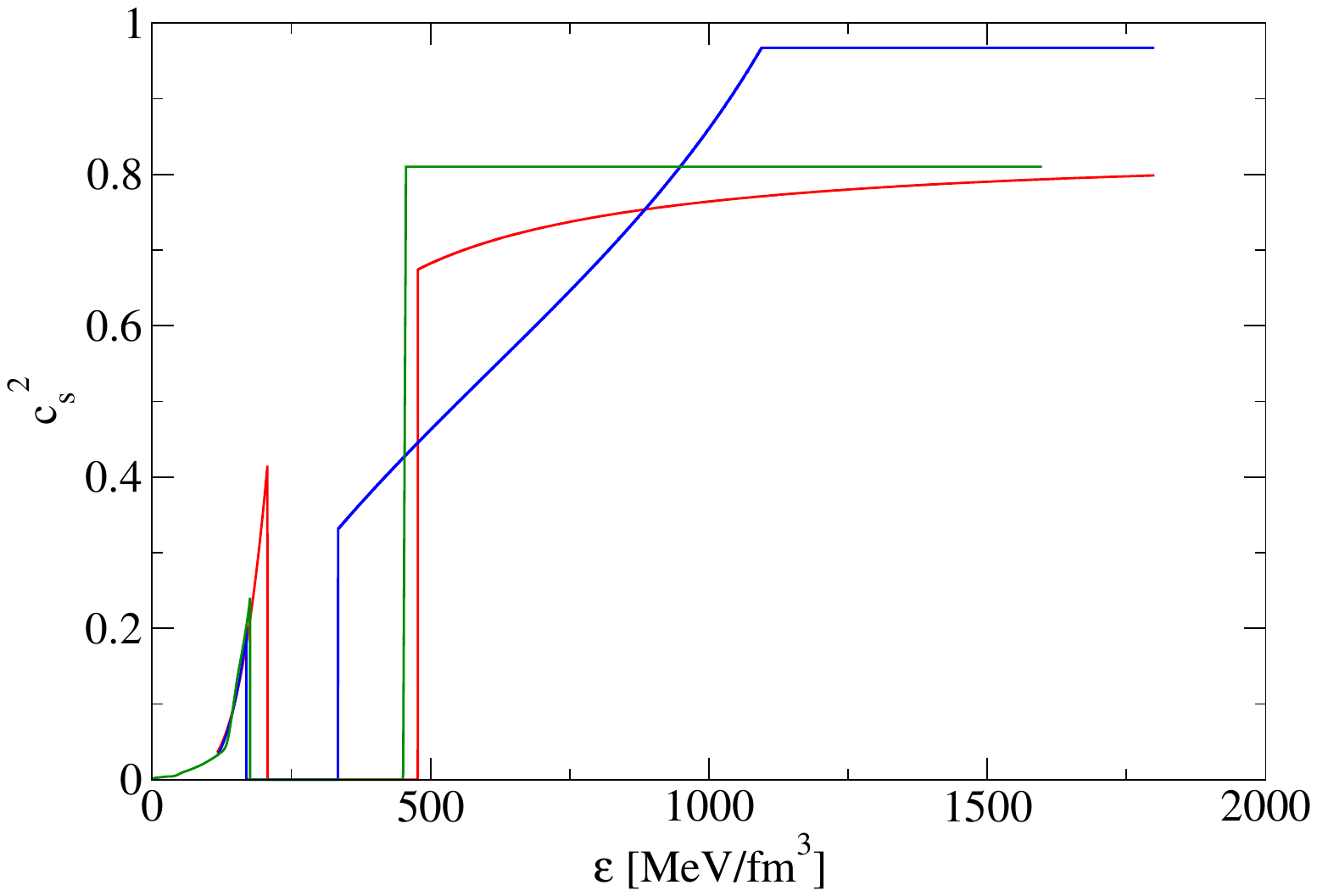} 
		\caption{\label{fig:eos}
			Left panel: Hybrid star equations of state with the hadronic branch from the RMF model DD2MEVp80 with a quark branch described by the CSS EoS with $c_s=0.9$ and  the hadronic DD2FMeVp80 with quark matter branch from the nonlocal chiral quark model nlNJL with a different onset of deconfinement and charaterized by $\mu_<=860$ MeV and $\mu_<=955$ MeV, see the text for details. The first-order phase transition from hadronic to deconfined quark matter is implemented by a Maxwell construction.  Right panel:  The square of the speed of sound $c_s$ as a function of the energy density for the EoS models on the left. 
		} 
	\end{figure}

The EoSs of hybrid stars in this work feature a Maxwell construction from hadronic to quark matter. In turn, the Maxwell construction features a plateau of constant energy density in the EoS pressure--energy diagram, see Figure~\ref{fig:eos}. As discussed below, it turns out that the length of this plateau plays an important role in the determination of the existence of compact star twins. A very important relation is the Seidov condition $\De\ep>\destab$~\cite{Seidov:1971}, which ensures the appearance of a disconnected branch in the mass-radius relation of compact stars. The critical values at the transition are related by
\beq
\frac{\destab}{\etrans} = \frac{1}{2} + \frac{3}{2}\frac{\ptrans}{\etrans} \ .
\label{eqn:stability}
\eeq	

\begin{figure}[H]
	
	\includegraphics[width=10.5 cm]{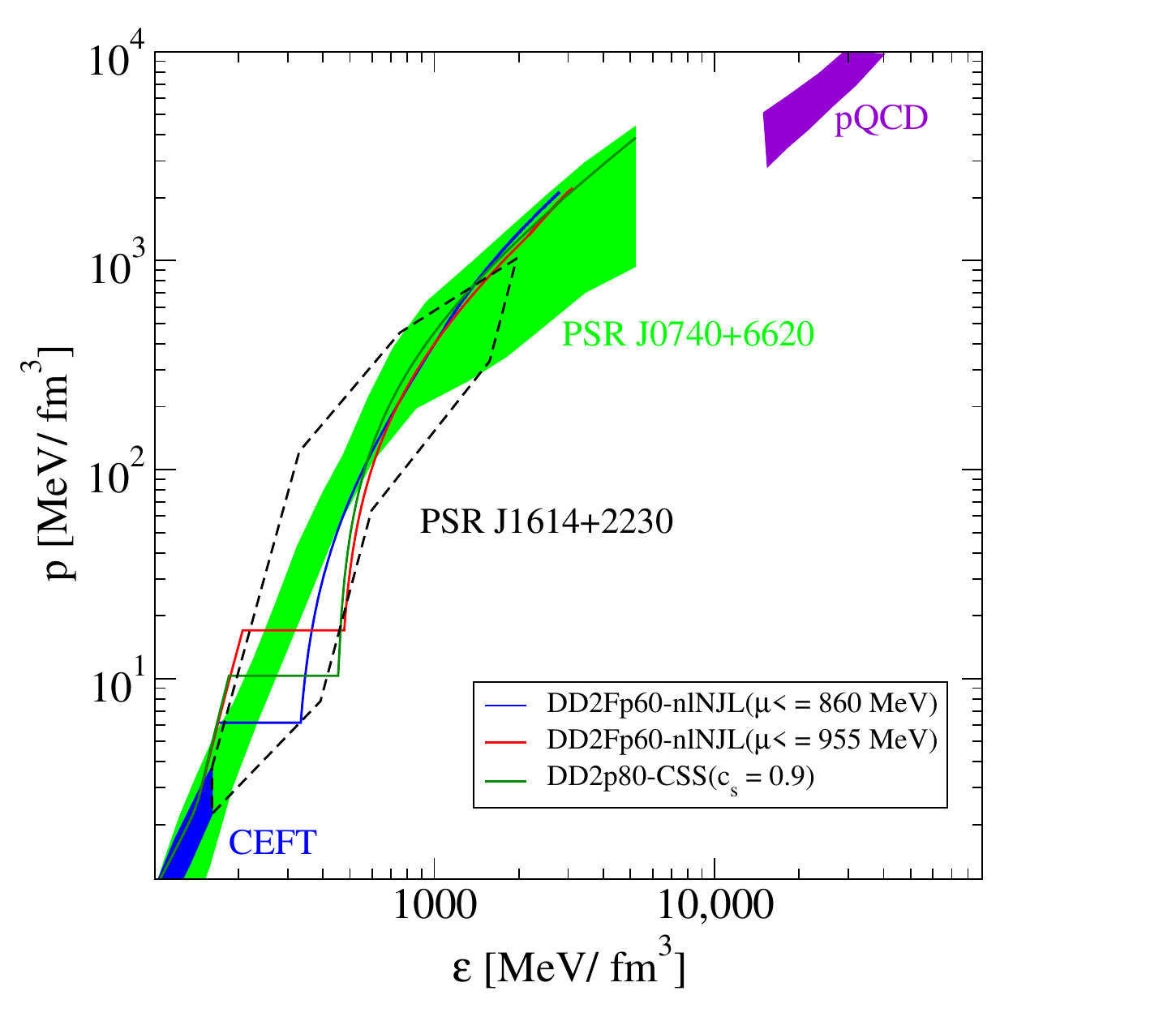}
	\caption{{Equation 
 of state of compact star matter in beta equilibrium for the description of the mass of the twins together estimates from various studies. The blue region near the origin corresponds to Chiral Effective Theory calculations that consider all many-body forces among nucleons~\cite{Kruger:2013kua}.  The area marked by the dashed contour corresponds to the derivation, which is based on Chiral Effective Field Theory together with the mass measurement of the object J1614-2230 reported as M = ($1.97\pm0.04$) 
 M$_{\odot}$~\cite{Hebeler:2013nza}. The green area is derived from a combined analysis of the NICER measurement of PSR J0740+6620 and XMM-Newton data~\cite{Miller:2021qha}, see the text for a discussion on its derivation and constraining power on the twins. At the far end of the plot, in the upper right corner, the constraint from perturbative QCD related to its conformal limit at very large density is shown~\cite{Fraga:2013qra}. \label{evsp_constraints}}}
	\end{figure}

Of the EoSs chosen to describe the HESS J1731-347 object as a twin compact start, one is described by the nlNJL QM+CSS model, whereas the second is fully described by the CSS approach. Those with nlNJL QM have a DD2FMEV hadronic component in contrast to the CSS QM bearing a hadronic DD2MEV description. Naming these EoSs by their most representative parameters, they are introduced as DD2FMEVp80-nlNJL ($\mu_<$ = 855), DD2FMEVp80-nlNJL ($\mu_<$ = 955), and DD2MEVp80-CSS ($c_s$ = 0.9), respectively. Figure~\ref{fig:eos} on its left panel shows the EoS as a relation between energy density $\epsilon$ and pressure $p$. The plateau is characteristic of a first-order phase transition that, in this case, marks the border between hadronic and deconfined quark matter. The right panel shows the squared speed of sound, which is related to the left panel by {$c_s^{2} = \frac{dp}{d\epsilon}$}.
{In addition, Figure~\ref{evsp_constraints} shows the pressure--energy density relation for the twins' EoS together with modern constraints. Those constraints include, on the one hand, Chiral Effective Theory predictions for the low density region derived from the many-body forces among nucleons~\cite{Kruger:2013kua} for the low density region, see~\cite{Drischler:2021kxf} for the extension towards higher density as well. On the other hand, constraints that include compact star observations are presented in the intermediate region as a closed dashed region~\cite{Hebeler:2013nza} and as in the green region~\cite{Miller:2021qha}. Regarding the latter, the combined analysis of the mass and radius of PSR J0740+6620 by NICER with XMM-Newton data has resulted in a very narrow green area. It is clearly seen that the Maxwell construction region extends beyond its energy density limits. The reason for this behavior is that the corresponding Bayesian analysis based in the aforementioned data also features agnostic EoS models (in contrast to physics-informed ones, where the physical parameters are in agreement with empirical data) with not many twin configurations giving low statistical weight to the analysis. Moreover, since there are no low-mass compact star measurements and the measurements were carried out before the discovery of HESS J1731-347, which is, therefore, not included, the green area does not favor strong first-order phase transitions. In fact, it cannot be used as a strong constraint for low-mass twin models. Finally, within the figure, the purple region corresponds to the conformal limit of QCD~\cite{Fraga:2013qra}.}

	\vspace{-10pt}
	
\subsubsection{QCD Conformality and Trace Anomaly Inside Twin Compact Stars}

\begin{figure}[H]
	\includegraphics[width=10.5 cm]{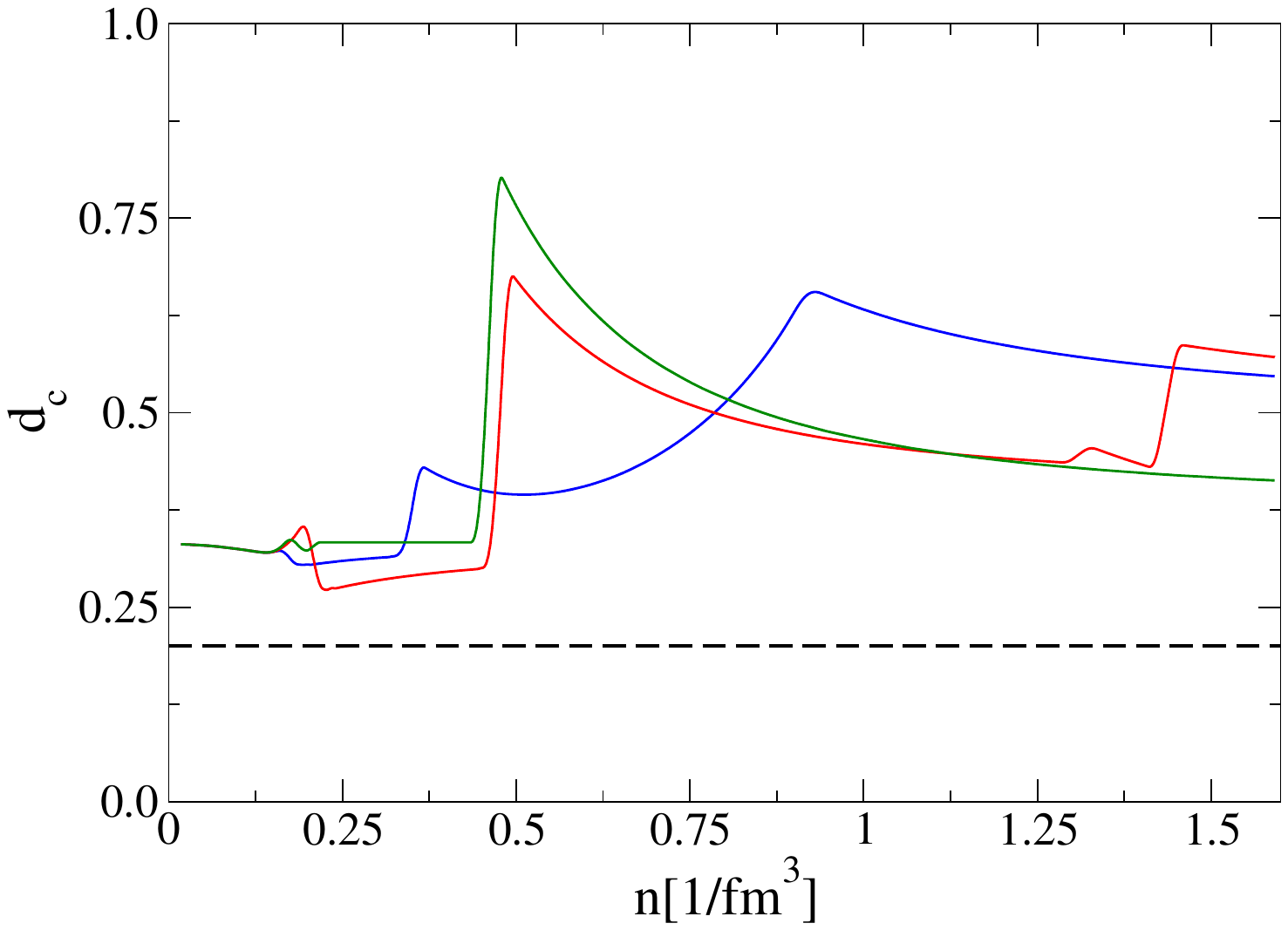}
	\caption{Conformality 
 factor $d_c$ as a function of baryon density, color codes for the lines are the same as in figure 1. The horizontal dashed line at $d_c = 0.2$ corresponds to 
the boundary between conformal
and non-conformal matter as introduced in~\cite{Annala:2023cwx}. The models for compact star twins stay above this limit, implying non-conformal matter everywhere inside the twins. \label{QCD-conformality}}
\end{figure}

\begin{figure}[H]
	\includegraphics[width=10.5 cm]{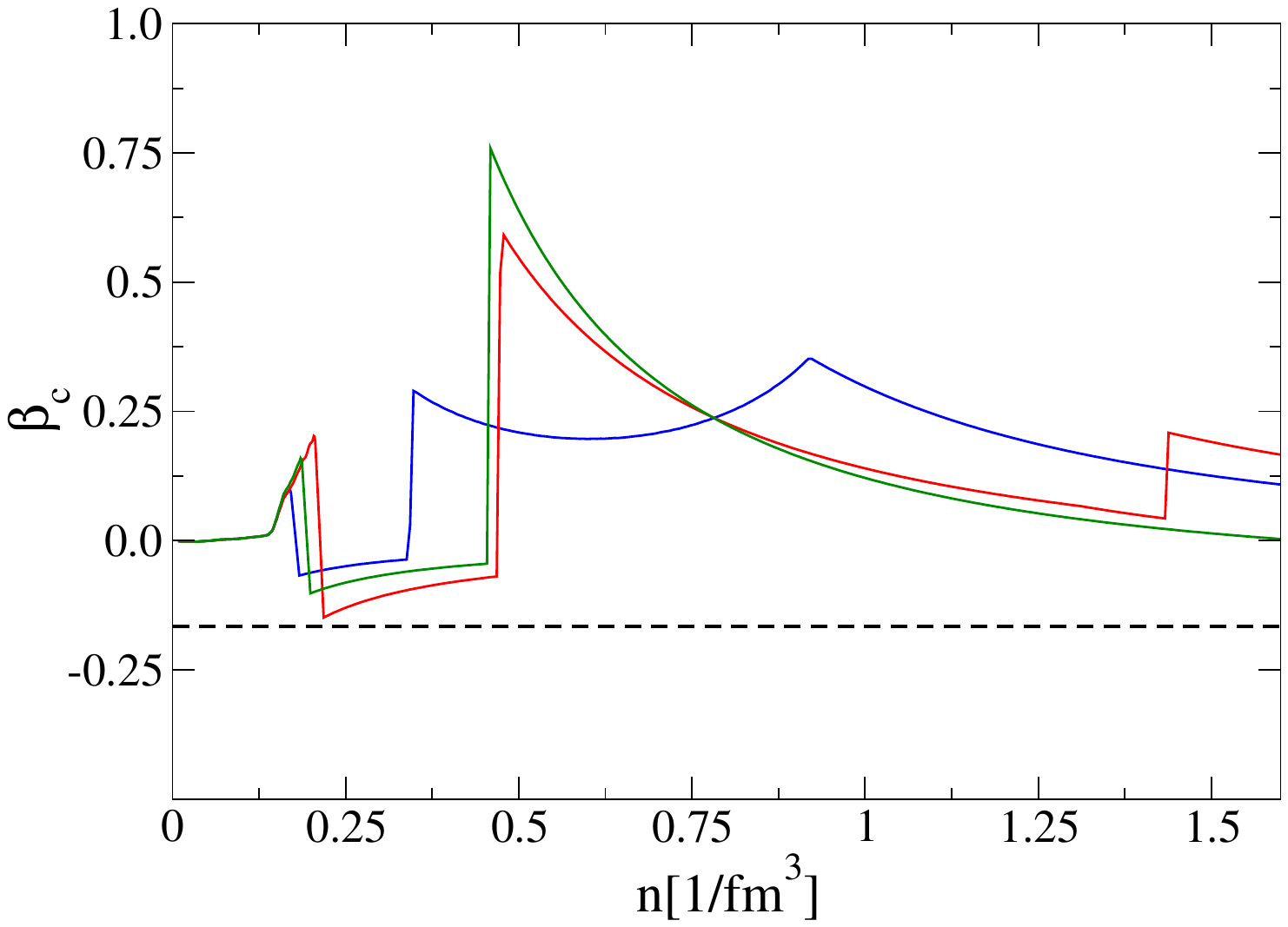}
	\caption{$\beta_c$ parameter 
 as function of baryon density. $\beta_c$~\cite{Marczenko:2024uit} becomes negative as conformality is restored. It is also seen that it becomes negative through and around the first-order phase transition baryon density region. Color codes for the lines are the same as in figure 1. The horizontal line corresponds to the conformal limit border. \label{beta-conformality}}
\end{figure}

The conformality of dense nuclear matter is an extensively studied topic regarding the physics of QCD.  A Conformal Quantum Field Theory (CFT) is expected to retain conformal invariance at the quantum level, implying that its energy--momentum tensor remains traceless. Even though the classical theory of QCD is conformally invariant, i.e., at the pure theoretical level with massless quarks, this symmetry is broken at the quantum level, implying the trace of the energy--momentum tensor would no longer be zero. This is the so called QCD trace anomaly. As QCD approaches asymptotic freedom, it behaves approximately like a CFT, with the square speed of sound in the medium $c_s^{2}$ approaching the value of 1/3.
An abrupt increase in the speed of sound crossing its conformal value in dense matter entails the disappearance
of the trace anomaly~\cite{Fujimoto:2022ohj}.  The vanishing of the trace anomaly is a necessary but not sufficient condition for the full restoration of conformal symmetry~\cite{Marczenko:2024uit}. In order to study the conformality of matter inside compact twin stars, some characteristic quantities can be derived, all of them related to $c_s$~\cite{Fujimoto:2022ohj,Marczenko:2022jhl}:
\begin{equation}
c_{s}^{2}=\frac{1}{3}-\Delta -\Delta',
\end{equation}
where  $\Delta$ is the trace anomaly scaled by the energy density
\begin{equation}
\Delta=\frac{\varepsilon-3p}{3\varepsilon},
\end{equation}
{also related to the average speed of sound $⟨c_{s}^{2}⟩ = \frac{1}{\varepsilon} \int\limits_0^\varepsilon c_s^2 \textrm{d} \varepsilon = p/\varepsilon$ computed within the interval $⟨0,\varepsilon⟩$ and obeying the causality condition {$0\leq ⟨c_{s}^{2}⟩ \leq 1$} so that the trace anomaly and its derivative $\Delta'$ = d$\Delta$/dln($\varepsilon$) read~\cite{Marczenko:2024uit}: }
\begin{equation}
\Delta=\frac{1}{3}-⟨c_{s}^{2}⟩,
\end{equation}
\begin{equation}
\Delta'=⟨c_{s}^{2}⟩-c_{s}^{2}.
\end{equation}
It is argued in~\cite{Annala:2023cwx} that the conformality parameter $d_c$ obeys
\begin{equation}
d_{c}\equiv \sqrt{\Delta^{2}+(\Delta')^{2}}<0.2,
\end{equation}
and the auxiliary parameter~\cite{Marczenko:2023txe,Marczenko:2025hsh}
\begin{equation}
\beta_c = c_{s}^2 - \frac{2p}{p+\varepsilon}
\end{equation}
must turn negative as conformality is restored. Interestingly, $\beta_c$ is related to the compressibility of nuclear matter~\cite{Ivanytskyi:2022bjc}:
\begin{equation}
K_{NM}=9\mu\left(c_{s}^2 - \frac{2p}{p+\varepsilon}\right)=9\mu\beta_c .
\end{equation}
 In the work of~\cite{Jimenez:2024hib}, the $d_c$ conformality parameter has been studied for different classes of compact star twins, as defined in~\cite{Christian:2017jni}. The results presented there for the conformality of matter inside compact star twins are in agreement with those found in this work. Figure~\ref{QCD-conformality} shows the behavior of the conformality parameter $d_c$ as a function of baryon density for the three EoSs considered here. In this figure, all the curves remain above the 0.2 limit, which is displayed as a dashed line, implying non-conformality. Similarly, Figure~\ref{beta-conformality} shows the $\beta_c$ parameter in the same baryon density range. It remains positive away from the phase transition region and is expected to turn negative $\beta_c \rightarrow -1/6$ as conformality is restored. The authors of~\cite{Jimenez:2024hib} also bring up the hypothesis that the discontinuities in the trace anomaly $\Delta$ are produced by discontinuities in the QCD running coupling $\alpha_s$ and point out the fact that there are still several unknown microphysical aspects that allow the construction of appropriate twin compact star EoSs. Extreme bounds for the speed of sound related to special relativity, relativistic kinetic theory, and conformality are presented in~\cite{Laskos-Patkos:2024otk}. Apart from the latter, the twin compact star EoSs presented here fall well within this bound. {All in all, the speed of sound of the compact star models presented in this manuscript strongly deviate from the conformal limit. The obtained results for the low-mass twins clearly point out this fact, whereas works that employ conformality as a constraint inside neutron stars seem not to be fully justified from astrophysical measurements.}

\vspace{-8pt}
   \vspace{-12pt}
\section{Results}
\subsection{Compact Star Structure}

In order to find compact stars configurations compatible with HESS J1731-347, the free parameters of the models have been varied, particularly the quark matter onset density, which fixes the most massive hadronic star. Three adequate sets, two of them describing the HESS object as a twin compact star, fulfill all state-of the art compact star constraints with their parameters chosen after inspection of the derived compact star configurations, which are simply static and spherical, general relativistic objects. Those configurations are computed by solving the Tolman--Oppenheimer--Volkoff Equations (\cite{Tolman:1939jz,Oppenheimer:1939ne}):
\begin{eqnarray}
 \label{TOV}
\frac{\dd p( r)}{\dd r}&=& 
-\frac{\left(\varepsilon( r)+p( r)\right)
\left(m( r)+ 4\pi r^3 p( r)\right)}{r\left(r- 2m( r)\right)},\\
\frac{\dd m( r)}{\dd r}&=& 4\pi r^2 \varepsilon( r).
\label{eq:TOVb}
 \end{eqnarray}
Integration of the above equations is carried out from the center of the star, where the pressure is at the maximum, towards the surface, where the pressure vanishes $p(r=R)=0$. In this way, the total mass of the compact star is defined as $M = m(r = R$) with $R$ as the stellar radius that defines the size of the star. In order to close the system, the complementary condition $m (r = 0) = 0$ must be included. In addition, the initial condition $p(r=R)=0$ determines the mass and radius of the star, which can be either pure hadronic or hybrid depending on whether the central density of the star $\varepsilon_c$ lies above the quark deconfinement density or not. Sequences or families of compact stars are derived by increasing such a central density of the star in consideration when integrating the TOV equations, each point in the so called mass-radius diagram representing a single star. The process is iterative and is stopped once the maximum compact star is reached, in the case of the EoS, meaning the iteration is stopped once the most massive hybrid compact star is found. This is because stars above the central density of the maximum hybrid compact star will become unstable, the instability being dictated by the condition $\partial M /\partial \varepsilon_c> 0${~\cite{1965gtgc.book.....H}.}

\begin{figure}[H]
		\centering
		\vspace{-5mm}
		\includegraphics[width=0.48\textwidth]{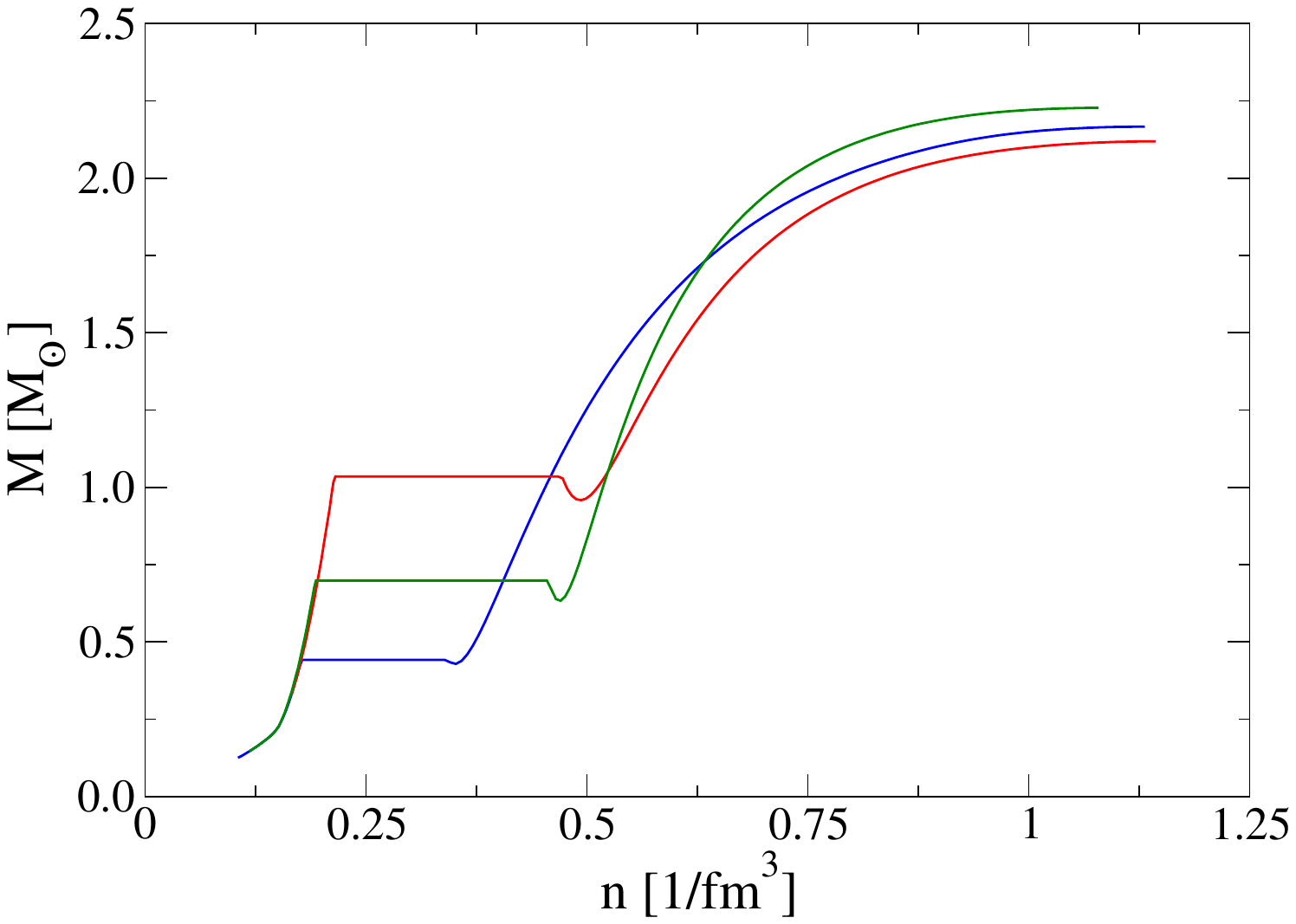} 
		\includegraphics[width=0.48\textwidth]{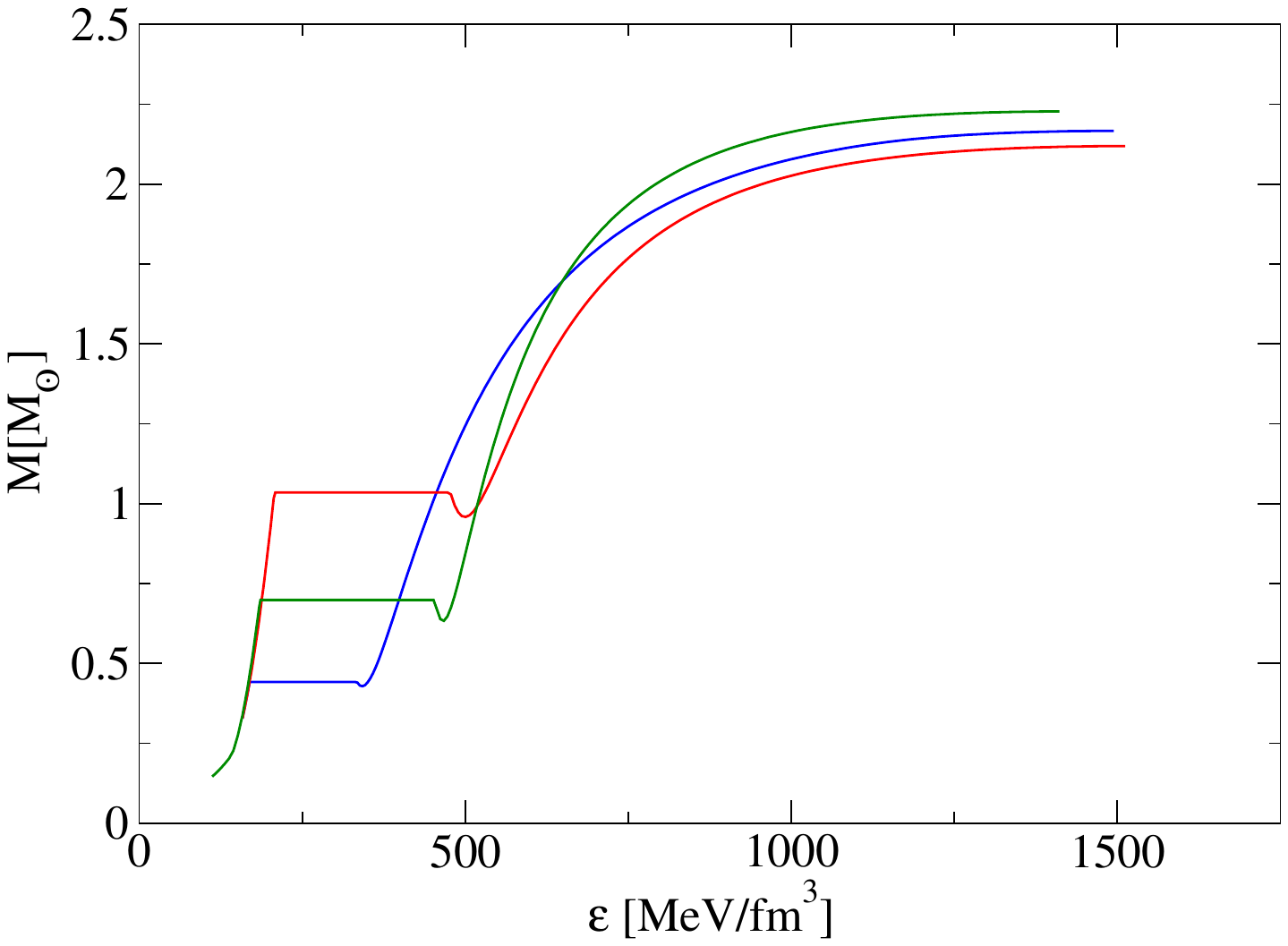}
		\caption{\label{Mvsn_c} Compact 
 star mass dependence on densities at the center of the star. The lines end at the value of the maximum hybrid star mass, their color codes corresponds to the ones of the EoS
in Fig. 1.\\
			Left panel:  mass dependence on central baryonic density. 
			Right panel: mass dependence on energy density.
		} 
	\end{figure}

Figure \ref{MvsR} shows the mass-radius diagram, which display sequences of compact stars as continuous lines for each of the chosen parameters of models considered in this work. It also includes multi-messenger astronomy compact star measurements and derived constraints which are described in the figure caption. In addition, Figure~\ref{Mvsn_c} shows the masses of the compact stars as function of their corresponding central baryon densities (left panel) and of their corresponding central energy densities (right panel).

\vspace{-6pt}
 	{\begin{figure}[H]
	
	\includegraphics[width=10.5 cm]{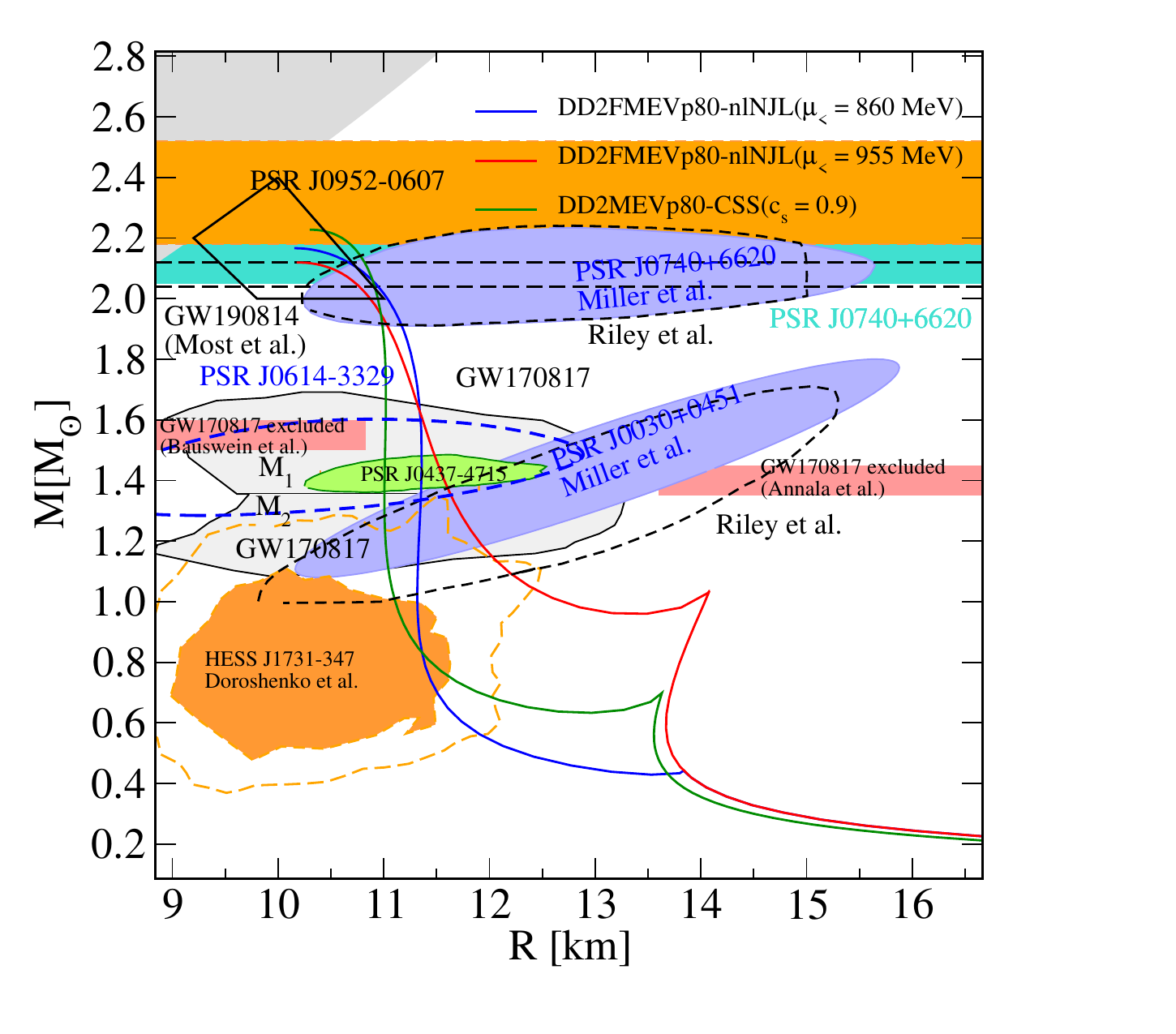}\vspace{-14pt}
	\caption{\label{MvsR} Mass-radius diagram with constraints derived from observations of compact stars. In the left upper corner there is a gray region where compact stars cannot be populated because of causality}
	\end{figure}  
	\captionof*{figure}{ violation in the EoS describing them.  The horizontal bands above 2 M$_{\odot}$ correspond to mass measurements of PSR J0740+6620~\cite{NANOGrav:2019jur} and to PSR J0952-0607~\cite{Bassa:2017zpe}, which falls within the category of~\textit{black widow
} compact star and is one of the most massive objects of this type detected. The horizontal dash lines define a band around 2.08 M$_{\odot}$, which is a  lower bound on the maximum mass derived from GW190814 under the assumption that one of the compact objects was a fast rotating neutron star involved in the corresponding merger~\cite{Most:2020bba}. The blue ellipses correspond to the 2$\sigma$ confidence level measurements of PSR J0030+0451~\cite{Miller:2019cac}
and PSR J0740+6620 by NICER~\cite{Miller:2021qha}, {and the black, dashed ellipses to an alternative analysis of the same objects~\cite{Riley:2019yda,Riley:2021pdl}. }  Similarly, the dashed, blue ellipse with mass measurement around $M = 1.44$ M$_{\odot}$ corresponds to the most recent NICER measurement of PSR J0614-3329~\cite{Mauviard:2025dmd}. Furthermore, from the  GW170817 event, an estimate of the properties of both components of the merger, labeled as M$_1$ and M$_2$ in the gray regions, was  derived from the analysis of the GW signal emitted during the inspiral phase~\cite{LIGOScientific:2018cki}. The small green region at the 2$\sigma$ confidence level that overlaps with the GW170817 components is another NICER measurement, this one of the object PSR J0437-4715~\cite{Choudhury:2024xbk}.
 Red bands are forbidden regions derived also from GW170817 by~\cite{Bauswein:2017vtn} and~\cite{Annala:2017llu}. Hybrid compact stars including the hybrid twins within this work satisfy the measurement of the very compact object HESS J1731-347 as reported in~\cite{Doroshenko:2022nwp} and are displayed in orange for the 1$\sigma$ and 2$\sigma$ confidence levels. The polygonal area above 2 M$_{\odot}$ in the left upper corner corresponds to allowed configurations if the remnant of GW170817 was short-lived, implying that those outside this region should be ruled out~\cite{Sneppen:2024jch}.}}
\vspace{8pt}

\subsubsection*{Moment of Inertia and Tidal Deformabilities}

\begin{figure}[H]
		
		\vspace{-4mm}
		\includegraphics[width=0.51\textwidth]{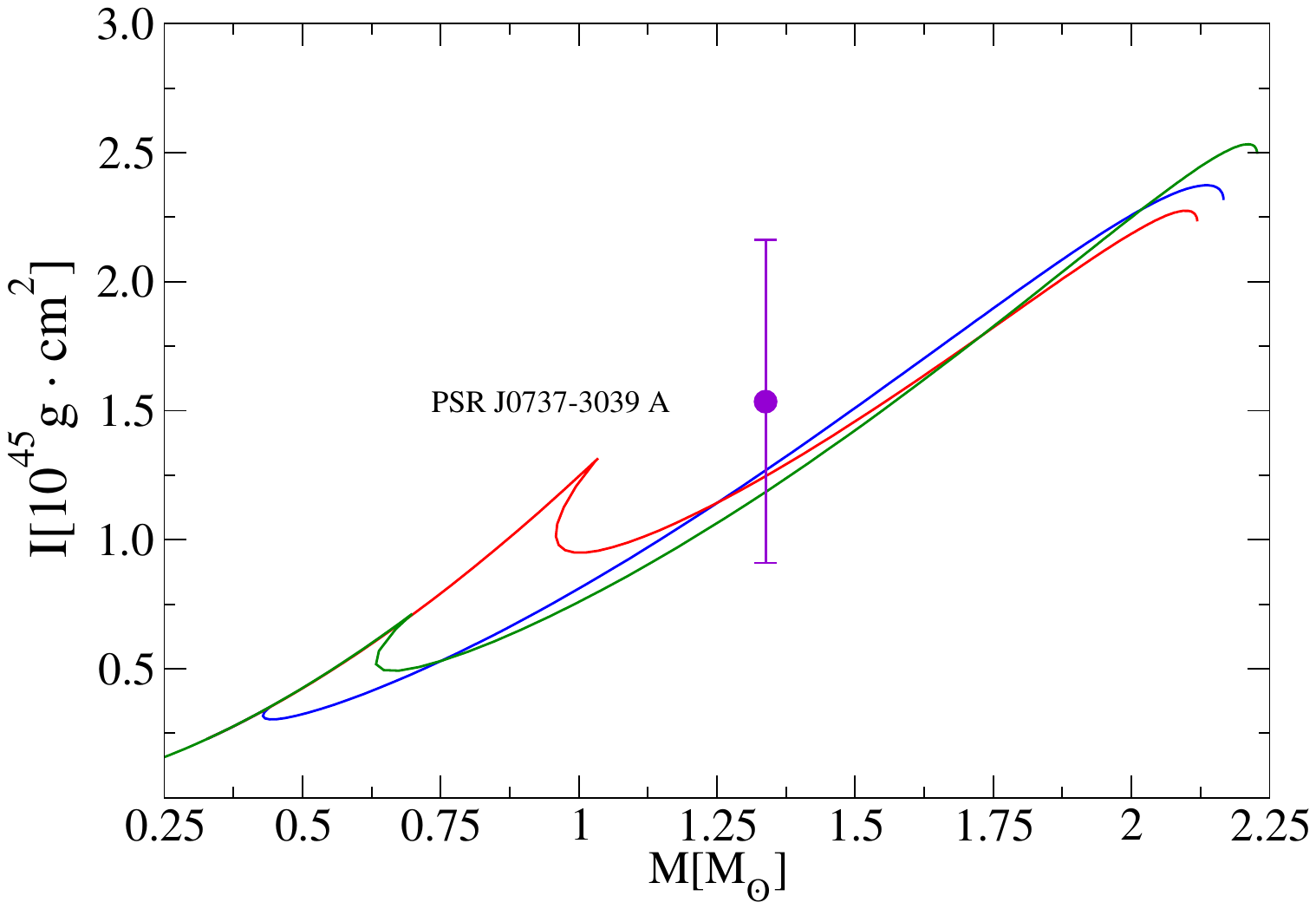} \hspace{-0.5cm}
		\includegraphics[width=0.51\textwidth]{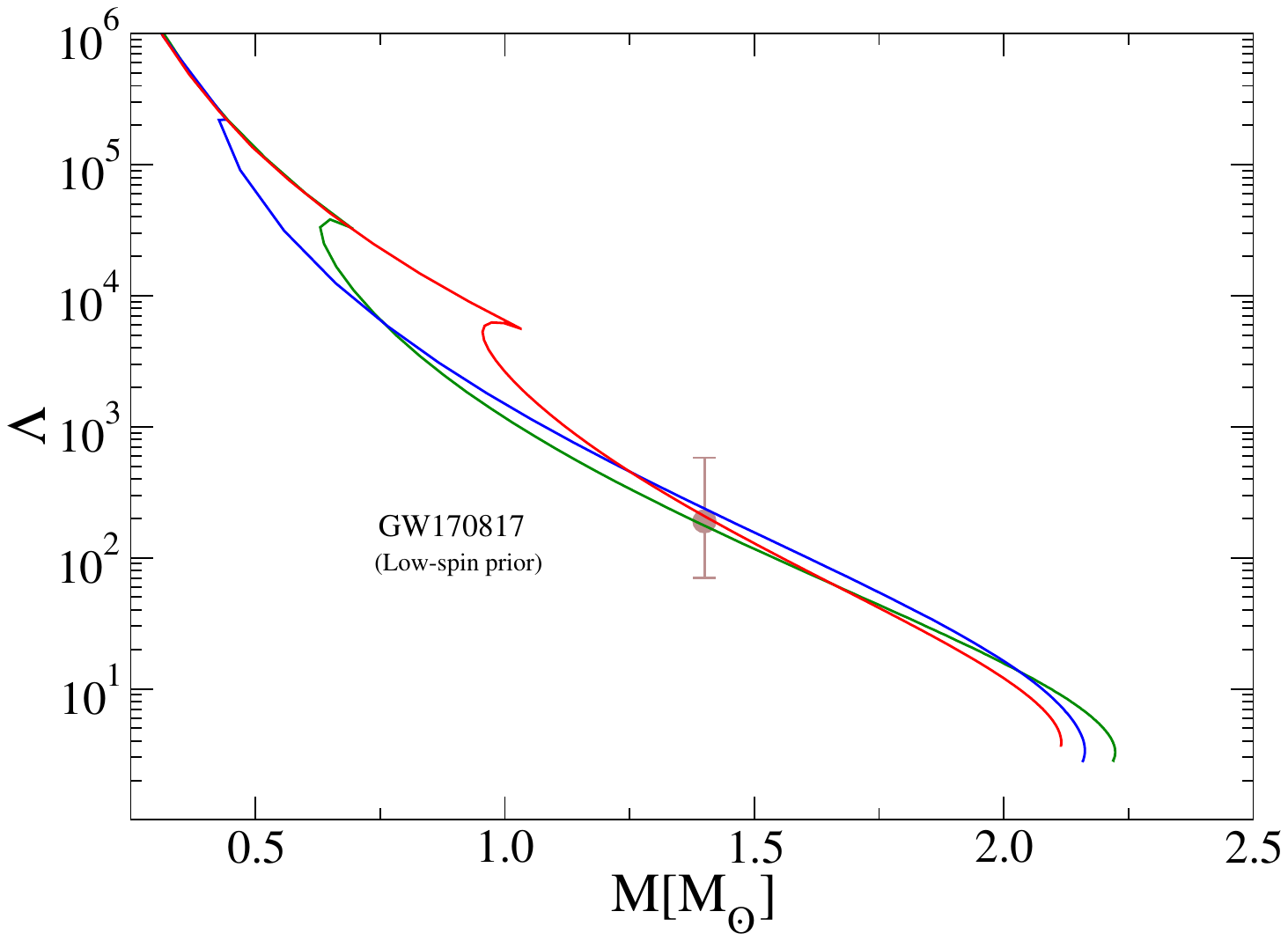} 
		\caption{\label{MoI_TD}
		Compact star measurements within the 2$\sigma$ confidence level. 
			Left panel:  Moments of inertia for compact stars together with the measurement of PSR J0737-3039 A~\cite{Landry:2018jyg,Miao:2021gmf,Silva:2020acr}. 
			Right panel: {Computed dimensionless tidal deformabilities for the equations of state in this work. The vertical error bar is an estimation from the GW170817 event under the assumption of stellar low-spin dynamics right before occurrence of the merger~\cite{LIGOScientific:2018cki}.}
		} 
	\end{figure}

The moment of inertia (MoI) is an important quantity that can potentially be {estimated} with observations of binary systems of pulsars. The work of~\cite{Kramer:2021jcw}, which consists of the analysis of 16 years of data from the object PSR J0737-3039 A, has resulted in a constraint for the MoI of the 1.338 M$_{\odot}$ star of the system, $I_{A} < 3.0 \times 10^{45}$g cm$^{2}$. By using data from other multi-messenger observations of compact stars like the measurement of LIGO/Virgo and NICER into universal relations insensitive to the EoS~\cite{Landry:2018jyg,Miao:2021gmf,Silva:2020acr}, derived estimations fall within $0.91\times 10^{45}$g cm$^{2}< I_{A} < 2.16.0 \times 10^{45}$g cm$^{2}$, which is shown as an error bar in the left panel of Figure \ref{MoI_TD} together with computed MoI values for the EoSs in this work. Thus, the relativistic moment of inertia is computed based on the approach presented in~\cite{Ravenhall:1994}
\begin{eqnarray}
 I &\simeq& \frac{J}{1+2J/R^{3}}~,\\
 J &=&\frac{8\pi}{3}\int_{0}^{R}  dr r^{4}\frac{\varepsilon(r)+p(r)}{1-2m(r)/r}. 
\end{eqnarray}
For a detailed discussion of the moment of inertia in the slow-rotation approximation, and for the hybrid star case, see, e.g.,~\cite{Chubarian:1999yn,Zdunik:2005kh,Bejger:2016emu}, and references therein. Interestingly, the system of PSR J0737-3039 suffers precession, which changes the direction of the pulsar beam; therefore, the effect is the appearance and disappearance of the pulsar as seen from Earth~\cite{Lattimer:2006xb}, which, in turn, provides the possibility of the first direct detection of a compact star moment of inertia~\cite{Kramer:2009zza}. By looking at the left panel of Figure~\ref{MoI_TD}, which shows the moments of inertia for the hybrid compact stars in this work, it is clear that the aforementioned constraint for PSR J0737-3039 A is fulfilled. See also~\cite{Li:2024sft} for similar results and discussion.

\begin{figure}[H]

\includegraphics[width=10.3 cm]{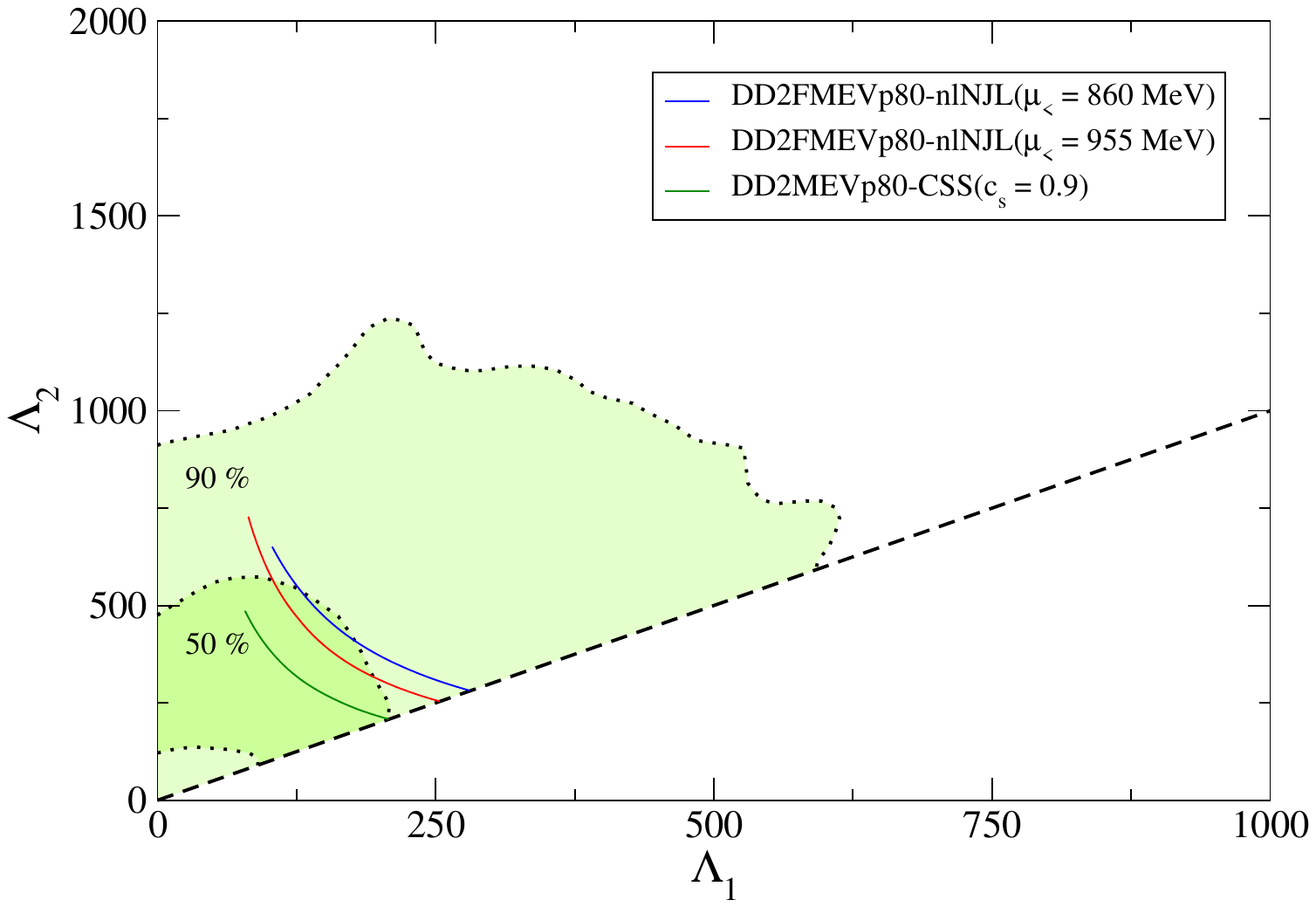}
\caption{GW170817 
 Tidal deformabilities diagram. The color codes for the lines are the same
as in figure 1. Green regions delimited by dotted lines correspond to 1$\sigma$ and 2$\sigma$ confidence levels for the measurement of the tidal deformabilities of the two stellar components $\Lambda_1$ and of $\Lambda_2$ of the merger~\cite{LIGOScientific:2018cki}. The hybrid EoS describing compact stars in this work can fulfill this observation.\label{Lambda1-Lambda2}}
\end{figure}

{The tidal deformability (TD) $\lambda$} of a compact star measures the effect of the deformation from the spherical shape of the star caused by an external gravitational field, which in the case of a binary system, is caused by its companion star. The level of deformation of the stars that merged in the GW170817 event has been derived from an analysis of the detected gravitational wave signal before the fusion of the stars; therefore, it has been taken as a constraint for neutron star matter. For a compact star described by a given EoS, the TD can be computed following the prescription derived in~\cite{Hinderer:2007mb,Damour:2009vw,Binnington:2009bb,Yagi:2013awa,Hinderer:2009ca}. The dimensionless TD $\Lambda$ depends on the stellar TD $\lambda$ and the stellar mass $M$, and $\Lambda=\lambda/M^{5}$,  and is computed for small tidal deformabilities. Moreover, $\lambda$ is related to the Love number $k_2$ 
\begin{equation} 
k_2 = \frac{3}{2} \lambda R^{-5}.
\label{k2def}
\end{equation}
The TD corresponds to a linear $l=2$ perturbation onto the spherically symmetric body representing the star,
\begin{eqnarray}
ds^2 &=& - e^{2\Phi(r)} \left[1 + H(r) Y_{20}(\theta,\varphi)\right]dt^2
\nonumber\\
& & + e^{2\Lambda(r)} \left[1 - H(r) Y_{20}(\theta,\varphi)\right]dr^2
\nonumber \\
& & + r^2 \left[1-K(r) Y_{20}(\theta,\varphi)\right] \left( d\theta^2+ \sin^2\theta
d\varphi^2 \right),
\end{eqnarray}
with $K'(r)=H'(r)+2 H(r) \Phi'(r)$, and primes denote derivatives with respect to $r$. 
$H(r)$ and $\beta(r) = dH/dr$ are determined by
\begin{eqnarray}
\frac{dH}{dr}&=& \beta\\
\frac{d\beta}{dr}&=&2 \left(1 - 2\frac{m(r)}{r}\right)^{-1} \nonumber\\
&& H\left\{-2\pi
  \left[5\varepsilon(r)+9 p(r)+f(\varepsilon(r)+p(r))\right]\phantom{\frac{3}{r^2}} \right. \nonumber\\
&& \quad \left. +\frac{3}{r^2}+2\left(1 - 2\frac{m(r)}{r}\right)^{-1}
  \left(\frac{m(r)}{r^2}+4\pi r p(r)\right)^2\right\}\nonumber\\
&&+\frac{2\beta}{r}\left(1 -
  2\frac{m(r)}{r}\right)^{-1}\nonumber\\
&&  \left\{-1+\frac{m(r)}{r}+2\pi r^2
  (\varepsilon(r)-p(r))\right\}~,
  \end{eqnarray}
where {$f = d\epsilon/dp = 1/c_s^2$} is a function derived from the EoSs.  These equations require the mass and pressure profiles derived from the TOV equations; therefore, they are to be solved simultaneously.  Just like the TOV equations, the integration of the above equations starts near the center of the star towards the stellar surface with the consideration of the expansions $H(r)=a_0 r^2$ and $\beta(r)=2a_0r$ as $r \to 0$. Here, $a_0$ is an arbitrary constant since it cancels in the expression for the Love number; however, it determines the level of stellar deformation. Using
\begin{equation}
y = \frac{ R\, \beta(R)} {H(R)},
\end{equation}
the Love number for $l=2$ is computed as
\begin{eqnarray}
k_2 &=& \frac{8C^5}{5}(1-2C)^2[2+2C(y-1)-y]\nonumber\\
      & & \times\bigg\{2C[6-3y+3C(5y-8)]\nonumber\\
      & & ~~~+4C^3[13-11y+C(3y-2)+2C^2(1+y)]\nonumber\\
      & & ~~~+3(1-2C)^2[2-y+2C(y-1)] \ln(1-2C)\bigg\}^{-1},
\label{eq:k2}
\end{eqnarray}
with $C=M/R$ denoting the compactness of the star. The right panel of Figure~\ref{MoI_TD} shows the tidal deformabilities as a function of mass for the hybrid stars that include mass twins. The display error bar corresponds to the derived TD value from the GW170817 event. In addition, Figure~\ref{Lambda1-Lambda2} shows the probability confidence regions for the tidal deformabilities of each of the two components of the binary system that merged during the GW170817 event. {In this plot, each colored line for the EoS is computed by using as input the posterior probabilities for the mass components of the merger $m_{1}$ and $m_{2}$~\cite{LIGOScientific:2018cki}, which are approximately linearly correlated as
\begin{equation}
m_{2} = 1.36 - 0.7917 (m_{1} - 1.36),
\end{equation}
where $1.36 < m_{1} < 1.60$ and $1.17 < m_{2} < 1.36$ and $(1.36 - 1.17)/(1.6 - 1.36) = 0.791667$.} It is clear that the probability regions in this diagram tend to favor very compact stars like the ones described by the models in this work. It is interesting to see that within these models both stars can only be hybrid stars; none can be described as a pure hadronic star due to the compact star low mass onset, as can be seen in the mass-radius diagram~\ref{MvsR}. An interesting discussion about the nature of the stars associated with GW170817 can be found in~\cite{Blaschke:2020qqj}.    
\vspace{-6pt}
%
\subsection{Energy Release at the Onset of Deconfinement Leading to the Compact Star Twins Transition.}

\begin{figure}[H]
	
	\includegraphics[width=10.5 cm]{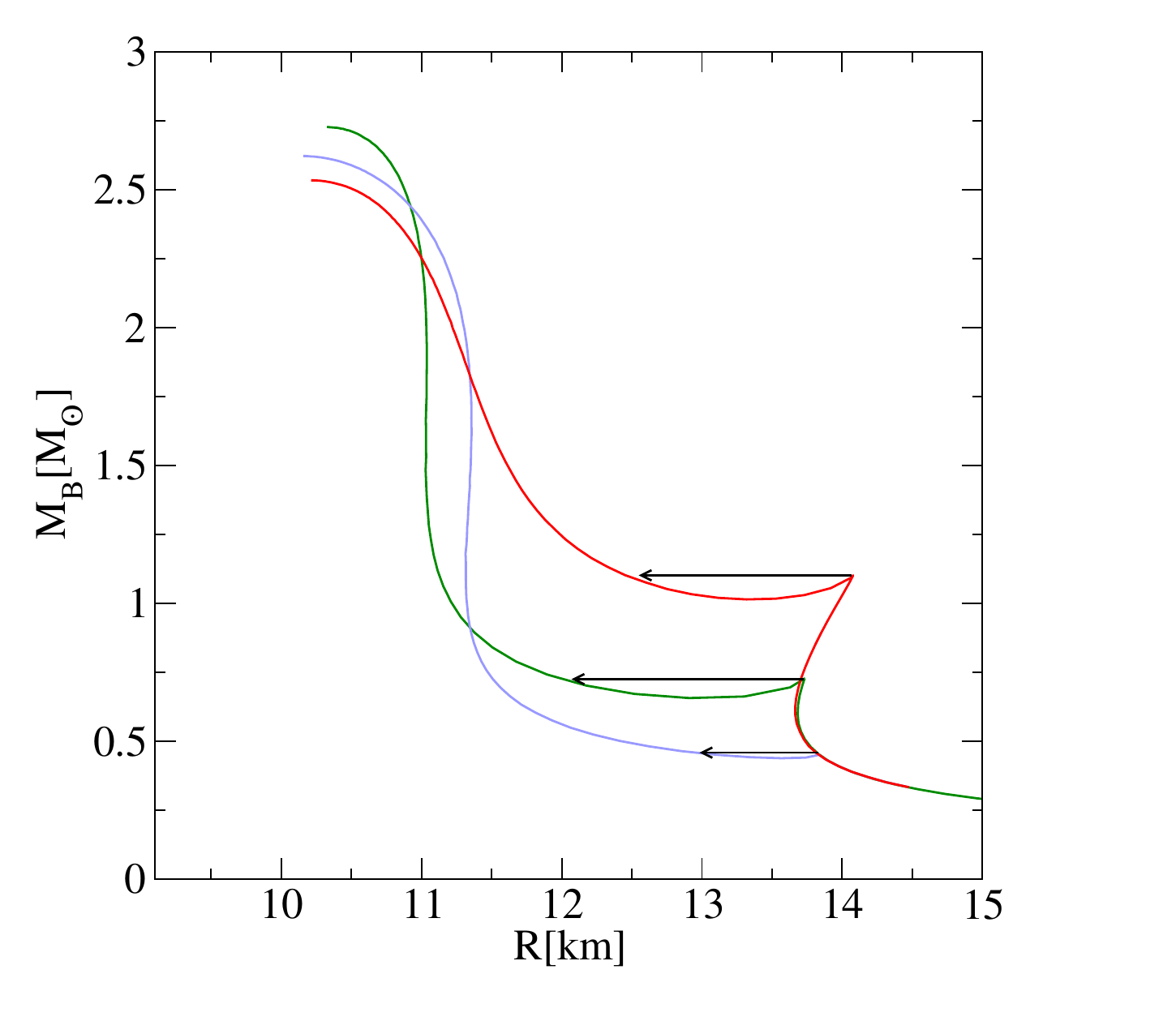}
	\caption{Baryonic 
 mass-radius diagram. The color codes for the lines are the same
as in figure 1. The arrows that point from right to left of between twin compact stars indicate an evolutionary path under baryon number conservation. \label{MbvsR}}
	\end{figure}

A pure hadronic compact star whose centrar density is slightly lower than the density of the deconfinement phase transition may reach such critical density value $n_c$ either by mass accretion from a companion or simply by slowing down its rotational frequency, which will rearrange its stellar interior density profile.  In this section, the twin EoSs are used to provide a raw estimate for the release of energy from stellar compactification due to a transition between the hadronic twin star into its hybrid twin by consideration of the conservation of baryon number density and under static conditions. Figure~\ref{MbvsR} shows the transition trajectories for a twin star in a stellar baryon mass vs. radius diagram.
A more extensive scan of the aforementioned transition under the same physical considerations can be found in~\cite{Alvarez-Castillo:2020nkp},  where a systematic change on the onset of deconfinement has been performed. The typical energy release values lie around $10^{51}$ ergs and depend on the radius difference between the two twin compact stars. Table~\ref{table_mass_defect} presents the change in radii, mass defect, and energy released following the stellar transition.

A more realistic study will consider the effect of the star rotation, possibly with considerations like the conservation of angular momentum. Following this philosophy, the calculations in~\cite{Alvarez-Castillo:2019apz,Chanlaridis:2024rov} show that this stellar transitional evolution can explain the existence of highly eccentric binary orbits featuring a millisecond pulsar (MSP) or, in some cases, leaving an isolated MSP after disruption of the orbit. The transition is expected to occur in low-mass X-ray binaries (LMXBs) triggered by accretion, possibly coupled with secondary pulsar kicks through neutrino or electromagnetic rocket effects. 
\begin{table}[H]
\caption{\label{table_mass_defect} Change in the compact star properties following the twin star transition.}
\begin{tabularx}{\textwidth}{Cccc}
\noalign{\hrule height 1pt} 
\textbf{Model} & \textbf{$\Delta $M} & \textbf{$\Delta$M} &\textbf{$\Delta$R}\\
& \textbf{[M$_{\odot}$]} & \textbf{[ergs]} & \textbf{[km]} \\
\midrule
DD2MEVp80-CSS ($c_s$ = 0.9) & {0.00057} &	$1.0275\times10^{51}$	& 1.72 \\
DD2FMEVp80-nlNJL ($\mu_{<} = $ 860 MeV) &{ 0.00003} &	$5.0461\times10^{49}$ &	0.75 \\
DD2FMEVp80-nlNJL ($\mu_{<} = $ 955 MeV) & {0.00114} & $2.0410\times10^{51}$ &	1.61\\ 
\noalign{\hrule height 1pt}
\end{tabularx}


\end{table}
%
%

\subsection{Rotating Compact Stars}

\begin{figure}[H]

	\includegraphics[width=10.5 cm]{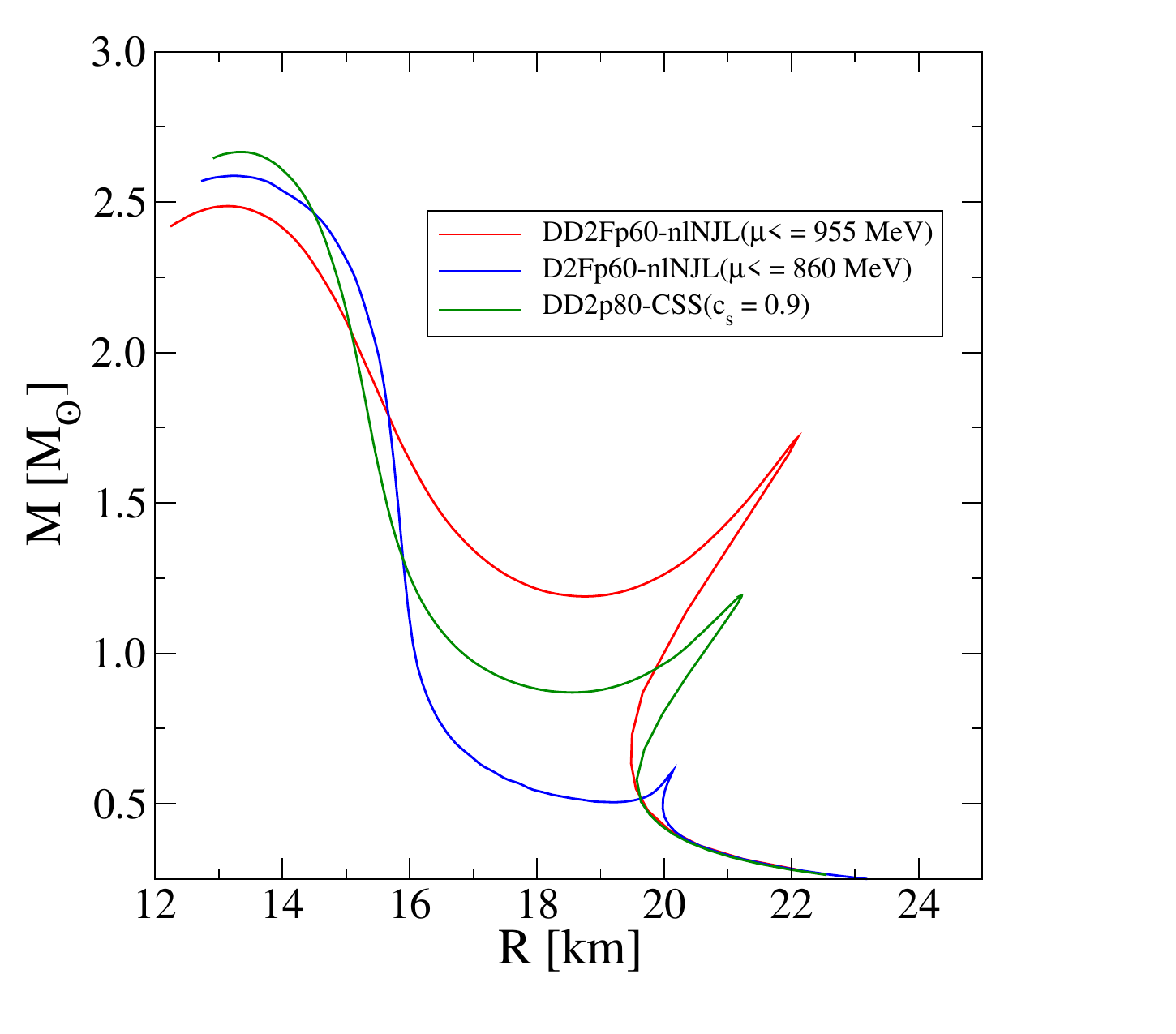}\vspace{-6pt}
	\caption{Compact stars rotating at the Kepler frequency. The color codes for the lines are the same
as in figure 1. \label{RotMvsR}}
	\end{figure}

The observed fast-rotating neutron stars represent a way of probing the extreme physics of nuclear matter. The mass-shedding frequency, the \textit{Kepler frequency, 
} is the maximum rotational frequency that a neutron star can have before its equatorial surface begins to shed mass due to centrifugal forces. In order to reproduce fast-rotating stars, the equation of state should be stiff enough to prevent matter from flying away. In addition to constraining the EoS, the spin rate of a neutron star can also provide clues about its formation history and its interactions with companion stars. Fast-rotating neutron stars have been detected, for instance, PSR J0952-060, which is the fastest  and heaviest known galactic object, spinning at a frequency of 707 Hz~\cite{Romani:2022jhd}, or PSR J1748-2446ad, which is spinning at a frequency of 716 Hz~\cite{Hessels:2006ze}. Other important aspects include the interplay of rotation with the magnetic field and the test of general relativity typically carried out with radio observations of pulsars. Importantly, compact stars rotating near their Kepler frequency are potential sources of gravitational waves. The properties of rotating compact stars for the EoS considered are derived using the RNS code~\cite{Nozawa:1998ak,Komatsu:1989zz,Cook:1993qj,Stergioulas:1994ea}, which solves the Einstein equations for an axisymmetric and stationary space-time, which is described by the metric
\begin{equation}
ds^{2} = -e^{\gamma+\rho}dt^{2}+e^{2\alpha}(dr^{2}+r^{2}d\theta^{2})+e^{\gamma-\rho}r^{2}\sin^{2}(\theta)(d\phi-\omega dt)^{2},
\end{equation}	
which include the potentials  $\gamma$, $\rho$, $\alpha$, and $\omega$, which are functions of the radial coordinate $r$ and the polar angle $\theta$. The method of solution for the rotating stellar configurations involve Green functions for these potentials, and the details can be found in the above references for the RNS code.
Figure~\ref{RotMvsR} shows compact stars rotating at the Kepler frequency. {Typical Keplerian frequencies range roughly from $10^3$s to $10^4$s, the highest values being found for the most massive hybrid compact stars in the corresponding mass-radius sequence. The properties of compact stars rotating at the Kepler frequencies for the EoS models here are presented in the Appendix A. 
} As a rule of the thumb, compact stars rotating at the Kepler frequency increase their mass by around 20$\%$~\cite{Breu:2016ufb}. Nevertheless, a recent work~\cite{Gartlein:2024cbj} has found that, for hybrid stars, such mass increases can vary between 0$\%$ and 30$\%$.   
For a discussion on the rotation properties of compact star twins, see~\cite{Blaschke:2019tbh}.

\subsection{$f$-Modes and Damping Times}

\begin{figure}[H]
		\centering
		\vspace{-5mm}
		\includegraphics[width=0.51\textwidth]{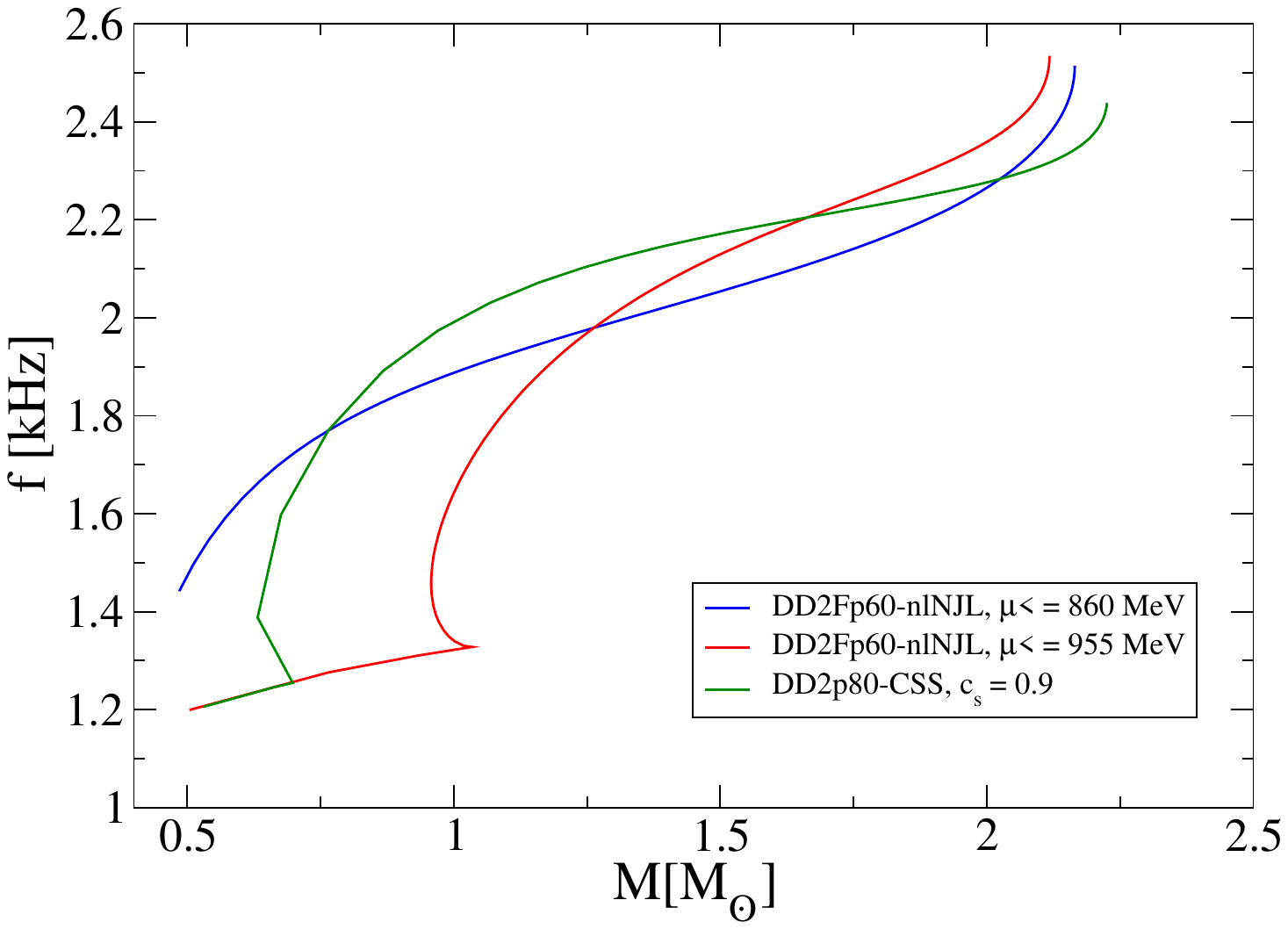} \hspace{-0.5cm}
		\includegraphics[width=0.51\textwidth]{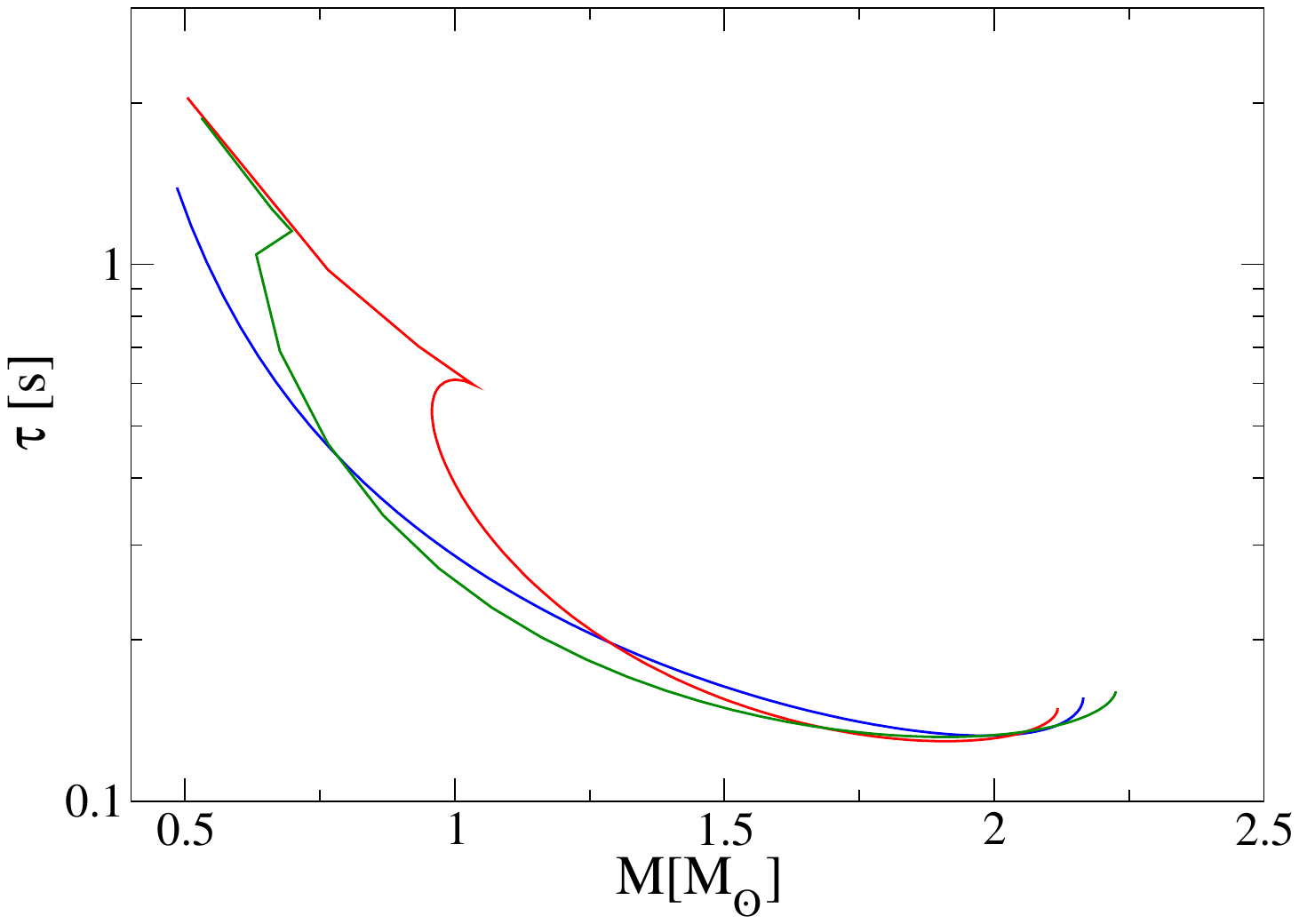}  
		\caption{\label{fig:f-modes} Compact star $f$-mode parameters, which can be investigated via detection of associated gravitational waves production. 
			Left panel:  Frequencies of the $f$-mode excitation for sequences of compact stars. 
			Right panel: Corresponding damping times of the $f$-modes. 
		} 
	\end{figure}

The fundamental modes of excitation of a compact star can release gravitational wave radiation and provide an imprint of the underlying EoS~\cite{Jyothilakshmi:2025wru}. According to~\cite{Kokkotas:1999mn,Ho:2020nhi},  the $f$-mode gravitational wave signal is given by
\begin{equation}\label{eqn:gwwaveform}
        h(t)=h_{0}e^{-(t-t_0)/\tau} \sin{\left[2\pi f (t-t_{0})+\phi\right]} ,\ \mathrm{for} \ t\geq t_{0}
\end{equation}
with
\begin{equation}\label{eqn:h0}
h_{0}= 4.85  \times 10^{-17} \sqrt{\frac{E_{\rm gw}}{M_{\odot}c^2}} \sqrt{\frac{0.1 {\rm sec}}{\tau}} \frac{1 \rm{kpc}}{d}\left(\frac{1 \rm{kHz}}{f}\right)~,
\end{equation}
where $f$ is the $f$-mode frequency, $\tau$ is its damping time, and $d$ is the distance to the compact star source of the GW. The excitation of the $f$-mode is expected to happen during the merger of compact stars or possibly during a pulsar glitch, for instance, under the assumption that most of the energy of the glitch is transformed into gravitational waves. If this is the case~\cite{Ho:2020nhi} ,
\begin{equation}\label{eqn:egw}
    E_{\text{gw}}=E_{\text{glitch}}=4\pi^2I\nu^2 (\frac{\Delta \nu}{\nu})~,
\end{equation}
the GW energy $E_{\text{gw}}$ will depend on the moments of inertia $I$ and spin frequency $\nu$ of the star,  and, most importantly, on  $\frac{\Delta \nu}{\nu}$, the relative change in spin frequency of the glitch. The corresponding mode parameters are derived by solving the perturbations in the full general relativistic treatment through the direct integration method developed in~\cite{Detweiler:1985zz,Sotani:2001bb,Pradhan:2022vdf}. In this way, it is possible to find the compact star $f$-mode frequency, the complex $f$ -mode frequency ($\omega = 2\pi f + \frac{1}{\tau}$), which corresponds to the outgoing wave solution to the Zerilli’s equation at infinity~\cite{Zerilli:1970se}. Thus, the real part of $\omega$ represents the $f$-mode angular frequency and the imaginary part represents the damping time $\tau$. Figure~\ref{fig:f-modes} shows both $f$ and $\tau$ for the EoSs of this work. The importance of measuring such quantities stem from the fact that thanks to universal relations insensitive to the EoS, it is possible to map $f$ and $\tau$ into $M$ and $R$:
\begin{equation}\label{eqn:remomega_compactness}
    \mathrm{Re}(M\omega)=a_0+a_1\  \left(\frac{M}{R}\right)+a_2\ \left(\frac{M}{R}\right)^2
\end{equation}
 \begin{equation}\label{eqn:immomega_compactness}
    \mathrm{Im}(M\omega)=b_0 \  \left(\frac{M}{R}\right)^4+b_1 \ \left(\frac{M}{R}\right)^5 +b_2 \ \left(\frac{M}{R}\right)^6~.
\end{equation}
The coefficients in the above equations have been derived in different studies and provide a measure of the level of accuracy. They can be read from the work of~\cite{Pradhan:2023zmg}, where a comparison with the values of other works is presented in the Appendix of that article. 
 Moreover, within that work we have shown that the level of accuracy of future GW interferometers like the Einstein Telescope or the Cosmic Explorer will allow to study the transition region near $M_{onset}$ in the mass-radius diagram in order to identify compact star twins.
\subsection{The Crust of Twin Compact Stars}

\begin{figure}[H]
	
	\includegraphics[width=10.5 cm]{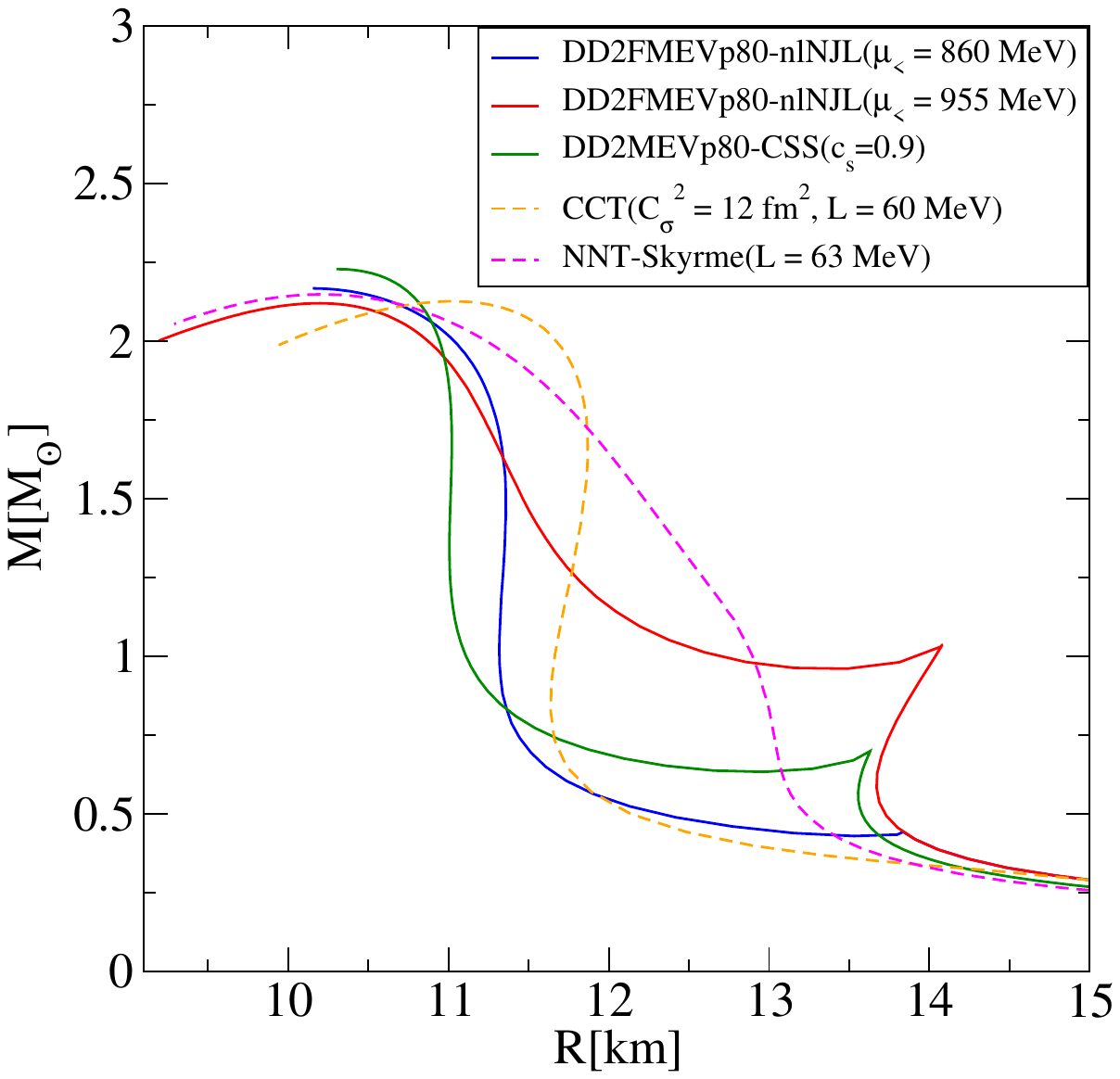}\vspace{-6pt}
	\caption{Compact star sequences, which include pure hadronic (dashed) and hybrid stars (solid) with similar symmetry energy values at saturation density $n_0$, see the text and Table~\ref{table_crust_parameters}.\label{MvsR_crust}}
	\end{figure}

\begin{figure}[H]
	
	\includegraphics[width=0.48\textwidth]{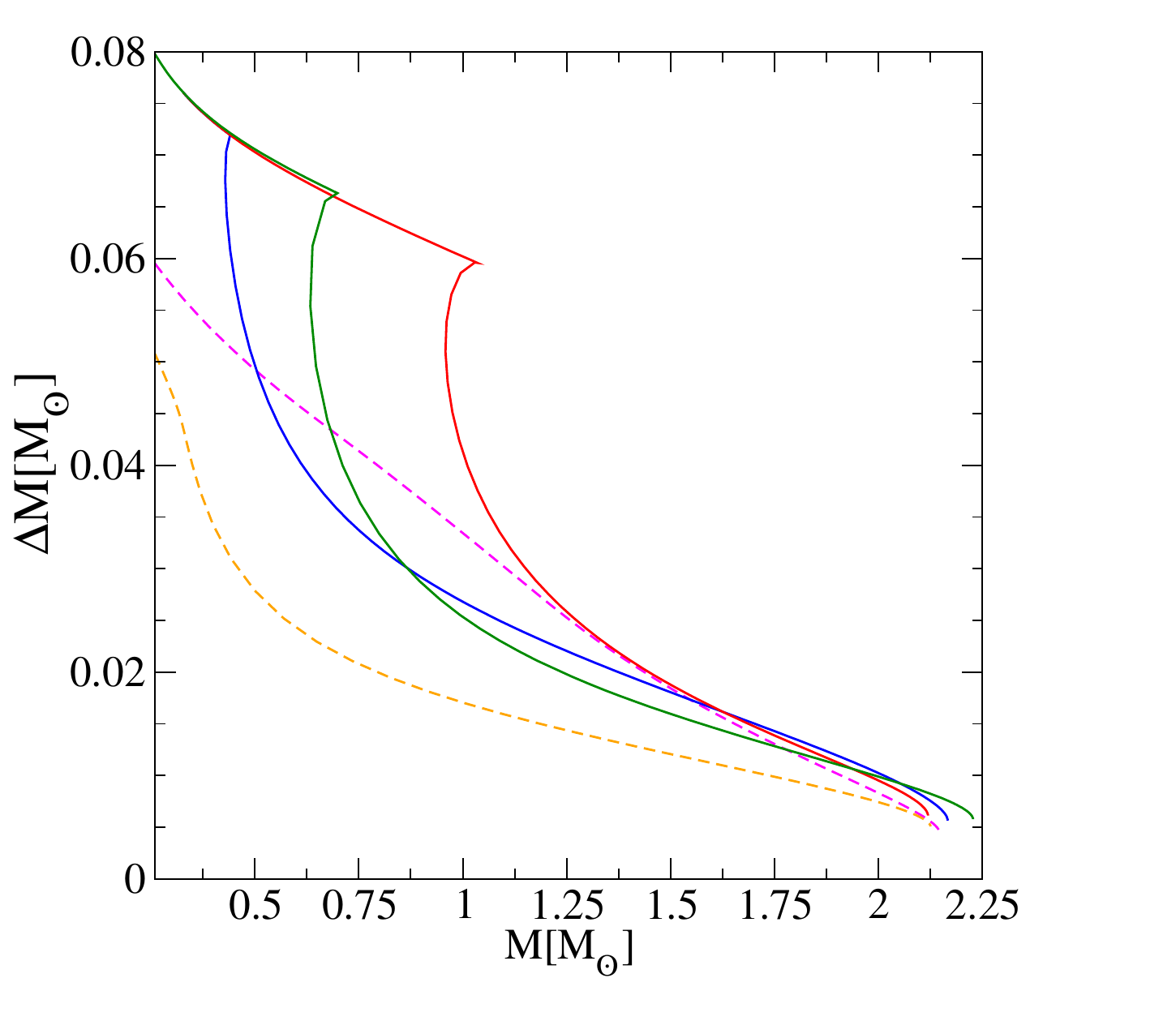}
	\includegraphics[width=0.48\textwidth]{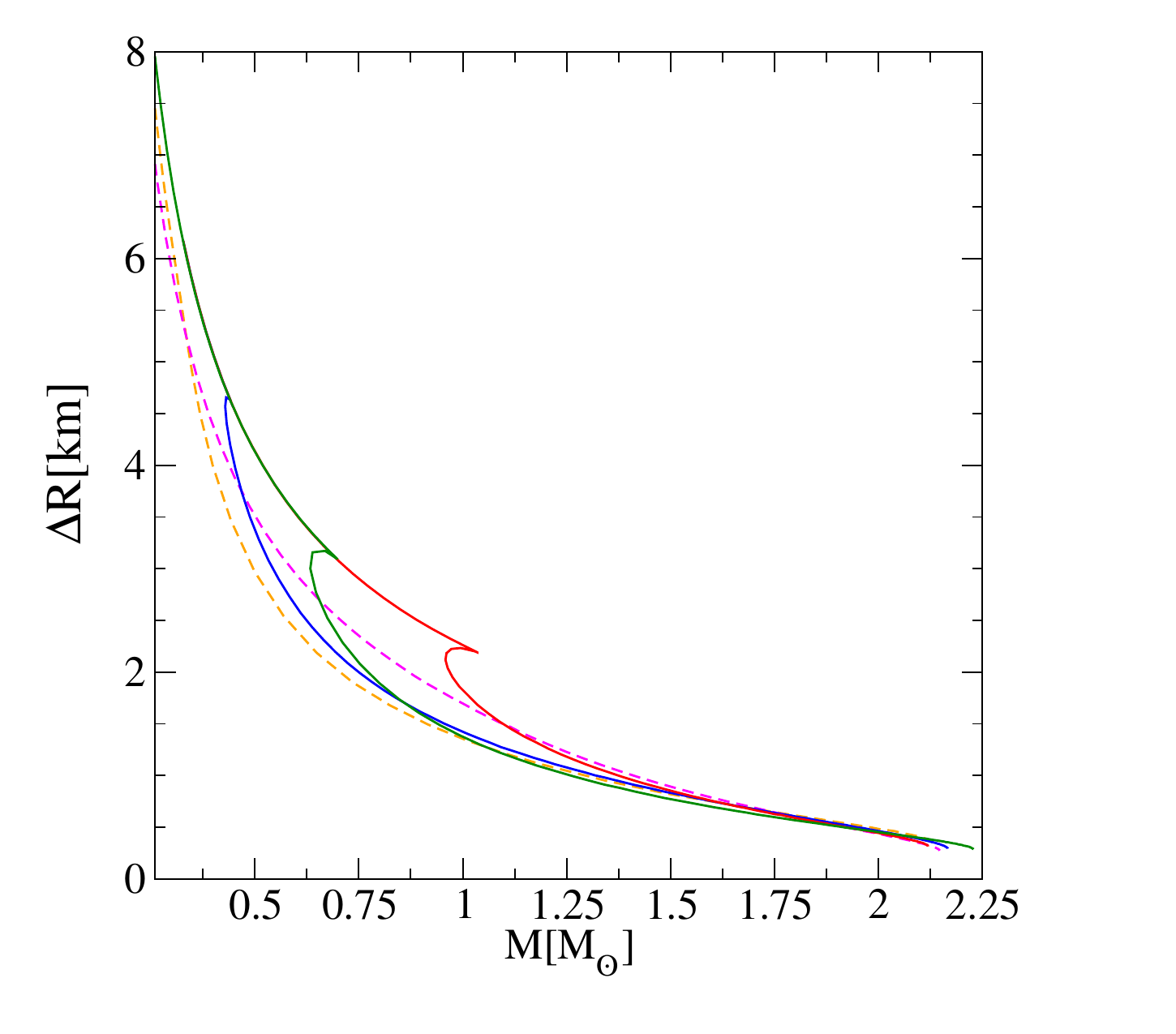} \\
	\includegraphics[width=0.48\textwidth]{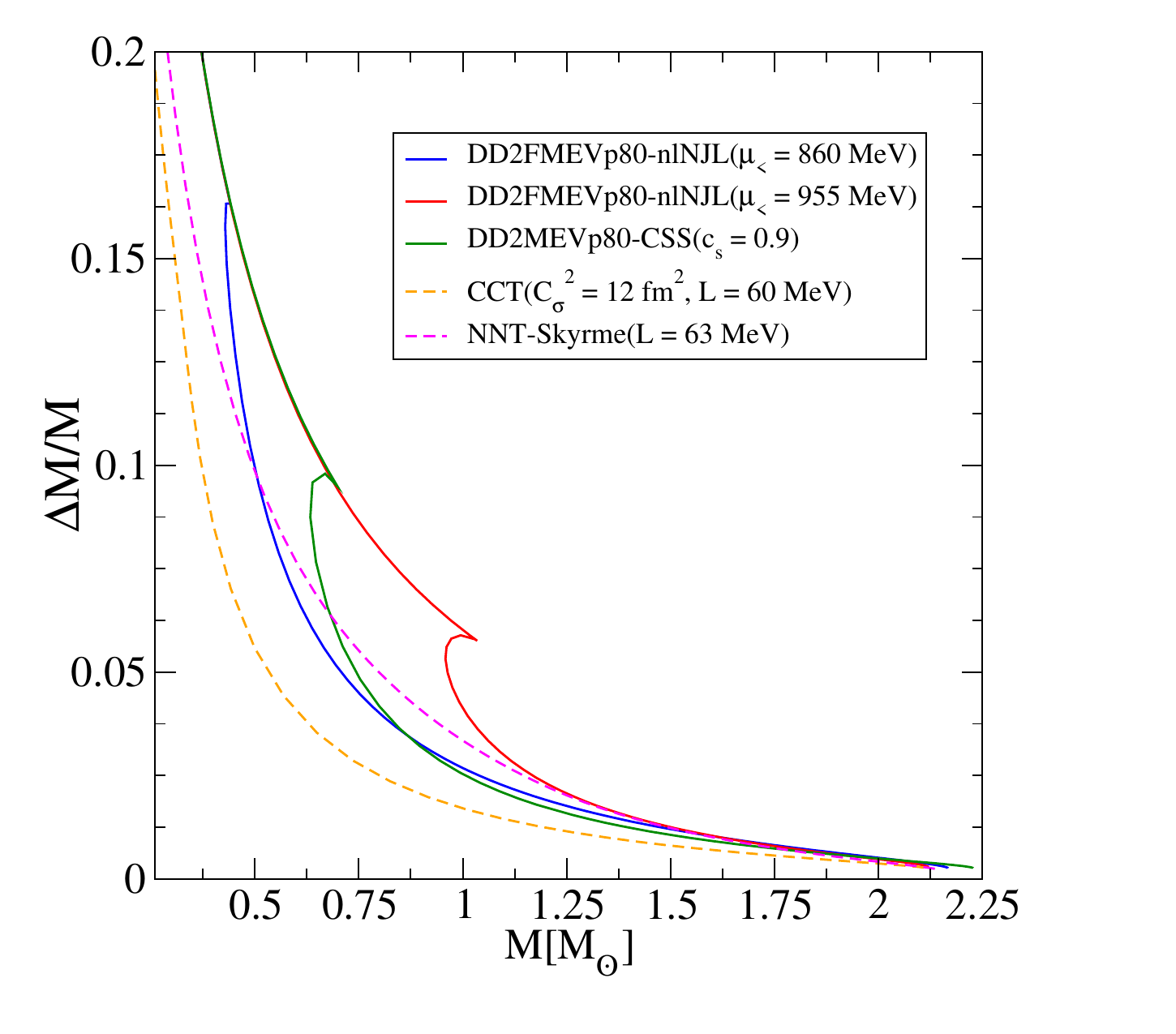}
	\includegraphics[width=0.48\textwidth]{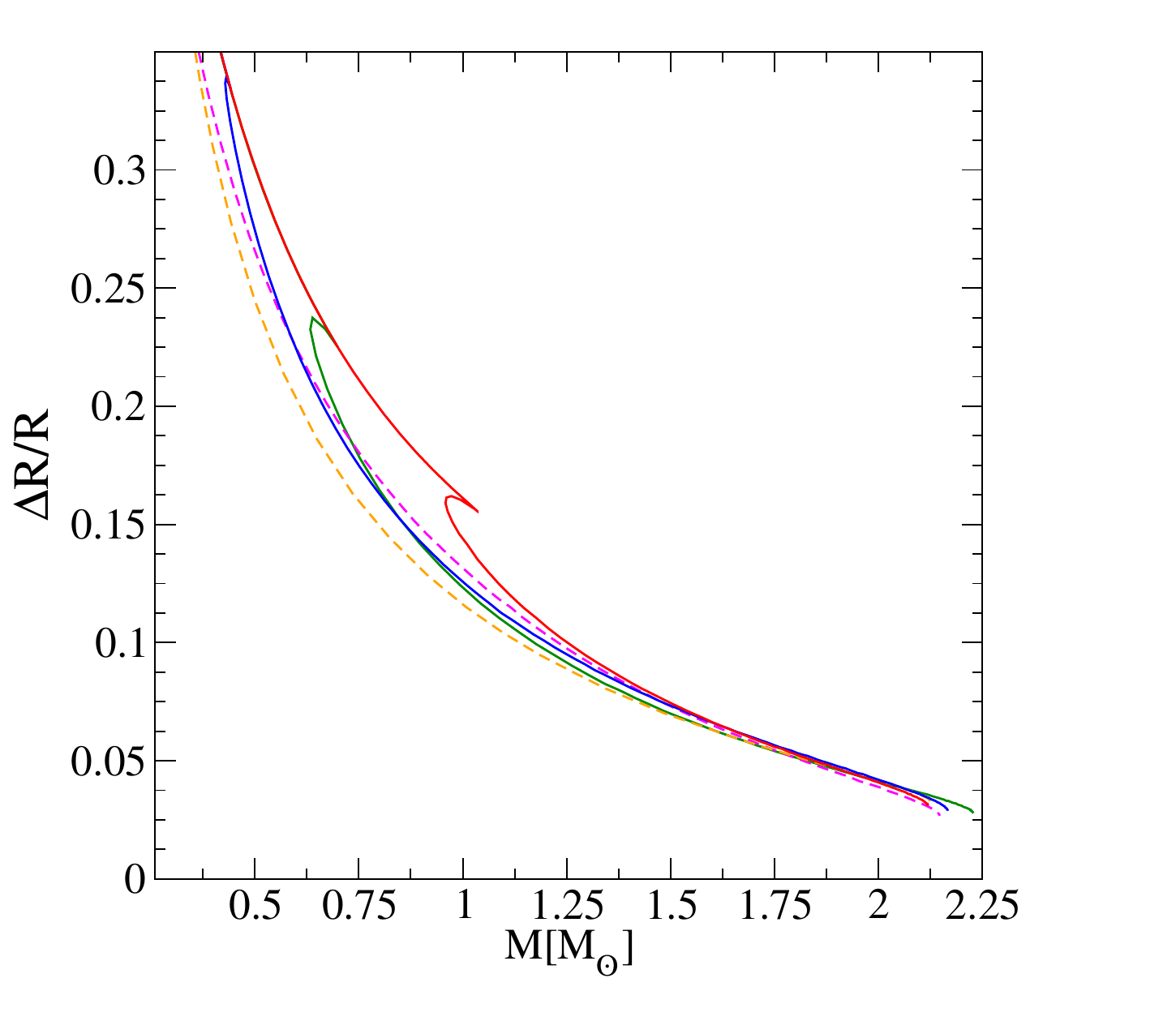}\vspace{-3pt}
	\caption{Properties of the crusts of compact stars for the EoS models in consideration. The upper figures show the mass $\Delta M$ and radius values $\Delta M$ for compact stars as a function of their total mass $M$, whereas the same quantities relative to either the total stellar mass $M$ or total stellar radius $R$ are displayed in the figures below. \label{Twins_crust_M_n_R}}
	\end{figure}

The crust of compact stars is composed of atomic nuclei in a crystallized structure and can be estimated to have a thickness of a few kilometers. As the pressure increases toward the center of the star, the bottom of the crust suffers a phase transition into a nuclear fluid, which defines the compact star core. Because of different internal density profiles of a compact star due the to the EoS model, hybrid neutron stars will posses crusts with different masses and thicknesses than those of pure hadronic stars.  The crust--core transition $n_{cc}$ can be defined in different ways depending on the criterion under consideration for the crust dissolution. For instance, the approach introduced in~\cite{Kubis:2006kb} considers the limit stability of nuclear structures by a thermodynamic method, i.e., the stability of homogeneous beta-equilibrated nuclear matter defined by the compressibility under constant chemical potential, which should obey the condition $K_{\mu} >0$ for matter to be stable
\begin{equation}
 K_\mu = n^2(E_s''\alpha^{2} +V'') + 2 n (E_s'\alpha^{2} +V') -
\frac{2 \alpha^{2} E_s'^{2} n^{2} }{E_s},
\label{Kmu}
\end{equation}
where $E_s$ is the symmetry energy, $V$ is the energy of symmetric nuclear matter, and $\alpha=(1-2x)$ is the isospin asymmetry with $x$ as the proton fraction. All these quantities are functions that depend on the baryon density $n$ with the primes denoting derivatives with respect to it. Alternatively, it is possible to also consider finite size effects like Coulomb and surface contributions of the nuclei. As shown in~\cite{Baym:1971ax}
\begin{equation}
v(Q)=v_{min}=v_0+2(4\pi e ^{2} \beta)^{1/2}-\beta k_{TF}^{2}
\end{equation}
is the minimal value of stable density modulations for the momentum $Q$, $k_{TF}$ being the Thomas--Fermi momentum and $\beta$ being a quantity that is related to the baryon number and chemical potential of the particle species $i$ in nuclear matter. Similarly to the thermodynamic method, the condition $v(Q)>0$ defines the stability of matter. The following relation between the above quantities hold~{\cite{Baym:1971ax} }
\begin{equation}
 v_0(n)=\frac{8K_{\mu}(n)E_s(n)}{n^{2}}\left(\frac{\partial \mu_i} {\partial n_i} \right)^{-1}.
\end{equation}
Thus, both of the above approaches consider stability of a one-phase system against density fluctuations. The general result is that the latter approach for determination of $n_{cc} $ gives lower values. Furthermore, in other works like~\cite{Fortin:2016hny}, a prescription for $n_{cc}$ has been derived taking optimal values in order to handle uncertainties on the variations of the value of $L$ that affect the stellar radius determination for non-unified EoS, or in~\cite{Canullan-Pascual:2025smm}, the authors introduce a thermodynamically and causally consistent formalism through an interpolation function. 
Nevertheless, $n_{cc}$ values must be the same for hybrid and pure hadronic stars with the same EoS for the hadronic mantle, because the case is that matter is described for the same model at such low densities. Importantly, the thickness of the crust for those two cases will be different, due to the distribution of matter, i.e., the density profile inside either the hadronic or hybrid star. Thus, the mass twins are perfect example of crust differences for two stars with the same mass $M$. Interestingly, the symmetry energy $E_{s}$ plays an important role in the determination of $n_{cc}$; therefore, in the figures of this section, two more EoSs of pure hadronic character bearing similar slopes of the symmetry energy parameter $L$ have been introduced to serve as a reference. Table \ref{table_crust_parameters} shows the symmetry energy values at saturation density $n_0$ together with  {$n_{cc}$}. The three twin star models in this work share the same equation of state at low densities up to saturation; therefore, they are listed only once. The two extra entries in the table below the DD2 models correspond to pure hadronic EoSs with similar $S$ and $L$ values to the rest. It can be seen that their { $n_{cc}$} lie either below or above than that for the DD2 models, a result that depends of the symmetry energy parameters. Those hadronic EoSs are the CCT-C$^{2}{_\sigma}12$-L60~\cite{Kubis:2023gxa}, which, for its chosen parameters, is soft enough within an intermediate density region in order to produce compact stars inside the HESS J1731-347 mass-radius region, as well as the NNT-SKyrme-L63 EoS described in~\cite{Neill:2020szr}, which, conversely, does not describe HESS J1731-347. Figure~\ref{MvsR_crust} shows the mass-radius diagram for the five EoSs considered in this section. In it, it can be seen that compact stars near the maximum mass have more or less the same properties, implying similar crust properties, as will be discussed below.
Figure~\ref{Twins_crust_M_n_R} shows a set of diagrams for all the EoSs that fully describe the derived properties of the crust of compact stars as a function of the total stellar mass $M$, the crust thickness $\Delta R$, and the mass of the crust $\Delta M$. Furthermore, as presented in the mass-radius plot, the change in values following the phase transition is clearly visible. Conversely, for the most massive compact stars near the maximum mass, all the equations of state that, in this case, share similar symmetry energy values at saturation display negligible differences in their crust properties in comparison to stars near the onset of deconfinement.

	\vspace{-6pt}
	
\begin{table}[H]
\caption{\label{table_crust_parameters}Parameters of a selection of hadronic EoSs that include those describing the mantle of twin stars in this work. The right-most column displays the crust--core transition density {$n_{cc}$}, which allows an estimation of the properties of the crust of both pure hadronic and twin hybrid stars, see the text for discussion.}
\begin{tabularx}{\textwidth}{Ccccc}
\noalign{\hrule height 1pt} 
\textbf{Hadronic Model} &	\boldmath{$S$} &	 \boldmath{$L$} & 	\boldmath{$n_{cc}$}\\
&					\textbf{[MeV]} &		\textbf{[MeV]}&	\textbf{[1/fm$^{3}$]}\\
\midrule
DD2MEVp80/DD2FMEVp80 & 34.74 &63.87 &0.078 \\
CCT-C$^{2}{_\sigma}12$-L60 & 30.00 &60.00 &0.0644 \\
NNT-SKyrme-L63 & 30.60 &63.12 &0.081 \\
\noalign{\hrule height 1pt}
\end{tabularx}


\end{table}

\begin{figure}[H]

\includegraphics[width=10cm]{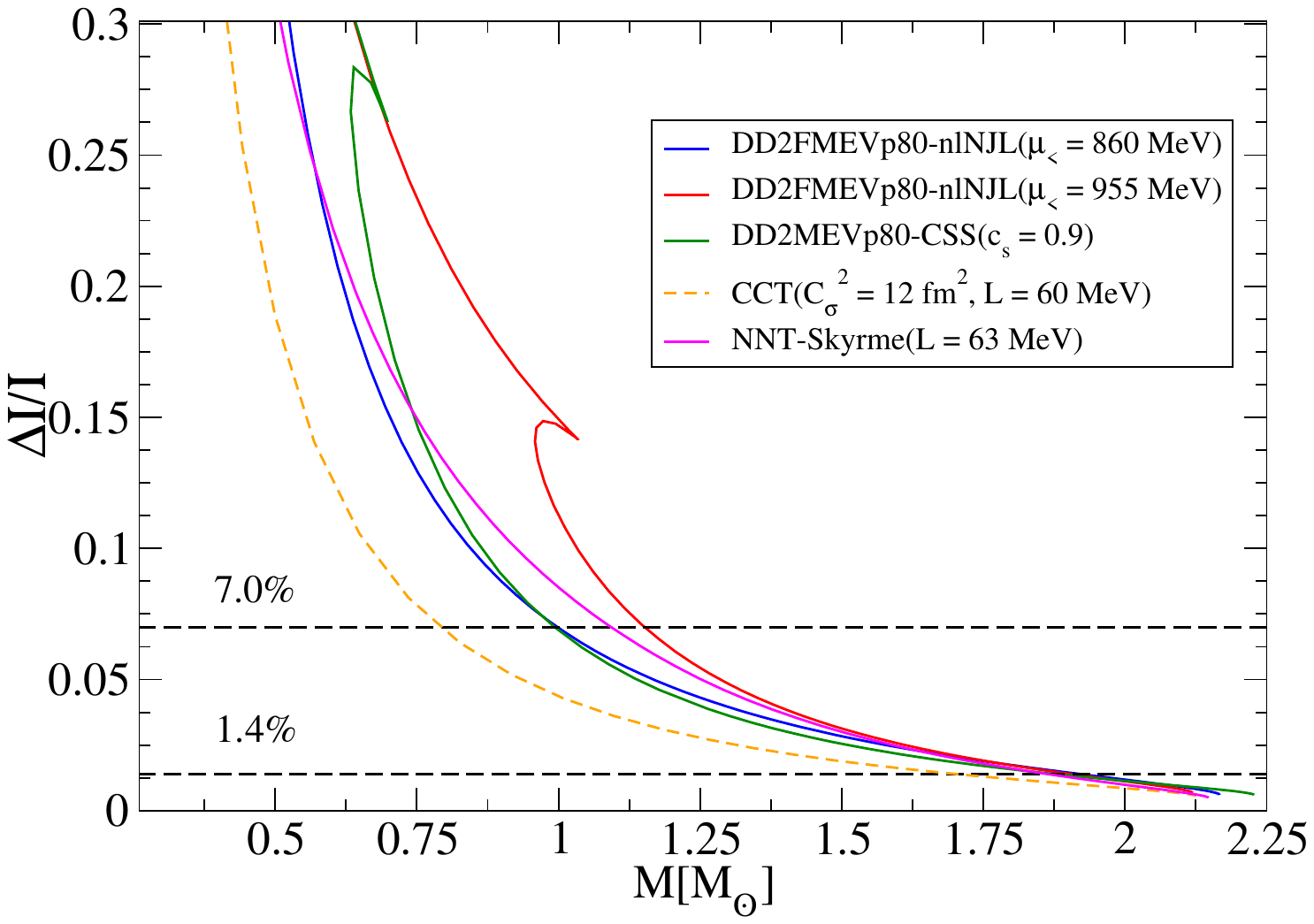}
\caption{Fraction of the stellar moment of inertia carried by the crust together with pulsar glitch constraints for the compact star mass $M$ of a glitcher derived by the Vela pulsar ($\Delta I/I$>1.4$\%$), and from consideration of entrainment of superfluid neutrons in the crust ($\Delta I/I$>7.0$\%$).\label{Frac-MoI-Crust}}
\end{figure}

In addition, Figure \ref{Frac-MoI-Crust} shows fraction of the moment of inertia carried by the crust $\Delta I/I$ together with some estimations of limiting values from the analysis of \textit{pulsar glitches}. A pulsar glitch is defined as a sudden change in the rotation frequency of the star leading to a spin-up that is observed in the pulsar radio signal~\cite{Antonelli:2023vpd}. The line at 1.4$\%$ excludes the region below it and has been derived by an analysis of the Vela pulsar together with six other glitchers with the characteristic that it is independent of the pulsar mass~\cite{Link:1999ca}, whereas a more recent study of a similar kind has lead to a constraint demanding values above 1.6$\%$~\cite{Andersson:2012iu}. A more stringent constraint that accounts for the entrainment of superfluid neutrons in the neutron star crust correspond to the line at 7.0$\%$~\cite{Chamel:2012zn,Andersson:2012iu}. These constraints provide an estimate of the value of the mass $M$ of the glitcher in consideration. It can be clearly seen in the figure that the dashed lines either rule out masses above 2.0 M$_{\odot}$ for the lower dashed line or above 1.2 M$_{\odot}$ for the higher dashed line, with the EoS leading to more compact configurations allowing for a higher mass range than the rest. For a recent discussion regarding the role of the symmetry energy parameters on $\Delta I/I$ for pure hadronic compact stars, see~\cite{Zhang:2024qwf}. Other recent studies also consider hybrid stars with an elastic quark core~\cite{Dong:2024lte}.
%
\section{Discussion}


During the last decade, the study of the nature of compact objects like pulsars and neutron stars has been intensively developed. Astrophysical observations resulting from new technologies like interferometers both for radio or gravitational wave signals as well as extraterrestrial X-rays detectors have revolutionized the understanding of the properties of compact stars. 

On the theoretical side, new developments like emulators, which are fast surrogate models capable of reliably approximating high-fidelity models, have been applied to computations of the low-energy nuclear physics in order to study the hadronic EoS~\cite{Drischler:2022ipa,Dietrich:2020efo}. Modern data analysis techniques of observational data, sometimes in conjunction with theoretical results, have been successfully applied to estimate the EoS. Those include Bayesian analyses~\cite{Lattimer:2013hma,Chatziioannou:2015uea,Lackey:2014fwa,Raithel:2017ity,Ayriyan:2021prr,Blaschke:2020qqj,Ayriyan:2018blj,Ayriyan:2017nhp,Alvarez-Castillo:2016oln,Ayriyan:2014yrb,Alvarez-Castillo:2020aku,Alvarez-Castillo:2020fyn,Ayriyan:2024zfw,Li:2025oxi,Li:2024pts,Guven:2020dok,Prakash:2023afe,Jiang:2022tps,Grundler:2025mcz}, neural networks~\cite{Morawski:2020izm}, machine learning approaches~\cite{Fujimoto:2017cdo,Farrell:2023ojk,Li:2025tku,Li:2025vhk,Magnall:2025zhm,Finch:2025bao}, deep learning inference~\cite{Fujimoto:2021zas,Ventagli:2024xsh},  analysis of correlations between the mass-radius relation and the dense matter EoS~\cite{Sun:2024nye}, quantification of the uncertainties when inverting the TOV Equations \cite{Lindblom:2025kjz}, and emulators of the TOV Equations \cite{Reed:2024urq}, among others. Furthermore, an engine for building neutrons called MUSES has been recently developed~\cite{ReinkePelicer:2025vuh,Pelicer:2025chz}. It consists of a collection of software modules that calculate equations of state using algorithms employing three different theories: Crust Density Functional Theory, Chiral Effective Field Theory, and the Chiral Mean Field Model, each of them applied at densities below, around, and above saturation density, respectively. 

Because of its relevance in the understanding of dense nuclear matter and the possibility of a deconfined quark matter phase inside neutron stars, many studies have been recently devoted to the study of the twin compact stars~\cite{Sharifi:2021ead,Shahrbaf:2022upc,Alvarez-Castillo:2022esq,Lyra:2022qmg,Jimenez:2024hib,Lindblom:2024dys,Carlomagno:2023nrc,Christian:2023hez,Zhang:2024npg,Naseri:2024rby,Carlomagno:2024vvr,Laskos-Patkos:2024zib,Christian:2021uhd,Mendes:2024hbn,Chen:2025esq,Christian:2025dhe}. 

{The low onset mass of hybrid compact stars in this work stems from rather low hadron--quark transition densities. These phase transition densities reach values as low as $n_{trans}$ = 0.19 fm$^{-3}$. Such low density values, which are just above saturation density for symmetric nuclear reactions, may appear to contradict nuclear reactions and the dense nuclear matter produced in relativistic heavy ion collisions; therefore, it is worth stressing the role of the isospin symmetry: matter in supernova explosions becomes highly isospin symmetric as the proto-neutron star is formed~\cite{Sagert:2008ka}. In \cite{Sagert:2008uq}, it was found that in supernova explosions, quark matter could be easily produced due to beta equilibrium, small proton fractions, and non-vanishing temperatures. A more recent work shows very good agreement with a low-density quark--hadron phase transition: quark deconfinement as a supernova explosion engine for massive blue supergiant stars~\cite{Fischer:2017lag}. A low critical density for the phase transition to quark matter is also compatible with present pulsar mass measurements. Typical values of the proton fraction $x$ in neutron stars in beta equilibrium are in agreement with the above estimations~\cite{Kubis:2020ysv}, which have a strong dependence on the functional form of the  the nuclear symmetry energy $E_s$ and, moreover, can trigger the activation of the compact star DUrca cooling, which is also being studied by X-ray observations. Recent papers like~\cite{Ivanytskyi:2025cnn} and~\cite{Kojo:2024sca} are devoted to the study of the QCD phase diagram in the finite isospin symmetry region, which features deconfined quark matter. For this reason, one cannot rule out the possibility of low density values of the hadron--quark phase transition in compact stars. Therefore, within this work, each model/set of parameters becomes relevant. Each indeed leads to a different value of the critical density for the phase transition $n_{trans}$ down to 0.19 fm$^{-3}$, a fact that shows the systematics in the variation in stiffness of both the quark and hadron EoSs. All in all, matter in relativistic heavy ion collisions has been estimated to be iso-spin asymmetric in the range x = [0.38, 0.5], which is quite symmetric with respect with neutron star matter~\cite{Yao:2023yda}.
}

In addition, the predictions from the EoS featuring the strong first-order phase transition scenario might be modified when considering the appearance of geometrical structures at the interface, commonly called \textit{pasta phases}. It has been found that sometimes the existence of these structures may hinder the compact star twins scenario, resulting in a single branch of compact stars. Several studies~\cite{Ayriyan:2017tvl,Ayriyan:2017nby,Alvarez-Castillo:2017xvu,Blaschke:2018pva,Blaschke:2020qqj,Maslov:2018ghi} have quantified this effect, whereas the work of~\cite{Pradhan:2023zmg} has assessed the possibility of probing if there is pasta at the quark--hadron interface with the future Einstein telescope or Cosmic Explorer interferometers for detection of gravitational waves, produced by the $f$-mode stellar excitation. More exotic, out of the canon, twin stars have been considered in~\cite{Lope-Oter:2024egz}, where the authors consider modeling such twin stars in the framework of extended theories of gravity, in order to contrast the results with those of general relativity.

{Interestingly, there exist a class of compact stars not covered here that are referred to as slow compact stars, in which a different mechanism of phase transition from the one of this work is considered. The main idea is that if the hadron--quark conversion is slow, the condition for stability holds even if  $\partial M /\partial \varepsilon_c<0$, as shown in~\cite{Pereira:2017rmp}. Several works have explored this alternative scenario~\cite{Rau:2021dds,Lugones:2021bkm}, in which, for the models presented here, additional regions of the third branch in the mass-radius diagram would be populated. }

Although the analysis concerning the methods of the observational modeling of HESS J1731-347~\cite{Doroshenko:2022nwp} has been questioned and not brought to a consensus, the concept of a very light and compact object is of valuable exploration; thus, this work has been devoted to frame it as a twin star. Moreover, there are other interesting objects that have been estimated to be very light, namely, PSR J1231-1411 with a mass of about 1.04 M$_{\odot}$ and a radius of about 13.5 km~\cite{Salmi:2024bss}, or the object XTE J1814-338~\cite{Kini:2024ggu,Laskos-Patkos:2024fdp} with an striking low mass of 1.21 M$_{\odot}$ and a radius of 7 km, which motivate the study of low-mass compact stars. 
{Is it clearly seen in Figure~\ref{MvsR} that the mass-radius curves barely graze the HESS J1731-347 detection region; therefore, the space of parameters within the models presented in this work are considerably narrowed, a fact that underlies the relevance of the detection of light and compact objects on the study of phase transitions in QCD matter.}  As one might expect, the predictions of this work are to be contrasted with future compact star observations.    

Of great relevant importance are the results for the properties of the crust of hybrid compact stars and in the particular case of twins, the differences and similarities between the hadronic twin and hybrid twin. These kinds of studies may potentially pave the way to measure the properties of compact star crusts to probe the existence and the nature of the quark--hadron phase transition in stellar interiors.  Further interesting aspects of the twins phenomenon are the behavior of the speed of sound in their interiors, the breaking of universal relations~\cite{Yagi:2016bkt,Largani:2021hjo,Wu:2025ips}, or their identification through r-mode instabilities~\cite{Laskos-Patkos:2024zib}. Works like~\cite{Christian:2021uhd,Li:2024sft} have assessed the possibility of probing twins by compact star observations of radii through the NICER X-ray detector or with next-generation gravitational-wave detectors~\cite{Landry:2022rxu}. Another possible way to reveal first-order phase transitions is through interface modes in the gravitational wave from inspiralling neutron stars, as recently shown in~\cite{Counsell:2025hcv}.

The compact star twins scenario allows a free-fall timescale transition, which turns out to be fast even when considering compact star rotation, see~\cite{Chanlaridis:2024rov} for a discussion of binary systems leading to either millisecond pulsars in eccentric orbits or isolated, explaining already available observations. This fast scenario is contrary to the one of~\cite{Glendenning:1997fy}, in which such evolutionary transition would take a period of time of the order of $10^{5}$ years; therefore, gaining support from the observational evidence of binary systems.

All in all, consideration of stellar events like resonant shattering flares~\cite{Neill:2020szr}, matter accretion leading to fundamental mode oscillations~\cite{Largani:2023kjx,Pradhan:2023zmg}, detailed cooling rate studies, or moment of inertia measurements may all together reveal the nature of matter in compact star cores.
\section{Conclusions}


Within this work has been presented a set of models of twin stars with a low stellar mass onset for the appearance of hybrid stars due to quark deconfinement in their interiors such that one the twins is a hybrid star. Astrophysical properties and scenarios have been also discussed based on quantitative results, in particular addressing the object HESS J1731-347. If such twins were found, it would be clear that this compact object is a hybrid star with an exotic core rather than a light hadronic neutron star, see~\cite{Tsaloukidis:2022rus} for a compatible analysis. 

It has also been shown here that the EoS of twin compact stars does not obey any conformality boundary at densities present in compact star interiors. Predictions for the crust of compact star twins, which are of great importance when considering gravitational wave emission from possible mountains in the stellar surface~\cite{Gittins:2020cvx,Morales:2023euv}, or by excitation of the stellar $f$-modes, have also been provided in this work. All in all, all present compact star constraints can be fulfilled by the {hybrid compact stars and the low-mass constraints by the hybrid twin}. In particular, measurement improvements leading to a narrowing the regions of HESS J1731-347 and the black widow  PSR J0952-0607 have the potential to rule out some of the models here or rule them all out in an extreme scenario. 
Conversely, if the mass difference between the aforementioned stellar objects is reduced, the twin compact stars presented in this work will gain observational support.

Future gravitational wave detectors like the Einstein Telescope or the Cosmic Explorer together with multi-messenger counterparts are expected provide strong support for testing the concept of twin compact stars, especially during violent and transient processes. Future work includes, for instance, modeling the cooling of compact stars, consideration of pasta phases at the hadron--quark interface, and modeling thermal twins, together with the addition of upcoming multi-messenger observations.

\vspace{6pt} 

\funding{The 
 author is supported by the program Excellence Initiative–Research University of the University of Wroclaw of the Ministry of Education and Science.}



\dataavailability{Data are available upon request.} 

\acknowledgments{The author acknowledges discussion and collaboration with B. Pradhan, W. Newton, S. Kubis,  M. Marczenko, O. Ivanytskyi, A. Ayriyan, and D. Blaschke.}

\conflictsofinterest{The author declares no conflicts of interest.} 



\abbreviations{Abbreviations}{
The following abbreviations are used in this manuscript:\\

\noindent 
\begin{tabular}{@{}ll}
QM & Quark Matter\\
RHIC & Relativistic Heavy Ion Collisions\\
EoS& Equation of State\\
GW & Gravitational Waves\\
NICER & Neutron Star Composition Explorer\\
QCD & Quantum Chromodynamics\\
RMF & Relativistic Mean Field\\
OGE & One Gluon Exchange\\
MFA& Mean Field Approximation\\
TOV & Tolman--Oppenheimer--Volkoff\\
MoI & Moment of Inertia\\
TD & Tidal deformability\\
MSP & Millisecond Pulsar\\
LMXB & Low-Mass X-ray Binary\\
\end{tabular}
}

%



\appendix
\appendixstart
\section{}

Properties of compact stars rotating at the Keplerian frequency computed using the RNS code~\cite{Nozawa:1998ak,Komatsu:1989zz,Cook:1993qj,Stergioulas:1994ea} are presented in the tables below. They correspond to the EoS models for hybrid stars introduced in this work. They are the following: 

%

\begin{itemize}
\item{ $\epsilon_c$ {\it central energy density;}}
\item{ $M$ {\it gravitational mass;}}
\item{ $M_0$ {\it rest mass;}}
\item{ $R_e$ {\it radius at the equator (circumferential, i.e., $2 \pi R_e$ is 
the proper circumference); }}
\item{ $\Omega$ {\it angular velocity;}}
\item{ $\Omega_p$ {\it angular velocity of a particle in circular orbit 
                      at the equator;}}
\item{ $T/W$ {\it rotational/gravitational energy;}}
\item{ $cJ/GM_{\odot}^2$ {\it angular momentum;}}
\item{ $I$ {\it moment of inertia;}}
\item{ $h_+$ {\it height from surface of last stable co-rotating circular 
               orbit in equatorial plane (circumferential);}}
\item{ $h_-$ {\it height from surface of last stable counter-rotating circular 
               orbit in equatorial plane (circumferential);}}
\item{ $Z_p$ {\it polar redshift;}}
\item{ $Z_b$ {\it backward equatorial redshift;}}
\item{ $Z_f$ {\it forward equatorial redshift;}}
\item{ $\omega_c/ \Omega$ {\it ratio of central value of potential $\omega$ to 
                             $\Omega$;}}
\item{ $r_e$ {\it coordinate equatorial radius; }}
\item{ $r_p/r_e$ {\it axes ratio (polar to equatorial).}}
\end{itemize} 

\begin{table}[H]

\caption {DD2FMEVp80-nlNJL ($\mu_{<} =$ 860 MeV).}

\begin{adjustwidth}{-\extralength}{0cm}

\resizebox{\fulllength}{!}{%
\begin{tabular}{ccccccccccccccccc}
\noalign{\hrule height 1pt} 
\boldmath{$\rho_c$} & \boldmath{$M$} & \boldmath{$M_0$} & \boldmath{$R_e$ }& \boldmath{$\Omega$} & \boldmath{$\Omega_p$} & \boldmath{$T/W$} & \boldmath{$cJ/GM_{\odot}^2$} & \boldmath{$I$} & \boldmath{$h_+$} & \boldmath{$h_-$} &\boldmath{ $Z_p$} & \boldmath{$Z_b$} & \boldmath{$Z_f$} & \boldmath{$\omega_c/ \Omega$} & \boldmath{$r_e$} & \boldmath{$r_p/r_e$} \\

\textbf{[$10^{14}$ g cm$^{-3}$]} & \boldmath{[$M/M_{\odot}$]} & \boldmath{[$M/M_{\odot}$]} &  \textbf{[km]} & \textbf{[$10^{4}$s$^{-1}$]}  & \textbf{[$10^{4}$s$^{-1}$]} & & & \textbf{[$10^{45}$g cm$^{2}$] } & \textbf{[km]} & \textbf{[km]} & & & & & \textbf{[km]} &  \\

\midrule
2.6 & 0.2466 & 0.2486 & 23.33 & 0.1608 & 0.1608 & 0.0179 & 0.02959 & 0.1617 & 8.927 & 0 & 0 & 0.02428 & 0.1545 & $-$0.1059 & 0.09017 & 0.6256 \\
2.716 & 0.2998 & 0.3034 & 21.57 & 0.1996 & 0.1996 & 0.02524 & 0.04851 & 0.2136 & 15.83 & 0 & 0 & 0.03232 & 0.1838 & $-$0.119 & 0.1063 & 0.5757 \\
2.838 & 0.371 & 0.3772 & 20.44 & 0.2411 & 0.2411 & 0.03431 & 0.08086 & 0.2947 & 27.73 & 0 & 0 & 0.04289 & 0.2193 & $-$0.1332 & 0.1263 & 0.542 \\
2.965 & 0.5681 & 0.5837 & 20.05 & 0.3088 & 0.3088 & 0.05714 & 0.2192 & 0.624 & 83.72 & 0 & 0 & 0.06974 & 0.3012 & $-$0.1606 & 0.1654 & 0.523 \\
3.098 & 0.5975 & 0.6147 & 20.09 & 0.3161 & 0.3161 & 0.06022 & 0.2456 & 0.683 & 94.45 & 0 & 0 & 0.07363 & 0.3125 & $-$0.1639 & 0.1743 & 0.5228 \\
3.237 & 0.5596 & 0.5747 & 19.95 & 0.3086 & 0.3087 & 0.05594 & 0.2103 & 0.5987 & 78.91 & 0 & 0 & 0.06893 & 0.2989 & $-$0.1599 & 0.1711 & 0.5207 \\
3.382 & 0.54 & 0.554 & 19.86 & 0.3052 & 0.3052 & 0.05359 & 0.1927 & 0.5551 & 71.03 & 0 & 0 & 0.0666 & 0.292 & $-$0.1578 & 0.1701 & 0.519 \\
3.533 & 0.532 & 0.5455 & 19.8 & 0.3041 & 0.3041 & 0.05256 & 0.1855 & 0.5361 & 67.64 & 0 & 0 & 0.0657 & 0.2894 & $-$0.157 & 0.1701 & 0.5177 \\
3.692 & 0.5277 & 0.541 & 19.77 & 0.3035 & 0.3035 & 0.052 & 0.1818 & 0.5263 & 65.91 & 0 & 0 & 0.06521 & 0.2879 & $-$0.1565 & 0.17 & 0.5171 \\
3.857 & 0.5285 & 0.5419 & 19.78 & 0.3036 & 0.3036 & 0.05211 & 0.1825 & 0.5283 & 66.25 & 0 & 0 & 0.06531 & 0.2882 & $-$0.1566 & 0.17 & 0.5173 \\
4.03 & 0.5337 & 0.5473 & 19.81 & 0.3043 & 0.3044 & 0.05277 & 0.187 & 0.54 & 68.32 & 0 & 0 & 0.06589 & 0.2899 & $-$0.1572 & 0.1701 & 0.518 \\
4.21 & 0.5395 & 0.5534 & 19.86 & 0.3051 & 0.3051 & 0.05352 & 0.1922 & 0.5538 & 70.8 & 0 & 0 & 0.06653 & 0.2918 & $-$0.1578 & 0.1701 & 0.5189 \\
4.399 & 0.5503 & 0.5648 & 19.91 & 0.3069 & 0.3069 & 0.05483 & 0.2018 & 0.5779 & 75.13 & 0 & 0 & 0.06781 & 0.2956 & $-$0.1589 & 0.1706 & 0.52 \\
4.596 & 0.5629 & 0.5782 & 19.97 & 0.3093 & 0.3093 & 0.05633 & 0.2133 & 0.606 & 80.24 & 0 & 0 & 0.06934 & 0.3001 & $-$0.1602 & 0.1714 & 0.5209 \\
4.802 & 0.578 & 0.5941 & 20.02 & 0.3122 & 0.3122 & 0.05805 & 0.2271 & 0.6394 & 86.37 & 0 & 0 & 0.07119 & 0.3054 & $-$0.1618 & 0.1726 & 0.5218 \\
5.017 & 0.5927 & 0.6096 & 20.07 & 0.3151 & 0.3151 & 0.05969 & 0.241 & 0.6722 & 92.44 & 0 & 0 & 0.07302 & 0.3108 & $-$0.1634 & 0.1739 & 0.5226 \\
5.242 & 0.6021 & 0.6196 & 20.1 & 0.317 & 0.317 & 0.06073 & 0.2501 & 0.6934 & 96.39 & 0 & 0 & 0.07421 & 0.3142 & $-$0.1644 & 0.1746 & 0.523 \\
5.477 & 0.6081 & 0.626 & 20.12 & 0.3182 & 0.3182 & 0.06138 & 0.256 & 0.7071 & 98.97 & 0 & 0 & 0.07496 & 0.3163 & $-$0.165 & 0.1749 & 0.5232 \\
5.722 & 0.6093 & 0.6272 & 20.13 & 0.3184 & 0.3184 & 0.0615 & 0.2571 & 0.7097 & 99.47 & 0 & 0 & 0.0751 & 0.3167 & $-$0.1651 & 0.175 & 0.5233 \\
5.978 & 0.5725 & 0.5882 & 20. & 0.3111 & 0.3111 & 0.05743 & 0.222 & 0.6272 & 84.11 & 0 & 0 & 0.07051 & 0.3035 & $-$0.1613 & 0.1721 & 0.5215 \\
6.246 & 0.507 & 0.5191 & 18.89 & 0.318 & 0.318 & 0.04736 & 0.1572 & 0.4344 & 50.18 & 0 & 0 & 0.06551 & 0.2886 & $-$0.1567 & 0.1817 & 0.4941 \\
6.526 & 0.5734 & 0.5896 & 17.67 & 0.3739 & 0.3739 & 0.05098 & 0.1943 & 0.4568 & 52.78 & 0 & 0 & 0.08079 & 0.3321 & $-$0.169 & 0.2132 & 0.4578 \\
6.818 & 0.6852 & 0.71 & 16.8 & 0.4411 & 0.4411 & 0.05863 & 0.2782 & 0.5543 & 68. & 0 & 0 & 0.1047 & 0.3974 & $-$0.1853 & 0.2534 & 0.4294 \\
7.124 & 0.82 & 0.8575 & 16.35 & 0.5028 & 0.5028 & 0.06766 & 0.4044 & 0.7069 & 89.71 & 0 & 0 & 0.1332 & 0.4727 & $-$0.2012 & 0.2953 & 0.4113 \\
7.443 & 0.9643 & 1.018 & 16.11 & 0.5568 & 0.5569 & 0.07659 & 0.5688 & 0.8978 & 115.8 & 0 & 0 & 0.1648 & 0.5538 & $-$0.2159 & 0.337 & 0.3987 \\
7.776 & 1.112 & 1.186 & 15.99 & 0.6045 & 0.6045 & 0.0851 & 0.7707 & 1.121 & 147. & 0 & 0 & 0.1986 & 0.6398 & $-$0.2293 & 0.3782 & 0.3888 \\
8.124 & 1.257 & 1.354 & 15.91 & 0.6469 & 0.6469 & 0.0926 & 1.001 & 1.36 & 179.5 & 0 & 1.107 & 0.2339 & 0.7292 & $-$0.2414 & 0.4174 & 0.3801 \\
8.488 & 1.396 & 1.518 & 15.86 & 0.6845 & 0.6845 & 0.09938 & 1.256 & 1.612 & 213.9 & 0 & 2.716 & 0.2702 & 0.8213 & $-$0.2523 & 0.4544 & 0.3722 \\
8.869 & 1.519 & 1.665 & 15.81 & 0.7155 & 0.7156 & 0.1047 & 1.498 & 1.839 & 239.3 & 0 & 4.067 & 0.3042 & 0.9078 & $-$0.2614 & 0.4863 & 0.3653 \\
9.266 & 1.646 & 1.821 & 15.76 & 0.7478 & 0.7478 & 0.11 & 1.786 & 2.099 & 273.6 & 0 & 5.581 & 0.3423 & 1.007 & $-$0.2709 & 0.5193 & 0.3578 \\
9.681 & 1.753 & 1.954 & 15.7 & 0.7746 & 0.7747 & 0.1139 & 2.036 & 2.31 & 294.9 & 0 & 6.788 & 0.3769 & 1.097 & $-$0.2784 & 0.5471 & 0.351 \\
10.11 & 1.862 & 2.092 & 15.63 & 0.8019 & 0.802 & 0.1182 & 2.326 & 2.549 & 325.8 & 0 & 8.121 & 0.4149 & 1.198 & $-$0.2864 & 0.5755 & 0.3439 \\
10.57 & 1.954 & 2.21 & 15.55 & 0.8261 & 0.8261 & 0.1212 & 2.576 & 2.74 & 345. & 0 & 9.21 & 0.4499 & 1.292 & $-$0.293 & 0.6 & 0.3372 \\
11.04 & 2.038 & 2.319 & 15.46 & 0.8494 & 0.8495 & 0.1237 & 2.814 & 2.912 & 360.9 & 0 & 10.22 & 0.4847 & 1.388 & $-$0.2992 & 0.6226 & 0.3304 \\
11.54 & 2.11 & 2.411 & 15.36 & 0.8705 & 0.8705 & 0.1257 & 3.017 & 3.046 & 370.4 & 0 & 11.07 & 0.517 & 1.477 & $-$0.3045 & 0.6428 & 0.3241 \\
12.05 & 2.174 & 2.497 & 15.26 & 0.891 & 0.891 & 0.1273 & 3.21 & 3.167 & 378.9 & 0 & 11.87 & 0.549 & 1.567 & $-$0.3095 & 0.6616 & 0.3179 \\
12.59 & 2.246 & 2.595 & 15.15 & 0.9137 & 0.9138 & 0.1297 & 3.461 & 3.329 & 400. & 0 & 12.84 & 0.5865 & 1.678 & $-$0.3155 & 0.6819 & 0.3112 \\
13.16 & 2.299 & 2.668 & 15.03 & 0.9337 & 0.9338 & 0.1309 & 3.635 & 3.421 & 406. & 0 & 13.54 & 0.6186 & 1.772 & $-$0.32 & 0.6987 & 0.305 \\
13.75 & 2.347 & 2.733 & 14.91 & 0.9533 & 0.9533 & 0.132 & 3.793 & 3.497 & 410.5 & 0.1641 & 14.18 & 0.6501 & 1.866 & $-$0.3244 & 0.7145 & 0.2991 \\
14.36 & 2.39 & 2.793 & 14.79 & 0.9726 & 0.9726 & 0.133 & 3.945 & 3.565 & 415.6 & 0.3427 & 14.78 & 0.6814 & 1.963 & $-$0.3285 & 0.7295 & 0.2932 \\
15. & 2.429 & 2.85 & 14.65 & 0.9922 & 0.9922 & 0.1338 & 4.089 & 3.622 & 419.7 & 0.5065 & 15.36 & 0.7132 & 2.062 & $-$0.3326 & 0.7439 & 0.2874 \\
15.68 & 2.46 & 2.893 & 14.52 & 1.011 & 1.011 & 0.1342 & 4.194 & 3.646 & 418.1 & 0.6379 & 15.8 & 0.7422 & 2.154 & $-$0.3361 & 0.7568 & 0.2818 \\
16.38 & 2.486 & 2.93 & 14.38 & 1.029 & 1.029 & 0.1345 & 4.283 & 3.658 & 415.4 & 0.7613 & 16.21 & 0.7706 & 2.245 & $-$0.3395 & 0.7689 & 0.2765 \\
17.11 & 2.507 & 2.961 & 14.24 & 1.047 & 1.047 & 0.1346 & 4.359 & 3.658 & 411.6 & 0.8775 & 16.56 & 0.7982 & 2.335 & $-$0.3427 & 0.7805 & 0.2712 \\

\noalign{\hrule height 1pt}
\end{tabular}
}

\end{adjustwidth}
\end{table}

\begin{table}[H]\ContinuedFloat

\caption{\textit{Cont}.}

\begin{adjustwidth}{-\extralength}{0cm}

\resizebox{\fulllength}{!}{%
\begin{tabular}{ccccccccccccccccc}
\noalign{\hrule height 1pt} 
\boldmath{$\rho_c$} & \boldmath{$M$} & \boldmath{$M_0$} & \boldmath{$R_e$ }& \boldmath{$\Omega$} & \boldmath{$\Omega_p$} & \boldmath{$T/W$} & \boldmath{$cJ/GM_{\odot}^2$} & \boldmath{$I$} & \boldmath{$h_+$} & \boldmath{$h_-$} &\boldmath{ $Z_p$} & \boldmath{$Z_b$} & \boldmath{$Z_f$} & \boldmath{$\omega_c/ \Omega$} & \boldmath{$r_e$} & \boldmath{$r_p/r_e$} \\

\textbf{[$10^{14}$ g cm$^{-3}$]} & \boldmath{[$M/M_{\odot}$]} & \boldmath{[$M/M_{\odot}$]} &  \textbf{[km]} & \textbf{[$10^{4}$s$^{-1}$]}  & \textbf{[$10^{4}$s$^{-1}$]} & & & \textbf{[$10^{45}$g cm$^{2}$] } & \textbf{[km]} & \textbf{[km]} & & & & & \textbf{[km]} &  \\

\midrule

17.88 & 2.526 & 2.989 & 14.09 & 1.066 & 1.066 & 0.1348 & 4.429 & 3.652 & 408.3 & 0.9884 & 16.9 & 0.8259 & 2.427 & $-$0.3458 & 0.7916 & 0.2661 \\

18.68 & 2.544 & 3.014 & 13.95 & 1.085 & 1.085 & 0.1349 & 4.498 & 3.645 & 405.9 & 1.099 & 17.23 & 0.8542 & 2.523 & $-$0.349 & 0.8026 & 0.261 \\

19.52 & 2.564 & 3.045 & 13.81 & 1.105 & 1.105 & 0.1355 & 4.596 & 3.656 & 409.4 & 1.227 & 17.64 & 0.8862 & 2.638 & $-$0.3526 & 0.8139 & 0.2557 \\

20.39 & 2.575 & 3.062 & 13.66 & 1.124 & 1.124 & 0.1355 & 4.641 & 3.63 & 405.2 & 1.32 & 17.89 & 0.9133 & 2.735 & $-$0.3555 & 0.8236 & 0.2508 \\
21.31 & 2.582 & 3.073 & 13.52 & 1.141 & 1.141 & 0.1353 & 4.671 & 3.596 & 399.8 & 1.398 & 18.1 & 0.9384 & 2.825 & $-$0.3581 & 0.8324 & 0.2463 \\
22.26 & 2.586 & 3.077 & 13.39 & 1.158 & 1.158 & 0.1351 & 4.684 & 3.555 & 394. & 1.464 & 18.25 & 0.9605 & 2.905 & $-$0.3604 & 0.8404 & 0.2423 \\
23.26 & 2.588 & 3.082 & 13.25 & 1.175 & 1.175 & 0.1347 & 4.696 & 3.512 & 387.4 & 1.523 & 18.4 & 0.9832 & 2.989 & $-$0.3626 & 0.8478 & 0.2382 \\
24.3 & 2.587 & 3.08 & 13.13 & 1.19 & 1.19 & 0.1343 & 4.691 & 3.464 & 381.1 & 1.573 & 18.48 & 1.002 & 3.058 & $-$0.3644 & 0.8545 & 0.2348 \\
25.39 & 2.584 & 3.075 & 13.02 & 1.205 & 1.205 & 0.1339 & 4.679 & 3.414 & 374.2 & 1.616 & 18.54 & 1.019 & 3.123 & $-$0.3661 & 0.8606 & 0.2315 \\
26.52 & 2.58 & 3.07 & 12.9 & 1.219 & 1.219 & 0.1334 & 4.667 & 3.365 & 367.9 & 1.656 & 18.6 & 1.036 & 3.187 & $-$0.3678 & 0.8665 & 0.2284 \\
27.71 & 2.575 & 3.06 & 12.8 & 1.232 & 1.232 & 0.1329 & 4.644 & 3.313 & 361.4 & 1.69 & 18.62 & 1.05 & 3.24 & $-$0.3691 & 0.8718 & 0.2256 \\
28.95 & 2.567 & 3.048 & 12.69 & 1.245 & 1.245 & 0.1323 & 4.614 & 3.258 & 354.1 & 1.719 & 18.62 & 1.063 & 3.288 & $-$0.3703 & 0.8767 & 0.2229 \\
30.25 & 2.56 & 3.037 & 12.59 & 1.257 & 1.258 & 0.1317 & 4.585 & 3.205 & 347.3 & 1.745 & 18.62 & 1.075 & 3.336 & $-$0.3715 & 0.8814 & 0.2204 \\

\noalign{\hrule height 1pt}
\end{tabular}
}

\end{adjustwidth}
\end{table}

\vspace{-8pt}
\begin{table}[H]

\caption {DD2FMEVp80-nlNJL ($\mu_{<} =$ 955 MeV).}

\begin{adjustwidth}{-\extralength}{0cm}

\resizebox{\fulllength}{!}{%
\begin{tabular}{ccccccccccccccccc}
\noalign{\hrule height 1pt} 
\boldmath{$\rho_c$} & \boldmath{$M$} & \boldmath{$M_0$} & \boldmath{$R_e$} & \boldmath{$\Omega$} & \boldmath{$\Omega_p$} & \boldmath{$T/W$} & \boldmath{$cJ/GM_{\odot}^2$} & \boldmath{$I$} & \boldmath{$h_+$} & \boldmath{$h_-$} & \boldmath{$Z_p$} & \boldmath{$Z_b$} & \boldmath{$Z_f$} & \boldmath{$\omega_c/ \Omega$ }& \boldmath{$r_e$} &\boldmath{ $r_p/r_e$ }\\

\textbf{[$10^{14}$ g cm$^{-3}$]} & \boldmath{[$M/M_{\odot}$]} & \boldmath{[$M/M_{\odot}$]} &  \textbf{[km]} & \textbf{[$10^{4}$s$^{-1}$]}  & \textbf{[$10^{4}$s$^{-1}$]} & & & \textbf{[$10^{45}$g cm$^{2}$]}  & \textbf{[km]} & \textbf{[km]} & & & & & \textbf{[km]} &  \\

\midrule
2.6 & 0.2466 & 0.2486 & 23.33 & 0.1608 & 0.1608 & 0.0179 & 0.02959 & 0.1617 & 8.927 & 0 & 0 & 0.02428 & 0.1545 & $-$0.1059 & 0.09017 & 0.6256 \\
2.673 & 0.2778 & 0.2807 & 22.16 & 0.1844 & 0.1845 & 0.02224 & 0.0401 & 0.1911 & 12.67 & 0 & 0 & 0.029 & 0.172 & $-$0.1139 & 0.09978 & 0.5925 \\
2.748 & 0.3159 & 0.32 & 21.24 & 0.2098 & 0.2098 & 0.02735 & 0.05507 & 0.2306 & 18.15 & 0 & 0 & 0.03471 & 0.1921 & $-$0.1225 & 0.111 & 0.5659 \\
2.825 & 0.3609 & 0.3666 & 20.56 & 0.2358 & 0.2358 & 0.03304 & 0.07568 & 0.2821 & 25.71 & 0 & 0 & 0.04139 & 0.2144 & $-$0.1314 & 0.1236 & 0.5455 \\
2.904 & 0.4141 & 0.4221 & 20.07 & 0.2622 & 0.2622 & 0.03935 & 0.1043 & 0.3496 & 36.2 & 0 & 0 & 0.04924 & 0.2395 & $-$0.1406 & 0.138 & 0.53 \\
2.985 & 0.4767 & 0.4877 & 19.74 & 0.2887 & 0.2887 & 0.04621 & 0.1438 & 0.4376 & 50.75 & 0 & 0 & 0.05841 & 0.2677 & $-$0.1501 & 0.1543 & 0.5185 \\
3.068 & 0.5495 & 0.5646 & 19.56 & 0.315 & 0.315 & 0.05349 & 0.1975 & 0.551 & 70.15 & 0 & 0 & 0.06902 & 0.299 & $-$0.1599 & 0.1725 & 0.5103 \\
3.154 & 0.634 & 0.6545 & 19.48 & 0.3409 & 0.3409 & 0.06116 & 0.2705 & 0.6974 & 96.06 & 0 & 0 & 0.08135 & 0.3343 & $-$0.1699 & 0.1928 & 0.5047 \\
3.243 & 0.7292 & 0.7566 & 19.49 & 0.3658 & 0.3658 & 0.06889 & 0.3661 & 0.8796 & 128.6 & 0 & 0 & 0.09527 & 0.373 & $-$0.1799 & 0.2148 & 0.501 \\
3.333 & 0.8694 & 0.9086 & 19.66 & 0.3951 & 0.3951 & 0.07933 & 0.5351 & 1.19 & 185.6 & 0 & 0 & 0.1156 & 0.4281 & $-$0.1927 & 0.2432 & 0.4995 \\
3.427 & 1.137 & 1.202 & 20.34 & 0.4308 & 0.4308 & 0.09799 & 0.9616 & 1.962 & 332. & 0 & 0 & 0.1531 & 0.5271 & $-$0.2123 & 0.2866 & 0.5064 \\
3.523 & 1.452 & 1.553 & 21.31 & 0.4555 & 0.4555 & 0.1178 & 1.643 & 3.171 & 567.6 & 0 & 1.277 & 0.1962 & 0.6395 & $-$0.2308 & 0.3313 & 0.5187 \\
3.621 & 1.66 & 1.787 & 21.95 & 0.4667 & 0.4667 & 0.1296 & 2.195 & 4.133 & 754.4 & 0 & 3.341 & 0.2245 & 0.7133 & $-$0.2412 & 0.3597 & 0.5265 \\
3.723 & 1.717 & 1.851 & 22.1 & 0.47 & 0.47 & 0.1326 & 2.356 & 4.405 & 805.6 & 0 & 3.901 & 0.2326 & 0.7343 & $-$0.2439 & 0.3702 & 0.5279 \\
3.827 & 1.709 & 1.842 & 22.06 & 0.4701 & 0.4701 & 0.1321 & 2.329 & 4.354 & 794.6 & 0 & 3.817 & 0.2317 & 0.7318 & $-$0.2436 & 0.3712 & 0.5272 \\
3.934 & 1.703 & 1.835 & 22.03 & 0.47 & 0.47 & 0.1317 & 2.31 & 4.319 & 787.4 & 0 & 3.753 & 0.2309 & 0.7298 & $-$0.2433 & 0.3712 & 0.5268 \\
4.044 & 1.702 & 1.835 & 22.03 & 0.47 & 0.47 & 0.1318 & 2.309 & 4.319 & 787.3 & 0 & 3.753 & 0.2309 & 0.7297 & $-$0.2433 & 0.3712 & 0.5268 \\
4.157 & 1.706 & 1.839 & 22.05 & 0.47 & 0.47 & 0.132 & 2.321 & 4.34 & 791.6 & 0 & 3.79 & 0.2313 & 0.731 & $-$0.2435 & 0.3712 & 0.527 \\
4.274 & 1.711 & 1.845 & 22.07 & 0.4701 & 0.4701 & 0.1323 & 2.337 & 4.368 & 797.5 & 0 & 3.841 & 0.232 & 0.7327 & $-$0.2437 & 0.3712 & 0.5274 \\
4.393 & 1.71 & 1.844 & 22.06 & 0.4701 & 0.4701 & 0.1322 & 2.334 & 4.363 & 796.3 & 0 & 3.832 & 0.2318 & 0.7323 & $-$0.2436 & 0.3712 & 0.5273 \\
4.516 & 1.709 & 1.842 & 22.06 & 0.4701 & 0.4701 & 0.1321 & 2.329 & 4.354 & 794.5 & 0 & 3.816 & 0.2316 & 0.7318 & $-$0.2436 & 0.3712 & 0.5272 \\
4.643 & 1.707 & 1.84 & 22.05 & 0.47 & 0.47 & 0.132 & 2.323 & 4.343 & 792.2 & 0 & 3.796 & 0.2314 & 0.7312 & $-$0.2435 & 0.3712 & 0.5271 \\
4.773 & 1.705 & 1.837 & 22.04 & 0.47 & 0.47 & 0.1319 & 2.317 & 4.332 & 790. & 0 & 3.776 & 0.2311 & 0.7305 & $-$0.2434 & 0.3712 & 0.5269 \\
4.906 & 1.703 & 1.835 & 22.03 & 0.47 & 0.47 & 0.1318 & 2.309 & 4.319 & 787.4 & 0 & 3.754 & 0.2309 & 0.7297 & $-$0.2433 & 0.3712 & 0.5268 \\
5.044 & 1.7 & 1.832 & 22.02 & 0.4699 & 0.4699 & 0.1316 & 2.302 & 4.305 & 784.4 & 0 & 3.727 & 0.2305 & 0.7289 & $-$0.2432 & 0.3712 & 0.5266 \\
5.185 & 1.697 & 1.829 & 22.01 & 0.4698 & 0.4699 & 0.1314 & 2.293 & 4.29 & 781.4 & 0 & 3.699 & 0.2302 & 0.728 & $-$0.2431 & 0.3711 & 0.5265 \\
5.33 & 1.694 & 1.825 & 22. & 0.4698 & 0.4698 & 0.1313 & 2.284 & 4.274 & 778.2 & 0 & 3.671 & 0.2298 & 0.727 & $-$0.2429 & 0.371 & 0.5263 \\
5.479 & 1.691 & 1.822 & 21.99 & 0.4697 & 0.4697 & 0.1311 & 2.275 & 4.256 & 774.6 & 0 & 3.638 & 0.2294 & 0.7259 & $-$0.2428 & 0.3709 & 0.5261 \\
5.633 & 1.688 & 1.818 & 21.98 & 0.4696 & 0.4696 & 0.1309 & 2.264 & 4.238 & 770.8 & 0 & 3.603 & 0.229 & 0.7247 & $-$0.2426 & 0.3708 & 0.526 \\
5.79 & 1.684 & 1.814 & 21.96 & 0.4695 & 0.4695 & 0.1307 & 2.253 & 4.218 & 766.9 & 0 & 3.567 & 0.2285 & 0.7235 & $-$0.2425 & 0.3707 & 0.5258 \\
5.952 & 1.68 & 1.809 & 21.95 & 0.4694 & 0.4694 & 0.1305 & 2.242 & 4.197 & 762.8 & 0 & 3.528 & 0.228 & 0.7222 & $-$0.2423 & 0.3705 & 0.5256 \\
6.119 & 1.676 & 1.805 & 21.94 & 0.4693 & 0.4693 & 0.1302 & 2.23 & 4.175 & 758.3 & 0 & 3.487 & 0.2275 & 0.7208 & $-$0.2421 & 0.3704 & 0.5254 \\
6.29 & 1.672 & 1.8 & 21.92 & 0.4691 & 0.4692 & 0.13 & 2.216 & 4.152 & 753.5 & 0 & 3.442 & 0.2269 & 0.7193 & $-$0.2419 & 0.3701 & 0.5251 \\
6.466 & 1.667 & 1.795 & 21.9 & 0.469 & 0.469 & 0.1297 & 2.202 & 4.127 & 748.5 & 0 & 3.396 & 0.2263 & 0.7177 & $-$0.2417 & 0.3699 & 0.5249 \\
6.648 & 1.662 & 1.789 & 21.88 & 0.4689 & 0.4689 & 0.1294 & 2.188 & 4.1 & 743.2 & 0 & 3.345 & 0.2256 & 0.716 & $-$0.2415 & 0.3697 & 0.5246 \\
6.834 & 1.657 & 1.783 & 21.87 & 0.4687 & 0.4687 & 0.1291 & 2.173 & 4.073 & 737.8 & 0 & 3.294 & 0.225 & 0.7143 & $-$0.2412 & 0.3694 & 0.5244 \\
7.025 & 1.652 & 1.777 & 21.85 & 0.4686 & 0.4686 & 0.1288 & 2.158 & 4.048 & 732.6 & 0 & 3.245 & 0.2243 & 0.7126 & $-$0.241 & 0.3692 & 0.5241 \\
7.222 & 1.648 & 1.773 & 21.83 & 0.4684 & 0.4685 & 0.1286 & 2.145 & 4.025 & 728. & 0 & 3.201 & 0.2238 & 0.7111 & $-$0.2408 & 0.369 & 0.5239 \\
7.424 & 1.639 & 1.762 & 21.8 & 0.4682 & 0.4682 & 0.128 & 2.119 & 3.978 & 718.7 & 0 & 3.11 & 0.2226 & 0.708 & $-$0.2404 & 0.3685 & 0.5234 \\
7.632 & 1.625 & 1.747 & 21.75 & 0.4677 & 0.4678 & 0.1272 & 2.081 & 3.91 & 704.8 & 0 & 2.975 & 0.2209 & 0.7035 & $-$0.2398 & 0.3677 & 0.5227 \\
7.845 & 1.612 & 1.732 & 21.7 & 0.4673 & 0.4673 & 0.1264 & 2.041 & 3.839 & 690.7 & 0 & 2.836 & 0.2191 & 0.6988 & $-$0.2391 & 0.3669 & 0.522 \\
8.065 & 1.597 & 1.715 & 21.65 & 0.4668 & 0.4669 & 0.1255 & 2. & 3.765 & 676. & 0 & 2.69 & 0.2172 & 0.6938 & $-$0.2384 & 0.366 & 0.5212 \\
8.291 & 1.584 & 1.7 & 21.6 & 0.4664 & 0.4664 & 0.1248 & 1.963 & 3.699 & 662.7 & 0 & 2.557 & 0.2155 & 0.6894 & $-$0.2378 & 0.3652 & 0.5204 \\
8.523 & 1.59 & 1.707 & 21.62 & 0.4666 & 0.4666 & 0.1251 & 1.979 & 3.728 & 668.5 & 0 & 2.616 & 0.2162 & 0.6913 & $-$0.2381 & 0.3656 & 0.5208 \\
8.761 & 1.31 & 1.392 & 20.35 & 0.4618 & 0.4618 & 0.1054 & 1.255 & 2.388 & 396.4 & 0 & 0 & 0.1819 & 0.6012 & $-$0.2245 & 0.3505 & 0.4988 \\
9.007 & 1.202 & 1.271 & 19.27 & 0.478 & 0.478 & 0.09317 & 0.9847 & 1.811 & 275.5 & 0 & 0 & 0.1742 & 0.5799 & $-$0.2207 & 0.3573 & 0.4744 \\
9.259 & 1.195 & 1.264 & 18.4 & 0.5092 & 0.5092 & 0.08745 & 0.922 & 1.591 & 225.8 & 0 & 0 & 0.1827 & 0.6006 & $-$0.2237 & 0.377 & 0.4511 \\
9.518 & 1.239 & 1.314 & 17.71 & 0.5479 & 0.5479 & 0.08575 & 0.9566 & 1.535 & 206.7 & 0 & 0 & 0.2004 & 0.6445 & $-$0.2301 & 0.4032 & 0.4303 \\
9.784 & 1.309 & 1.395 & 17.19 & 0.5878 & 0.5878 & 0.08644 & 1.045 & 1.563 & 202. & 0 & 0.5569 & 0.2231 & 0.7013 & $-$0.2378 & 0.4314 & 0.4129 \\
10.06 & 1.39 & 1.489 & 16.79 & 0.6262 & 0.6263 & 0.0884 & 1.167 & 1.637 & 204.9 & 0 & 1.588 & 0.2486 & 0.765 & $-$0.2457 & 0.4596 & 0.3983 \\
10.34 & 1.475 & 1.59 & 16.46 & 0.6629 & 0.6629 & 0.09098 & 1.311 & 1.738 & 212. & 0 & 2.665 & 0.2762 & 0.834 & $-$0.2536 & 0.4874 & 0.3856 \\
10.63 & 1.56 & 1.693 & 16.21 & 0.6967 & 0.6967 & 0.09394 & 1.471 & 1.856 & 221.8 & 0 & 3.741 & 0.3047 & 0.906 & $-$0.2611 & 0.5139 & 0.3747 \\
10.93 & 1.643 & 1.794 & 16. & 0.7275 & 0.7275 & 0.09688 & 1.637 & 1.978 & 231.3 & 0 & 4.766 & 0.3335 & 0.9789 & $-$0.2681 & 0.5384 & 0.3651 \\
11.23 & 1.723 & 1.894 & 15.81 & 0.7573 & 0.7574 & 0.09981 & 1.815 & 2.106 & 243.2 & 0 & 5.797 & 0.3634 & 1.056 & $-$0.2749 & 0.5624 & 0.356 \\
11.55 & 1.796 & 1.985 & 15.65 & 0.7832 & 0.7832 & 0.1025 & 1.982 & 2.224 & 252.5 & 0 & 6.708 & 0.3917 & 1.129 & $-$0.2809 & 0.5836 & 0.3483 \\
11.87 & 1.867 & 2.075 & 15.51 & 0.8086 & 0.8086 & 0.1052 & 2.159 & 2.347 & 264.4 & 0 & 7.631 & 0.4209 & 1.206 & $-$0.2868 & 0.6042 & 0.3408 \\
12.2 & 1.931 & 2.156 & 15.37 & 0.8316 & 0.8316 & 0.1075 & 2.324 & 2.457 & 273.7 & 0 & 8.455 & 0.4489 & 1.28 & $-$0.2922 & 0.6227 & 0.3339 \\
12.54 & 1.987 & 2.229 & 15.25 & 0.852 & 0.852 & 0.1095 & 2.474 & 2.552 & 280.6 & 0 & 9.176 & 0.4751 & 1.35 & $-$0.2968 & 0.639 & 0.3279 \\

\noalign{\hrule height 1pt}
\end{tabular}
}

\end{adjustwidth}
\end{table}

\begin{table}[H]\ContinuedFloat

\caption{\textit{Cont}.}

\begin{adjustwidth}{-\extralength}{0cm}

\resizebox{\fulllength}{!}{%
\begin{tabular}{ccccccccccccccccc}
\noalign{\hrule height 1pt} 
\boldmath{$\rho_c$} & \boldmath{$M$} & \boldmath{$M_0$} & \boldmath{$R_e$} & \boldmath{$\Omega$} & \boldmath{$\Omega_p$} & \boldmath{$T/W$} & \boldmath{$cJ/GM_{\odot}^2$} & \boldmath{$I$} & \boldmath{$h_+$} & \boldmath{$h_-$} & \boldmath{$Z_p$} & \boldmath{$Z_b$} & \boldmath{$Z_f$} & \boldmath{$\omega_c/ \Omega$ }& \boldmath{$r_e$} &\boldmath{ $r_p/r_e$ }\\

\textbf{[$10^{14}$ g cm$^{-3}$]} & \boldmath{[$M/M_{\odot}$]} & \boldmath{[$M/M_{\odot}$]} &  \textbf{[km]} & \textbf{[$10^{4}$s$^{-1}$]}  & \textbf{[$10^{4}$s$^{-1}$]} & & & \textbf{[$10^{45}$g cm$^{2}$]}  & \textbf{[km]} & \textbf{[km]} & & & & & \textbf{[km]} &  \\

\midrule

12.9 & 2.04 & 2.298 & 15.14 & 0.8716 & 0.8716 & 0.1115 & 2.625 & 2.647 & 289.2 & 0 & 9.887 & 0.5013 & 1.422 & $-$0.3014 & 0.6548 & 0.3221 \\
13.26 & 2.092 & 2.367 & 15.03 & 0.8911 & 0.8912 & 0.1134 & 2.781 & 2.742 & 298.7 & 0 & 10.59 & 0.5283 & 1.497 & $-$0.3059 & 0.6701 & 0.3164 \\
13.63 & 2.138 & 2.428 & 14.92 & 0.9091 & 0.9091 & 0.1151 & 2.921 & 2.823 & 306. & 0 & 11.22 & 0.5537 & 1.568 & $-$0.31 & 0.684 & 0.3111 \\
14.01 & 2.178 & 2.482 & 14.82 & 0.9256 & 0.9257 & 0.1166 & 3.046 & 2.893 & 311.5 & 0 & 11.77 & 0.5777 & 1.636 & $-$0.3136 & 0.6965 & 0.3062 \\
14.4 & 2.214 & 2.531 & 14.73 & 0.9412 & 0.9413 & 0.1178 & 3.16 & 2.951 & 315.6 & 0 & 12.27 & 0.6006 & 1.702 & $-$0.3169 & 0.708 & 0.3016 \\
14.8 & 2.248 & 2.576 & 14.63 & 0.9562 & 0.9562 & 0.1189 & 3.267 & 3.003 & 319.2 & 0 & 12.73 & 0.6229 & 1.767 & $-$0.3202 & 0.7188 & 0.2973 \\
15.22 & 2.279 & 2.619 & 14.54 & 0.9709 & 0.971 & 0.12 & 3.374 & 3.054 & 323.8 & 0.08803 & 13.19 & 0.6451 & 1.833 & $-$0.3233 & 0.7293 & 0.293 \\
15.64 & 2.308 & 2.661 & 14.45 & 0.9853 & 0.9854 & 0.1211 & 3.48 & 3.104 & 329.1 & 0.2284 & 13.63 & 0.6674 & 1.9 & $-$0.3264 & 0.7394 & 0.2889 \\
16.08 & 2.334 & 2.696 & 14.36 & 0.9989 & 0.9989 & 0.122 & 3.572 & 3.143 & 332.8 & 0.3501 & 14.02 & 0.6882 & 1.964 & $-$0.3291 & 0.7486 & 0.285 \\
16.53 & 2.356 & 2.728 & 14.28 & 1.012 & 1.012 & 0.1228 & 3.653 & 3.173 & 335.4 & 0.456 & 14.37 & 0.7078 & 2.024 & $-$0.3317 & 0.7571 & 0.2814 \\
17. & 2.376 & 2.755 & 14.19 & 1.024 & 1.024 & 0.1234 & 3.724 & 3.197 & 337. & 0.5515 & 14.68 & 0.7264 & 2.082 & $-$0.3341 & 0.765 & 0.278 \\
17.47 & 2.394 & 2.78 & 14.11 & 1.036 & 1.036 & 0.124 & 3.789 & 3.215 & 338.1 & 0.641 & 14.96 & 0.7443 & 2.138 & $-$0.3363 & 0.7725 & 0.2747 \\
17.96 & 2.409 & 2.802 & 14.03 & 1.047 & 1.047 & 0.1245 & 3.848 & 3.23 & 338.9 & 0.7252 & 15.22 & 0.7615 & 2.193 & $-$0.3383 & 0.7795 & 0.2715 \\
18.46 & 2.423 & 2.822 & 13.95 & 1.058 & 1.058 & 0.1249 & 3.9 & 3.239 & 339.4 & 0.8033 & 15.46 & 0.7782 & 2.246 & $-$0.3403 & 0.7862 & 0.2685 \\
18.98 & 2.435 & 2.839 & 13.87 & 1.069 & 1.069 & 0.1253 & 3.947 & 3.246 & 339.3 & 0.8741 & 15.67 & 0.794 & 2.297 & $-$0.3422 & 0.7925 & 0.2656 \\
19.51 & 2.446 & 2.854 & 13.8 & 1.079 & 1.079 & 0.1256 & 3.987 & 3.247 & 338.9 & 0.9389 & 15.87 & 0.809 & 2.346 & $-$0.344 & 0.7985 & 0.2628 \\
20.06 & 2.454 & 2.866 & 13.72 & 1.089 & 1.089 & 0.1258 & 4.021 & 3.244 & 337.9 & 0.9984 & 16.04 & 0.8233 & 2.394 & $-$0.3456 & 0.8041 & 0.2602 \\
20.62 & 2.462 & 2.877 & 13.65 & 1.099 & 1.099 & 0.126 & 4.05 & 3.239 & 336.7 & 1.052 & 16.19 & 0.837 & 2.439 & $-$0.3472 & 0.8094 & 0.2576 \\
21.2 & 2.468 & 2.886 & 13.58 & 1.109 & 1.109 & 0.1261 & 4.077 & 3.232 & 335.6 & 1.104 & 16.33 & 0.8502 & 2.483 & $-$0.3487 & 0.8145 & 0.2552 \\
21.79 & 2.474 & 2.895 & 13.5 & 1.118 & 1.118 & 0.1262 & 4.101 & 3.223 & 334.3 & 1.153 & 16.47 & 0.863 & 2.526 & $-$0.3501 & 0.8194 & 0.2528 \\
22.4 & 2.478 & 2.902 & 13.43 & 1.127 & 1.127 & 0.1262 & 4.121 & 3.213 & 332.9 & 1.199 & 16.59 & 0.8753 & 2.568 & $-$0.3515 & 0.8241 & 0.2505 \\
23.03 & 2.482 & 2.907 & 13.36 & 1.136 & 1.136 & 0.1262 & 4.138 & 3.2 & 331.4 & 1.242 & 16.7 & 0.8871 & 2.609 & $-$0.3527 & 0.8286 & 0.2483 \\
23.67 & 2.484 & 2.91 & 13.3 & 1.145 & 1.145 & 0.1262 & 4.15 & 3.185 & 329.7 & 1.282 & 16.79 & 0.8983 & 2.648 & $-$0.354 & 0.8329 & 0.2462 \\
24.33 & 2.486 & 2.913 & 13.23 & 1.154 & 1.154 & 0.1262 & 4.16 & 3.169 & 327.8 & 1.319 & 16.88 & 0.909 & 2.685 & $-$0.3551 & 0.8371 & 0.2441 \\
25.02 & 2.487 & 2.914 & 13.16 & 1.162 & 1.162 & 0.126 & 4.165 & 3.149 & 325.4 & 1.352 & 16.95 & 0.919 & 2.72 & $-$0.3562 & 0.8409 & 0.2421 \\
25.72 & 2.486 & 2.914 & 13.1 & 1.17 & 1.17 & 0.1259 & 4.167 & 3.129 & 322.9 & 1.382 & 17. & 0.9283 & 2.753 & $-$0.3572 & 0.8447 & 0.2403 \\
26.44 & 2.486 & 2.913 & 13.03 & 1.178 & 1.179 & 0.1257 & 4.165 & 3.106 & 320.2 & 1.41 & 17.05 & 0.9372 & 2.784 & $-$0.3581 & 0.8482 & 0.2384 \\
27.18 & 2.484 & 2.911 & 12.97 & 1.186 & 1.186 & 0.1255 & 4.161 & 3.083 & 317.4 & 1.435 & 17.09 & 0.9455 & 2.813 & $-$0.359 & 0.8516 & 0.2367 \\
27.94 & 2.482 & 2.907 & 12.91 & 1.194 & 1.194 & 0.1252 & 4.154 & 3.059 & 314.4 & 1.457 & 17.12 & 0.9532 & 2.84 & $-$0.3598 & 0.8549 & 0.2351 \\
28.72 & 2.479 & 2.902 & 12.85 & 1.201 & 1.201 & 0.125 & 4.144 & 3.032 & 311.2 & 1.478 & 17.13 & 0.9603 & 2.865 & $-$0.3606 & 0.858 & 0.2335 \\
29.52 & 2.476 & 2.897 & 12.79 & 1.208 & 1.208 & 0.1247 & 4.132 & 3.006 & 307.9 & 1.496 & 17.15 & 0.967 & 2.889 & $-$0.3612 & 0.861 & 0.2319 \\
30.35 & 2.471 & 2.891 & 12.73 & 1.215 & 1.215 & 0.1244 & 4.117 & 2.978 & 304.6 & 1.513 & 17.15 & 0.9732 & 2.91 & $-$0.3619 & 0.8639 & 0.2305 \\
31.2 & 2.467 & 2.885 & 12.67 & 1.222 & 1.222 & 0.124 & 4.102 & 2.951 & 301.2 & 1.528 & 17.14 & 0.9789 & 2.93 & $-$0.3625 & 0.8667 & 0.2291 \\
32.07 & 2.462 & 2.877 & 12.62 & 1.228 & 1.229 & 0.1237 & 4.084 & 2.921 & 297.7 & 1.542 & 17.13 & 0.9842 & 2.949 & $-$0.363 & 0.8694 & 0.2277 \\
32.97 & 2.457 & 2.869 & 12.56 & 1.235 & 1.235 & 0.1233 & 4.065 & 2.893 & 294.2 & 1.554 & 17.12 & 0.9891 & 2.966 & $-$0.3636 & 0.8719 & 0.2264 \\
33.89 & 2.452 & 2.861 & 12.51 & 1.241 & 1.241 & 0.1229 & 4.044 & 2.864 & 290.6 & 1.565 & 17.1 & 0.9935 & 2.981 & $-$0.364 & 0.8744 & 0.2252 \\
34.84 & 2.445 & 2.851 & 12.45 & 1.247 & 1.248 & 0.1225 & 4.021 & 2.833 & 287. & 1.575 & 17.07 & 0.9976 & 2.995 & $-$0.3644 & 0.8768 & 0.2239 \\
35.82 & 2.438 & 2.84 & 12.4 & 1.254 & 1.254 & 0.1221 & 3.996 & 2.801 & 283.3 & 1.584 & 17.03 & 1.001 & 3.008 & $-$0.3648 & 0.879 & 0.2227 \\
36.82 & 2.433 & 2.832 & 12.35 & 1.26 & 1.26 & 0.1215 & 3.973 & 2.772 & 279.5 & 1.592 & 17. & 1.005 & 3.02 & $-$0.3652 & 0.8815 & 0.2216 \\
37.85 & 2.425 & 2.82 & 12.29 & 1.266 & 1.266 & 0.1211 & 3.945 & 2.739 & 275.6 & 1.599 & 16.95 & 1.008 & 3.03 & $-$0.3655 & 0.8837 & 0.2204 \\
38.91 & 2.419 & 2.811 & 12.24 & 1.272 & 1.272 & 0.1204 & 3.92 & 2.708 & 271.6 & 1.608 & 16.91 & 1.011 & 3.041 & $-$0.3658 & 0.8862 & 0.2193 \\
\noalign{\hrule height 1pt}
\end{tabular}
}

\end{adjustwidth}
\end{table}
\vspace{-8pt}
\begin{table}[H]

\caption {DD2MEVp80-CSS ($c_s$ = 0.9).}

\begin{adjustwidth}{-\extralength}{0cm}

\resizebox{\fulllength}{!}{%
\begin{tabular}{ccccccccccccccccc}
\noalign{\hrule height 1pt} 
\boldmath{$\rho_c$} & \boldmath{$M$} & \boldmath{$M_0$} & \boldmath{$R_e$} & \boldmath{$\Omega$} & \boldmath{$\Omega_p$} & \boldmath{$T/W$} & \boldmath{$cJ/GM_{\odot}^2$} & \boldmath{$I$} & \boldmath{$h_+$} & \boldmath{$h_-$} & \boldmath{$Z_p$} & \boldmath{$Z_b$} & \boldmath{$Z_f$} & \boldmath{$\omega_c/ \Omega$} & \boldmath{$r_e$} & \boldmath{$r_p/r_e$} \\

\textbf{[$10^{14}$ g cm$^{-3}$]} & \boldmath{[$M/M_{\odot}$]} & \boldmath{[$M/M_{\odot}$]} &  \textbf{[km]} & \textbf{[$10^{4}$s$^{-1}$]}  & \textbf{[$10^{4}$s$^{-1}$]} & & & \textbf{[$10^{45}$g cm$^{2}$]}  & \textbf{[km]} & \textbf{[km]} & & & & & \textbf{[km]} &  \\

\midrule
2.636 & 0.263 & 0.2646 & 22.55 & 0.1748 & 0.1748 & 0.02052 & 0.03515 & 0.1768 & 11.05 & 0 & 0 & 0.02689 & 0.1643 & $-$0.1105 & 0.09535 & 0.6039 \\
2.709 & 0.3007 & 0.3033 & 21.48 & 0.2012 & 0.2013 & 0.02569 & 0.0491 & 0.2144 & 16.05 & 0 & 0 & 0.03257 & 0.1847 & $-$0.1194 & 0.1066 & 0.5731 \\
2.785 & 0.3455 & 0.3494 & 20.7 & 0.2283 & 0.2283 & 0.0315 & 0.06857 & 0.264 & 23.18 & 0 & 0 & 0.03923 & 0.2073 & $-$0.1286 & 0.1193 & 0.5499 \\
2.862 & 0.3988 & 0.4047 & 20.14 & 0.2558 & 0.2558 & 0.03799 & 0.09601 & 0.3299 & 33.33 & 0 & 0 & 0.04709 & 0.2328 & $-$0.1382 & 0.1338 & 0.5326 \\
2.942 & 0.4625 & 0.4712 & 19.78 & 0.2836 & 0.2836 & 0.04515 & 0.1348 & 0.4177 & 47.68 & 0 & 0 & 0.05641 & 0.2616 & $-$0.1482 & 0.1505 & 0.52 \\
3.024 & 0.5798 & 0.5947 & 19.57 & 0.3235 & 0.3235 & 0.05702 & 0.2242 & 0.609 & 80.76 & 0 & 0 & 0.07323 & 0.3112 & $-$0.1635 & 0.1773 & 0.5094 \\
3.108 & 0.7982 & 0.8279 & 19.97 & 0.3699 & 0.3699 & 0.07636 & 0.4584 & 1.089 & 170.9 & 0 & 0 & 0.103 & 0.3943 & $-$0.1851 & 0.2184 & 0.511 \\
3.195 & 1.032 & 1.082 & 20.72 & 0.3998 & 0.3998 & 0.09382 & 0.8093 & 1.779 & 307.1 & 0 & 0 & 0.1336 & 0.4765 & $-$0.203 & 0.2574 & 0.5212 \\
3.284 & 1.166 & 1.229 & 21.15 & 0.4125 & 0.4126 & 0.1024 & 1.055 & 2.247 & 400.6 & 0 & 0 & 0.151 & 0.5224 & $-$0.2117 & 0.2799 & 0.5272 \\
3.375 & 1.193 & 1.259 & 21.22 & 0.4155 & 0.4155 & 0.1039 & 1.107 & 2.341 & 418.7 & 0 & 0 & 0.1548 & 0.5323 & $-$0.2135 & 0.2865 & 0.5277 \\
3.469 & 1.193 & 1.258 & 21.21 & 0.4158 & 0.4158 & 0.1037 & 1.104 & 2.335 & 416.9 & 0 & 0 & 0.1548 & 0.5322 & $-$0.2135 & 0.2875 & 0.5274 \\
3.566 & 1.192 & 1.257 & 21.2 & 0.4158 & 0.4158 & 0.1036 & 1.102 & 2.33 & 415.8 & 0 & 0 & 0.1547 & 0.532 & $-$0.2134 & 0.2877 & 0.5273 \\
3.665 & 1.192 & 1.258 & 21.2 & 0.4158 & 0.4158 & 0.1037 & 1.103 & 2.332 & 416.4 & 0 & 0 & 0.1547 & 0.5321 & $-$0.2135 & 0.2876 & 0.5273 \\
3.767 & 1.193 & 1.259 & 21.21 & 0.4157 & 0.4157 & 0.1038 & 1.106 & 2.338 & 417.7 & 0 & 0 & 0.1548 & 0.5324 & $-$0.2135 & 0.2873 & 0.5275 \\
3.872 & 1.194 & 1.259 & 21.22 & 0.4156 & 0.4157 & 0.1038 & 1.107 & 2.34 & 418.3 & 0 & 0 & 0.1548 & 0.5324 & $-$0.2135 & 0.287 & 0.5276 \\
3.98 & 1.193 & 1.259 & 21.21 & 0.4157 & 0.4157 & 0.1038 & 1.106 & 2.339 & 417.9 & 0 & 0 & 0.1548 & 0.5324 & $-$0.2135 & 0.2872 & 0.5275 \\
4.091 & 1.193 & 1.259 & 21.21 & 0.4158 & 0.4158 & 0.1037 & 1.105 & 2.337 & 417.4 & 0 & 0 & 0.1548 & 0.5323 & $-$0.2135 & 0.2874 & 0.5274 \\
4.205 & 1.192 & 1.258 & 21.2 & 0.4158 & 0.4158 & 0.1037 & 1.104 & 2.333 & 416.7 & 0 & 0 & 0.1547 & 0.5322 & $-$0.2135 & 0.2876 & 0.5273 \\
4.322 & 1.192 & 1.257 & 21.2 & 0.4158 & 0.4158 & 0.1036 & 1.102 & 2.329 & 415.7 & 0 & 0 & 0.1546 & 0.5319 & $-$0.2134 & 0.2877 & 0.5273 \\
4.442 & 1.191 & 1.256 & 21.19 & 0.4158 & 0.4158 & 0.1035 & 1.1 & 2.325 & 414.7 & 0 & 0 & 0.1546 & 0.5317 & $-$0.2134 & 0.2879 & 0.5271 \\
4.566 & 1.19 & 1.255 & 21.19 & 0.4158 & 0.4158 & 0.1034 & 1.097 & 2.319 & 413.5 & 0 & 0 & 0.1544 & 0.5313 & $-$0.2133 & 0.288 & 0.527 \\
4.693 & 1.188 & 1.253 & 21.18 & 0.4158 & 0.4158 & 0.1033 & 1.094 & 2.313 & 412.2 & 0 & 0 & 0.1543 & 0.531 & $-$0.2133 & 0.2881 & 0.5268 \\
4.824 & 1.187 & 1.252 & 21.17 & 0.4157 & 0.4157 & 0.1032 & 1.091 & 2.306 & 410.6 & 0 & 0 & 0.1541 & 0.5305 & $-$0.2132 & 0.2882 & 0.5267 \\
4.958 & 1.185 & 1.25 & 21.16 & 0.4157 & 0.4157 & 0.103 & 1.088 & 2.3 & 409.2 & 0 & 0 & 0.1539 & 0.53 & $-$0.2131 & 0.2882 & 0.5265 \\
5.096 & 1.184 & 1.248 & 21.15 & 0.4156 & 0.4156 & 0.1029 & 1.084 & 2.292 & 407.7 & 0 & 0 & 0.1537 & 0.5295 & $-$0.213 & 0.2883 & 0.5264 \\
5.238 & 1.181 & 1.246 & 21.14 & 0.4155 & 0.4156 & 0.1027 & 1.08 & 2.283 & 405.7 & 0 & 0 & 0.1535 & 0.5289 & $-$0.2129 & 0.2883 & 0.5262 \\
5.384 & 1.179 & 1.243 & 21.13 & 0.4155 & 0.4155 & 0.1025 & 1.074 & 2.273 & 403.5 & 0 & 0 & 0.1532 & 0.5282 & $-$0.2127 & 0.2883 & 0.5259 \\

\noalign{\hrule height 1pt}
\end{tabular}
}

\end{adjustwidth}
\end{table}

\begin{table}[H]\ContinuedFloat

\caption{\textit{Cont}.}

\begin{adjustwidth}{-\extralength}{0cm}

\resizebox{\fulllength}{!}{%
\begin{tabular}{ccccccccccccccccc}
\noalign{\hrule height 1pt} 
\boldmath{$\rho_c$} & \boldmath{$M$} & \boldmath{$M_0$} & \boldmath{$R_e$} & \boldmath{$\Omega$} & \boldmath{$\Omega_p$} & \boldmath{$T/W$} & \boldmath{$cJ/GM_{\odot}^2$} & \boldmath{$I$} & \boldmath{$h_+$} & \boldmath{$h_-$} & \boldmath{$Z_p$} & \boldmath{$Z_b$} & \boldmath{$Z_f$} & \boldmath{$\omega_c/ \Omega$} & \boldmath{$r_e$} & \boldmath{$r_p/r_e$} \\

\textbf{[$10^{14}$ g cm$^{-3}$]} & \boldmath{[$M/M_{\odot}$]} & \boldmath{[$M/M_{\odot}$]} &  \textbf{[km]} & \textbf{[$10^{4}$s$^{-1}$]}  & \textbf{[$10^{4}$s$^{-1}$]} & & & \textbf{[$10^{45}$g cm$^{2}$]}  & \textbf{[km]} & \textbf{[km]} & & & & & \textbf{[km]} &  \\

\midrule

5.534 & 1.176 & 1.241 & 21.12 & 0.4154 & 0.4154 & 0.1023 & 1.069 & 2.261 & 401.1 & 0 & 0 & 0.1529 & 0.5274 & $-$0.2126 & 0.2883 & 0.5257 \\

5.688 & 1.174 & 1.237 & 21.1 & 0.4152 & 0.4153 & 0.1021 & 1.063 & 2.249 & 398.5 & 0 & 0 & 0.1526 & 0.5265 & $-$0.2124 & 0.2883 & 0.5254 \\

5.847 & 1.171 & 1.234 & 21.09 & 0.4151 & 0.4151 & 0.1019 & 1.056 & 2.236 & 395.8 & 0  & 0 & 0.1522 & 0.5255 & $-$0.2123 & 0.2883 & 0.5252 \\

6.01 & 1.167 & 1.23 & 21.07 & 0.415 & 0.415 & 0.1016 & 1.05 & 2.223 & 392.9 & 0 & 0. & 0.1518 & 0.5245 & $-$0.2121 & 0.2882 & 0.5249 \\
6.177 & 1.164 & 1.227 & 21.06 & 0.4148 & 0.4149 & 0.1014 & 1.042 & 2.208 & 389.8 & 0 & 0 & 0.1514 & 0.5234 & $-$0.2119 & 0.2881 & 0.5246 \\
6.349 & 1.16 & 1.223 & 21.04 & 0.4147 & 0.4147 & 0.1011 & 1.035 & 2.193 & 386.6 & 0 & 0 & 0.151 & 0.5223 & $-$0.2117 & 0.288 & 0.5243 \\

6.526 & 1.156 & 1.218 & 21.02 & 0.4145 & 0.4145 & 0.1008 & 1.027 & 2.177 & 383.2 & 0 & 0 & 0.1505 & 0.521 & $-$0.2114 & 0.2879 & 0.5239 \\
6.708 & 1.152 & 1.214 & 21. & 0.4143 & 0.4143 & 0.1004 & 1.018 & 2.16 & 379.7 & 0 & 0 & 0.15 & 0.5197 & $-$0.2112 & 0.2878 & 0.5236 \\
6.895 & 1.148 & 1.209 & 20.98 & 0.4141 & 0.4141 & 0.1001 & 1.009 & 2.142 & 375.9 & 0 & 0 & 0.1495 & 0.5184 & $-$0.2109 & 0.2876 & 0.5232 \\
7.087 & 1.143 & 1.204 & 20.96 & 0.4139 & 0.4139 & 0.09976 & 0.9999 & 2.123 & 372.1 & 0 & 0 & 0.149 & 0.5169 & $-$0.2107 & 0.2875 & 0.5228 \\
7.284 & 1.138 & 1.199 & 20.93 & 0.4136 & 0.4137 & 0.09938 & 0.9903 & 2.104 & 368. & 0 & 0 & 0.1484 & 0.5154 & $-$0.2104 & 0.2873 & 0.5224 \\
7.487 & 1.133 & 1.193 & 20.91 & 0.4134 & 0.4134 & 0.09898 & 0.9802 & 2.084 & 363.8 & 0 & 0 & 0.1478 & 0.5138 & $-$0.2101 & 0.2871 & 0.522 \\
7.695 & 1.152 & 1.214 & 21. & 0.4143 & 0.4143 & 0.1005 & 1.018 & 2.16 & 379.8 & 0 & 0 & 0.1501 & 0.5198 & $-$0.2112 & 0.2878 & 0.5236 \\
7.91 & 1.154 & 1.216 & 21.01 & 0.4144 & 0.4144 & 0.1006 & 1.023 & 2.169 & 381.7 & 0 & 0 & 0.1503 & 0.5205 & $-$0.2113 & 0.2879 & 0.5238 \\
8.13 & 1.063 & 1.116 & 20.57 & 0.4099 & 0.4099 & 0.09335 & 0.8447 & 1.811 & 307.4 & 0 & 0 & 0.1395 & 0.4917 & $-$0.2058 & 0.2837 & 0.5157 \\
8.356 & 0.8885 & 0.9258 & 19.21 & 0.4127 & 0.4128 & 0.07483 & 0.5352 & 1.14 & 169.3 & 0 & 0 & 0.1216 & 0.4436 & $-$0.1958 & 0.28 & 0.4865 \\
8.589 & 0.8933 & 0.9322 & 17.78 & 0.4633 & 0.4633 & 0.06748 & 0.4941 & 0.9373 & 125.2 & 0 & 0 & 0.1337 & 0.4742 & $-$0.2017 & 0.3095 & 0.4473 \\
8.828 & 0.9929 & 1.044 & 16.88 & 0.5271 & 0.5272 & 0.06968 & 0.5896 & 0.983 & 125.6 & 0. & 0 & 0.1614 & 0.545 & $-$0.2144 & 0.3506 & 0.4183 \\
9.074 & 1.124 & 1.193 & 16.33 & 0.5883 & 0.5883 & 0.07514 & 0.7518 & 1.123 & 140.3 & 0. & 0 & 0.196 & 0.6323 & $-$0.2281 & 0.3937 & 0.3976 \\
9.327 & 1.263 & 1.354 & 15.99 & 0.6425 & 0.6425 & 0.08153 & 0.9571 & 1.309 & 161.1 & 0 & 0.7999 & 0.2337 & 0.7273 & $-$0.241 & 0.4353 & 0.382 \\
9.587 & 1.399 & 1.516 & 15.76 & 0.6895 & 0.6895 & 0.08786 & 1.19 & 1.517 & 184.3 & 0 & 2.432 & 0.2728 & 0.8257 & $-$0.2526 & 0.4738 & 0.3696 \\
9.854 & 1.529 & 1.672 & 15.61 & 0.7302 & 0.7302 & 0.09385 & 1.442 & 1.736 & 208.9 & 0 & 3.984 & 0.3123 & 0.926 & $-$0.2631 & 0.509 & 0.3593 \\
10.13 & 1.649 & 1.819 & 15.48 & 0.766 & 0.766 & 0.09922 & 1.702 & 1.952 & 232.5 & 0 & 5.425 & 0.3515 & 1.027 & $-$0.2725 & 0.5408 & 0.3502 \\
10.41 & 1.758 & 1.955 & 15.38 & 0.7971 & 0.7971 & 0.1039 & 1.958 & 2.159 & 253.9 & 0 & 6.732 & 0.3896 & 1.126 & $-$0.2807 & 0.5691 & 0.3421 \\
10.7 & 1.859 & 2.083 & 15.29 & 0.8255 & 0.8255 & 0.1083 & 2.218 & 2.361 & 276.2 & 0 & 7.975 & 0.4276 & 1.227 & $-$0.2883 & 0.5952 & 0.3347 \\
11. & 1.948 & 2.197 & 15.2 & 0.8501 & 0.8502 & 0.1119 & 2.457 & 2.54 & 293.9 & 0 & 9.056 & 0.4632 & 1.323 & $-$0.2949 & 0.6179 & 0.328 \\
11.3 & 2.03 & 2.302 & 15.12 & 0.8729 & 0.8729 & 0.1152 & 2.691 & 2.709 & 311.2 & 0 & 10.07 & 0.4981 & 1.418 & $-$0.301 & 0.6386 & 0.3217 \\
11.62 & 2.105 & 2.402 & 15.04 & 0.8945 & 0.8946 & 0.1182 & 2.923 & 2.871 & 328.7 & 0 & 11.03 & 0.5328 & 1.516 & $-$0.3068 & 0.6579 & 0.3156 \\
11.94 & 2.173 & 2.492 & 14.96 & 0.9144 & 0.9144 & 0.1208 & 3.141 & 3.019 & 344.6 & 0 & 11.9 & 0.5662 & 1.612 & $-$0.3121 & 0.6754 & 0.31 \\
12.28 & 2.231 & 2.57 & 14.89 & 0.932 & 0.9321 & 0.123 & 3.329 & 3.139 & 356. & 0. & 12.64 & 0.5969 & 1.701 & $-$0.3166 & 0.6907 & 0.3048 \\
12.62 & 2.283 & 2.64 & 14.81 & 0.9483 & 0.9483 & 0.1249 & 3.502 & 3.246 & 365.8 & 0. & 13.31 & 0.626 & 1.787 & $-$0.3207 & 0.7046 & 0.3 \\
12.97 & 2.331 & 2.705 & 14.74 & 0.9641 & 0.9641 & 0.1267 & 3.672 & 3.347 & 376.4 & 0.1583 & 13.95 & 0.6549 & 1.874 & $-$0.3247 & 0.7177 & 0.2954 \\
13.33 & 2.375 & 2.767 & 14.67 & 0.9792 & 0.9792 & 0.1284 & 3.835 & 3.442 & 386.9 & 0.3244 & 14.55 & 0.6831 & 1.961 & $-$0.3286 & 0.7299 & 0.291 \\
13.7 & 2.415 & 2.823 & 14.6 & 0.9937 & 0.9938 & 0.1299 & 3.986 & 3.525 & 396.3 & 0.4739 & 15.1 & 0.7104 & 2.047 & $-$0.3321 & 0.7413 & 0.2867 \\
14.08 & 2.451 & 2.873 & 14.52 & 1.007 & 1.007 & 0.1311 & 4.123 & 3.597 & 404.1 & 0.6053 & 15.6 & 0.7364 & 2.13 & $-$0.3353 & 0.7517 & 0.2827 \\
14.48 & 2.481 & 2.916 & 14.45 & 1.02 & 1.02 & 0.1322 & 4.244 & 3.656 & 410.5 & 0.7246 & 16.04 & 0.761 & 2.211 & $-$0.3383 & 0.7612 & 0.2788 \\
14.88 & 2.51 & 2.956 & 14.38 & 1.033 & 1.033 & 0.1331 & 4.354 & 3.706 & 415.3 & 0.8299 & 16.44 & 0.7844 & 2.288 & $-$0.3411 & 0.77 & 0.2753 \\
15.29 & 2.533 & 2.99 & 14.31 & 1.044 & 1.044 & 0.1339 & 4.448 & 3.743 & 418.8 & 0.9236 & 16.79 & 0.8062 & 2.361 & $-$0.3436 & 0.7781 & 0.2719 \\
15.72 & 2.554 & 3.02 & 14.24 & 1.055 & 1.055 & 0.1346 & 4.531 & 3.774 & 421.3 & 1.007 & 17.1 & 0.827 & 2.431 & $-$0.3459 & 0.7856 & 0.2687 \\
16.16 & 2.572 & 3.045 & 14.17 & 1.066 & 1.066 & 0.1351 & 4.605 & 3.796 & 423.1 & 1.083 & 17.38 & 0.8467 & 2.499 & $-$0.3482 & 0.7926 & 0.2656 \\
16.61 & 2.588 & 3.069 & 14.1 & 1.076 & 1.077 & 0.1356 & 4.67 & 3.813 & 424.2 & 1.153 & 17.64 & 0.8655 & 2.564 & $-$0.3502 & 0.7991 & 0.2627 \\
17.07 & 2.602 & 3.089 & 14.03 & 1.086 & 1.087 & 0.136 & 4.727 & 3.824 & 424.9 & 1.218 & 17.87 & 0.8834 & 2.627 & $-$0.3521 & 0.8053 & 0.2599 \\
17.54 & 2.615 & 3.108 & 13.96 & 1.096 & 1.096 & 0.1363 & 4.783 & 3.835 & 425.6 & 1.279 & 18.08 & 0.9008 & 2.689 & $-$0.3539 & 0.8112 & 0.2573 \\
18.03 & 2.626 & 3.124 & 13.89 & 1.106 & 1.106 & 0.1366 & 4.831 & 3.84 & 426.1 & 1.336 & 18.28 & 0.9175 & 2.749 & $-$0.3557 & 0.8168 & 0.2547 \\
18.54 & 2.635 & 3.137 & 13.83 & 1.115 & 1.115 & 0.1369 & 4.873 & 3.841 & 426.2 & 1.389 & 18.46 & 0.9334 & 2.808 & $-$0.3573 & 0.8221 & 0.2522 \\
19.05 & 2.644 & 3.15 & 13.76 & 1.124 & 1.124 & 0.1371 & 4.911 & 3.84 & 426.1 & 1.438 & 18.62 & 0.9486 & 2.864 & $-$0.3589 & 0.8271 & 0.2499 \\
19.58 & 2.65 & 3.159 & 13.7 & 1.133 & 1.133 & 0.1372 & 4.941 & 3.834 & 425.5 & 1.482 & 18.76 & 0.9629 & 2.917 & $-$0.3603 & 0.8318 & 0.2476 \\
20.13 & 2.656 & 3.168 & 13.64 & 1.141 & 1.141 & 0.1373 & 4.967 & 3.826 & 424.4 & 1.521 & 18.88 & 0.9765 & 2.969 & $-$0.3617 & 0.8363 & 0.2455 \\
20.69 & 2.66 & 3.174 & 13.58 & 1.149 & 1.149 & 0.1374 & 4.987 & 3.814 & 423.1 & 1.558 & 18.99 & 0.9893 & 3.017 & $-$0.3629 & 0.8406 & 0.2434 \\
21.26 & 2.663 & 3.178 & 13.52 & 1.157 & 1.157 & 0.1374 & 5.003 & 3.8 & 421.5 & 1.591 & 19.09 & 1.001 & 3.064 & $-$0.3642 & 0.8447 & 0.2414 \\
21.86 & 2.666 & 3.181 & 13.46 & 1.165 & 1.165 & 0.1373 & 5.015 & 3.784 & 419.5 & 1.622 & 19.17 & 1.013 & 3.109 & $-$0.3653 & 0.8485 & 0.2395 \\
22.47 & 2.666 & 3.183 & 13.39 & 1.173 & 1.173 & 0.1373 & 5.022 & 3.764 & 417.2 & 1.651 & 19.24 & 1.024 & 3.151 & $-$0.3664 & 0.8522 & 0.2377 \\
23.09 & 2.667 & 3.183 & 13.34 & 1.18 & 1.18 & 0.1371 & 5.025 & 3.743 & 414.6 & 1.677 & 19.3 & 1.034 & 3.191 & $-$0.3673 & 0.8557 & 0.2359 \\
23.73 & 2.666 & 3.182 & 13.28 & 1.187 & 1.187 & 0.137 & 5.025 & 3.72 & 411.8 & 1.701 & 19.35 & 1.044 & 3.228 & $-$0.3683 & 0.8591 & 0.2342 \\
24.4 & 2.665 & 3.18 & 13.22 & 1.194 & 1.194 & 0.1368 & 5.021 & 3.695 & 408.7 & 1.723 & 19.38 & 1.053 & 3.264 & $-$0.3691 & 0.8623 & 0.2326 \\
25.08 & 2.663 & 3.177 & 13.16 & 1.201 & 1.201 & 0.1365 & 5.014 & 3.668 & 405.4 & 1.743 & 19.41 & 1.061 & 3.297 & $-$0.37 & 0.8653 & 0.231 \\
25.77 & 2.66 & 3.173 & 13.11 & 1.208 & 1.208 & 0.1363 & 5.004 & 3.64 & 401.9 & 1.761 & 19.42 & 1.069 & 3.327 & $-$0.3707 & 0.8682 & 0.2295 \\
26.49 & 2.657 & 3.168 & 13.05 & 1.215 & 1.215 & 0.1359 & 4.991 & 3.611 & 398.1 & 1.777 & 19.43 & 1.076 & 3.356 & $-$0.3714 & 0.8711 & 0.2281 \\
27.23 & 2.653 & 3.162 & 13. & 1.221 & 1.221 & 0.1356 & 4.975 & 3.581 & 394.2 & 1.791 & 19.43 & 1.083 & 3.382 & $-$0.372 & 0.8737 & 0.2267 \\
27.99 & 2.648 & 3.154 & 12.94 & 1.227 & 1.227 & 0.1353 & 4.956 & 3.549 & 390.3 & 1.805 & 19.43 & 1.089 & 3.407 & $-$0.3726 & 0.8763 & 0.2254 \\
\noalign{\hrule height 1pt}

\end{tabular}
}

\end{adjustwidth}

\end{table}

\begin{adjustwidth}{-\extralength}{0cm}

\reftitle{References}






\PublishersNote{}

\end{adjustwidth}
\end{document}